\documentclass[nohyper,12pt,letterpaper]{JHEP3}
\usepackage{graphicx,epsfig,youngtab}

\author{Robert de Mello Koch$^{1,2},$ Matthias Dessein$^1$, Dimitrios Giataganas$^1$ and Christopher Mathwin$^{1}$\\
$^{1}$ National Institute for Theoretical Physics,\\
Department of Physics and Centre for Theoretical Physics,\\ 
University of the Witwatersrand,\\ 
Wits, 2050, South Africa\\
\qquad\\
$^{2}$Stellenbosch Institute for Advanced Studies,\\
Stellenbosch, South Africa\\
\qquad\\
E-mail: \email{robert@neo.phys.wits.ac.za, matthias.dessein@students.wits.ac.za, dimitrios.giataganas@wits.ac.za, christopher.mathwin@students.wits.ac.za}}

\abstract{
We study the action of the dilatation operator on restricted Schur polynomials
labeled by Young diagrams with $p$ long columns or $p$ long rows. A new version 
of Schur-Weyl duality provides a powerful approach to the computation
and manipulation of the symmetric group operators appearing in the restricted Schur 
polynomials. Using this new technology, we are able to evaluate the action of the 
one loop dilatation operator. The result has a direct and natural connection
to the Gauss Law constraint for branes with a compact world volume. We find considerable
evidence that the dilatation operator reduces to a decoupled set of harmonic
oscillators. This strongly suggests that integrability in ${\cal N}=4$ super Yang-Mills
theory is not just a feature of the planar limit, but extends to other large $N$ but 
non-planar limits.
}

\preprint{WITS-CTP-078}

\title{Giant Graviton Oscillators}


\begin{document}

\section{Introduction and Conclusions}

Integrability has proven to be a powerful tool in analyzing ${\cal N} = 4$ super Yang-Mills theory 
in the planar limit\cite{Minahan:2002ve,Beisert:2010jr}. 
An interesting question is whether or not integrability is present in other large $N$ limits of the theory. 

Our focus in this article is on operators that have a bare dimension of order $N$. For these operators the large $N$ limit
of correlation functions is not captured by summing the planar diagrams. Indeed, huge combinatoric factors (arising from
the number of ways one can form the Feynman diagrams out of so many fields) enhance the non-planar contributions and 
completely overpower the usual ${1\over N^2}$ suppression of non-planar diagrams\cite{Balasubramanian:2001nh}. 
One is faced with the daunting task of
having to sum a lot more than just the planar diagrams. In an inspired article, \cite{Corley:2001zk} have shown how all possible diagrams
can be summed, at least in the free field theory and in a ${1\over 2}$-BPS sector. 
By changing from the trace basis to the basis of Schur polynomials one
finds that the two point function of the theory is diagonal in the labels of the Schur polynomial and that the higher
point correlators of Schur polynomials have an extremely simple form, being expressed in terms of quantities that are
familiar from representation theory. Soon after this initial work, an elegant explanation of the results of \cite{Corley:2001zk} were given
in terms of projection operators\cite{SS}. One of the basic observations made in \cite{SS} is the fact that two point functions of operators
of the form
$$
  \hat{A}_n\equiv  A^{i_1\, i_2\cdots i_n}_{j_1\, j_2\cdots j_n}Z^{j_1}_{i_1}Z^{j_2}_{i_2}\cdots Z^{j_n}_{i_n}={\rm Tr}
     (AZ^{\otimes n})
$$
are given by
$$
  \left\langle \hat{A}_n\hat{B}_n^\dagger\right\rangle=\sum_{\sigma\in S_n}{\rm Tr}(\sigma A\sigma^{-1}B^\dagger)\, .
$$
By choosing $A$ and $B$ to be projection operators projecting onto irreducible representations of the symmetric group,
they clearly commute with $\sigma$ (rendering the above sum trivial) and are orthogonal. With this choice for $A$,
$\hat{A}_n$ is nothing but a Schur polynomial, so that we obtain a rather simple understanding of how and why the
Schur polynomials diagonalize the two point function. 

According to the AdS/CFT correspondence\cite{Maldacena:1997re}, 
these operators in the ${\cal N}=4$ super Yang-Mills theory will have a dual
interpretation in IIB string theory on asymptotically AdS$_5\times$S$^5$ backgrounds. Certain Schur polynomials containing 
order $N$ $Z$s were quickly identified\cite{Balasubramanian:2001nh,Corley:2001zk,Balasubramanian:2002sa,Berenstein:2003ah} 
with giant gravitons\cite{McGreevy:2000cw}, while Schur polynomials with order $N^2$ fields were identified 
with ${1\over 2}$-BPS geometries\cite{Lin:2004nb,Berenstein:2004kk}. Giant gravitons 
are $D3$ branes with a spherical world volume, stabilized by their angular momentum \cite{McGreevy:2000cw}. 
Excited $D$-brane states can be described in terms of open strings which end on the $D$-brane. Operators dual to excited giant 
gravitons were proposed in \cite{Balasubramanian:2004nb}. Since giant gravitons have a compact world volume, Gauss' Law forces the 
total charge on the worldvolume to vanish\cite{Sadri:2003mx}. A highly non-trivial test of the proposal of \cite{Balasubramanian:2004nb} is that the
number of operators that can be defined matches the number of states obeying 
this Gauss Law constraint. The operators of \cite{Balasubramanian:2004nb} are defined in terms of 
symmetric group operators that project from the carrier space of some irreducible representation of the symmetric group 
to a subspace defined using the carrier space of an irreducible representation of a subgroup. 
Although the construction of the operators proposed in \cite{Balasubramanian:2004nb}
is a highly non-trivial problem in the representation theory of the symmetric group, the two point functions of these operators, the 
{\it restricted Schur polynomials}, were computed exactly, in the free field theory limit, in \cite{Bhattacharyya:2008rb}, by exploiting
the technology developed in \cite{de Mello Koch:2007uu,de Mello Koch:2007uv,Bekker:2007ea}. 
It was also shown that the restricted Schur polynomials provide a basis for the gauge invariant
local operators built using only scalar (adjoint Higgs) fields\cite{Bhattacharyya:2008rc}. 
Further, it is a convenient description.  
Indeed, the restricted Schur basis diagonalizes the two point function in the free field theory limit and it mixes
weakly at one loop level\cite{de Mello Koch:2007uv,Bekker:2007ea}. Numerical studies of the dilatation operator, when acting on decoupled
sectors of the theory that have a sphere giant graviton number equal to two showed that the spectrum of the dilatation operator is that of a 
set of decoupled harmonic oscillators \cite{Koch:2010gp,VinceKate}. Using insights gained from these numerical studies, an analytic
study of the dilatation operator in the sector of the theory with either two sphere giants or two AdS giants has been carried out in
\cite{bhw}. The crucial new ingredient in \cite{bhw} is the realization that the problem of 
computing the symmetric group operators needed to define
the restricted Schur polynomial can be performed using an auxiliary spin chain. This is essentially an application of Schur-Weyl duality.
The suggestion that Schur-Weyl duality may play an important role in the study of 
gauge theory/gravity duality was first made in \cite{Ramgoolam:2008yr}.

In this article we will recover the two giant graviton results of \cite{bhw} by clarifying the role of Schur-Weyl duality. An auxiliary 
spin chain will not be used. The advantage of the new approach is that it will allow us to study the $p$ giant graviton sector of the
theory. This generalization is highly non-trivial as we now explain. The two giant graviton problem is too simple to see the full complexity
of the problem. Indeed, the symmetric group operators needed to define the restricted Schur polynomials
in this case are simple because the subspaces they project to appear 
without multiplicity. For $p>2$ giant gravitons, this multiplicity problem must be solved. Our present approach, based on Schur Weyl duality,
allows us to
\begin{itemize}
\item{} Construct the restricted Schur polynomials for the $p$ giant graviton problem using the representation theory of $U(p)$. For the
        case of $p$ sphere giant gravitons we obtain an example of Schur-Weyl duality that is, as far as we know, novel.
\item{} Organize the multiplicity of $S_n\times S_m$ irreducible representations subduced from a given $S_{n+m}$ irreducible representation by
        mapping it into the {\it inner multiplicity} appearing in $U(p)$ representation theory. As far as we know, this connection has not 
        been pointed out in the maths literature, although it follows as a rather simple consequence of the Schur-Weyl duality we have found.
\item{} Evaluate the action of the dilatation operator in terms of known Clebsch-Gordan coefficients of $U(p)$.
\end{itemize}
Thus, we achieve a complete generalization of the results of \cite{bhw} together with a much clearer understanding of the general problem. 
One noteworthy feature of our results is that the action of the one loop dilatation operator has a direct and natural connection
to the Gauss Law constraint we discussed above.
We have not managed to solve the problem of diagonalizing the large $N$ dilatation operator for this class of operators in general.
For the problems that we do manage to solve, we again reproduce the spectrum of a set of decoupled oscillators. This leads us to 
conjecture that the specific large $N$ limit of the dilatation operator that we consider is again integrable.

Although we have focused on the restricted Schur polynomials in this article, they are not the only basis for local gauge invariant 
operators of a matrix model. Another interesting basis to consider is the Brauer basis\cite{Kimura:2007wy,Kimura:2009wy}. This basis 
is built using elements of the Brauer algebra. The structure constants of the Brauer algebra are $N$ dependent. There is an elegant 
construction of a class of BPS operators \cite{yusuke} in which the natural $N$ dependence appearing in the definition of the 
operator\cite{heslop} is reproduced by the Brauer algebra projectors\cite{yusuke}. Alternatively, another natural approach to the problem, 
is to adopt a basis that has sharp quantum numbers for the global symmetries of the theory\cite{Brown:2007xh,Brown:2008rr}. The action of 
the anomalous dimension operator in this sharp quantum number basis is very similar to the action in the restricted Schur basis: again 
operators which mix can differ at most by moving one box around on the Young diagram labeling the operator\cite{Brown:2008rs}.
For further related interesting work see \cite{tomyusuke,Huang:2010ne}.
Finally, for a rather general approach which correctly counts and constructs the weak coupling BPS operators see\cite{jurgis}.
The results obtained in \cite{jurgis} can be translated into any of the bases we have considered.

This article is organized as follows: In section 2 we explain our construction of restricted Schur polynomials. This includes
a detailed description of Schur-Weyl duality and its implications for the study of the dilatation operator of 
${\cal N}=4$ super Yang-Mills theory. In section 3 we describe in detail the action of the dilatation operator. This action
is used in section 4 to write the problem of diagonalizing the dilatation operator as a set of recursion relations.
Section 5 is used for discussion of our results. In particular, in this section we explain how the action of the one loop
dilatation operator is related to the Gauss Law constraint.
We have made an attempt to make the article self contained. For this reason, Appendices \ref{Unrep} and \ref{Snrep} review the 
background representation theory need to develop our construction. Detailed examples which demonstrate how Shur-Weyl duality can 
be used to construct the restricted Schur projectors are given in Appendix \ref{examplesofprojectors}. We give the details of the evaluation
of the dilatation operator in Appendix \ref{dilatationoperator} in general and give the details for specific examples in Appendix
\ref{explicitdilatationoperator}. Useful recursion relations are summarized in Appendix \ref{labelrelations}.
In Appendix \ref{gausslawexample} we report the result of the computation of the action of the dilatation operator 
for an example that demonstrates the link to the Gauss Law constraint very clearly. Finally, in Appendix \ref{continuumlimit}
we study a continuum limit of the dilatation operator. In this limit the dilatation operator reduces to a set of decoupled
oscillators.

\section{Constructing Restricted Schur Polynomials}

In this article we will diagonalize the dilatation operator within large sectors of decoupled states.
Each sector comprises restricted Schur polynomials with a fixed number $p$ of rows or columns.
Mixing with restricted Schur polynomials that have $n\ne p$ 
rows or columns (or of even more general shape) is suppressed at least by a factor of order\footnote{Mixing 
at the quantum level. There is no mixing in the free theory\cite{Bhattacharyya:2008rb}.}\cite{Koch:2010gp}. 
To achieve this a key new idea is needed: Schur-Weyl duality is used to construct the restricted Schur polynomials. 
In this section we will explain how Schur-Weyl duality arises and how it is exploited.

\subsection{Why it is difficult to build a Restricted Schur Polynomial}

There are six scalar fields $\phi^i{}_{ab}$ taking values in the adjoint of $u(N)$ in ${\cal N}=4$ super Yang Mills theory.
Assemble these scalars into the three complex combinations
$$
  Z=\phi_1+i\phi_2 ,\qquad Y=\phi_3+i\phi_4 ,\qquad X=\phi_5+i\phi_6 \, .
$$
We will study restricted Schur polynomials built using $n\sim O(N)$ $Z$ and $m\sim O(N)$ $Y$ fields and will often refer 
to the $Y$ fields as ``impurities''.
These operators have a large ${\cal R}$-charge and belong to the $SU(2)$ sector of the theory.
The definition of the restricted Schur polynomial is 
\begin{equation}
\chi_{R,(r,s)jk}(Z,Y)={1\over n!m!}\sum_{\sigma\in S_{n+m}}\chi_{R,(r,s),jk}(\sigma )
Y^{i_1}_{i_{\sigma (1)}}\cdots Y^{i_m}_{i_{\sigma (m)}}
Z^{i_{m+1}}_{i_{\sigma (m+1)}}\cdots Z^{i_{n+m}}_{i_{\sigma (n+m)}}\, .
\label{restrictedschur}
\end{equation}
In this definition $R$ is a Young diagram with $n+m$ boxes and hence labels an irreducible representation of $S_{n+m}$,
$r$ is a Young diagram with $n$ boxes and labels an irreducible representation of $S_{n}$ and
$s$ is a Young diagram with $m$ boxes and labels an irreducible representation of $S_{m}$. The group
$S_{n+m}$ has an $S_n\times S_m$ subgroup. Taken together $r$ and $s$ label an irreducible representation of this subgroup.
A single irreducible representation $R$ will in general subduce many possible representations of the $S_n\times S_m$
subgroup. A particular irreducible representation of the subgroup may be subduced more than once in which case we
must introduce a multiplicity label to keep track of the different copies subduced. The indices $j$ and $k$ appearing above are 
these multiplicity labels. The object $\chi_{R,(r,s)jk}(\sigma )$ is called a restricted 
character\cite{de Mello Koch:2007uu}. To compute the character of group element $\sigma$ in representation $R$, we take the trace of the matrix 
representing $\sigma$ in irreducible representation $R$, $\chi_R(\sigma)={\rm Tr}\left(\Gamma_R(\sigma)\right)$. To compute the 
restricted character $\chi_{R,(r,s),jk}(\sigma )$ trace the row index of $\Gamma_R(\sigma)$ only over the subspace associated to the
$j^{\rm th}$ copy of $(r,s)$ and the column index over the subspace associated to the
$k^{\rm th}$ copy of $(r,s)$. It is now clear why two multiplicity labels appear: when performing the ``trace'' over the
carrier space of $(r,s)$ the row and column indices can come from different copies of $(r,s)$ so that if $i\ne j$ we are not
in fact summing diagonal elements of $\Gamma_R(\sigma)$. Operators constructed by summing these ``off diagonal'' elements
are needed to obtain a complete basis of local operators\cite{Bhattacharyya:2008rc}. 
In terms of the symmetric group operator $P_{R\to (r,s)jk}$ which obeys
$$
\Gamma_{(r,s)j} (\sigma) P_{R\to (r,s)jk} = P_{R\to (r,s)jk}\Gamma_{(r,s)k} (\sigma) \qquad \sigma\in S_n\times S_m
$$
$$
\Gamma_{(r,s)l} (\sigma) P_{R\to (r,s)jk} = 0 = P_{R\to (r,s)jk}\Gamma_{(r,s)q} (\sigma) \quad \sigma\in S_n\times S_m
\quad l\ne j,\quad k\ne q,
$$
we can write the restricted character as
$$
  \chi_{R,(r,s),ji}(\sigma ) = {\rm Tr}\left(P_{R\to (r,s)ji}\Gamma_R(\sigma)\right)\, .
$$
When there are no multiplicities, $P_{R\to (r,s)jk}=P_{R\to (r,s)}$ is a projection operator which projects from the carrier
space of $R$ to the $(r,s)$ subspace. When there are multiplicities $P_{R\to (r,s)jk}$ is an 
intertwiner\cite{OVApproach}. However,
it is constructed in essentially the same way as a projector and satisfies very similar identities. For these reasons
we will sometimes be guilty of an abuse of language and refer to $P_{R\to (r,s)jk}$ simply as a projector even when there are multiplicities.    

{\vskip 0.25cm}

\noindent
{\bf Key Idea:} {\sl It is not easy to construct the operator $P_{R\to (r,s)jk}$ explicitly. This is the most serious obstacle in working with 
restricted Schur polynomials. An important result of this article is the use of a new version of Schur-Weyl duality to 
provide an efficient, transparent construction of this operator.}

{\vskip 0.25cm}

\noindent
Our construction is not quite completely general, but it does capture many interesting situations and should be a useful
tool to explore semi-classical physics dual to the restricted Schur polynomials.

The restricted Schur polynomials are a very convenient basis for gauge invariant operators in the theory built using only the
adjoint scalars. This follows because
\begin{itemize}
\item{}   The restricted Schur polynomials are complete in the sense that any multitrace operator or linear combination
          of multitrace operators can be written as a linear combination of restricted Schur polynomials\cite{Bhattacharyya:2008rc}.
\item{}   The free theory two point function of the restricted Schur polynomial has been computed exactly\cite{Bhattacharyya:2008rb}
          \begin{equation}
                \langle\chi_{R,(r,s)jk}(Z,Y)\chi_{T,(t,u)lm}(Z,Y)^\dagger\rangle =
                \delta_{R,(r,s)\,T,(t,u)}\delta_{kl}\delta_{jm}f_R {{\rm hooks}_R\over {\rm hooks}_{r}\, {\rm hooks}_s}\, .
          \label{restschurtwopoint}
          \end{equation}
          In this expression $f_R$ is the product of the factors in Young diagram $R$ and ${\rm hooks}_R$ is the product of the hook
          lengths of Young diagram $R$\footnote{See section \ref{lastremarks}
          for a definition of factors and hook lengths for a Young diagram.}.
          The fact that this two point function is known exactly as a function of $N$, implies that all Feynman diagrams (not just the
          planar diagrams) have been summed and this is what allows one to go beyond the planar limit.
\item{}   Restricted Schur polynomials have highly constrained mixing at the quantum level\cite{de Mello Koch:2007uv,Bekker:2007ea}.          
\end{itemize}
Our goal for the rest of this section is to build a basis from the carrier space of an $S_{n+m}$ irreducible 
representation $R$ for the carrier space of an $S_n\times S_m$ irreducible representation $(r,s)j$.
It is then a small step to build $P_{R\to (r,s)jk}$.
We accomplish the construction in two steps: First we project from $S_{n+m}$ to $S_n\times (S_1)^m$
(this is easy) and second, we assemble the $S_n\times (S_1)^m$ representations into $S_n\times S_m$ representations
(this is the trying step). It is this second step that is accomplished using Schur-Weyl duality. As a consequence we learn that
the multiplicity index can be organized using $U(p)$ representations, with $p$ the number of rows or columns in $R$. 
The background material from representation theory
needed to understand this section is collected in Appendices \ref{Unrep} and \ref{Snrep}.

\subsection{From $S_{n+m}$ to $S_n\times (S_1)^m$}

Start from the carrier space for an irreducible representation $R$ of $S_{n+m}$. If we restrict ourselves to an
$S_n\times (S_1)^m$ subgroup this space will decompose into a direct sum of invariant subspaces, each of which is
the carrier space of a particular irreducible representation of the subgroup. In this subsection we will explain how 
to extract a particular $S_n\times (S_1)^m$ invariant subspace from the full carrier space of $R$.

Since $S_1$ has only a single irreducible representation, we need not include it in our labels for the 
irreducible representation of the subgroup. Consequently, to specify an irreducible representation of the 
$S_n\times (S_1)^m$ subgroup, we only need to specify an irreducible representation of $S_n$, that is, a
Young diagram $r$ with $n$ boxes. The only representations $r$ that are subduced by $R$ are those with
Young diagrams that can be obtained by removing $m$ boxes from $R$. Pulling the same set of $m$ boxes off in 
different orders leads to different subspaces which all carry the same irreducible representation $r$.
To resolve this multiplicity, we only need to specify the order in which the boxes are removed.
To specify this order, label the boxes to be removed from $R$ 
with a label ranging from 1 to $m$, such that box 1 is removed first, then box 2 
and so on until box $m$ is removed. Thus, by labeling any given set of boxes in such a way that if we were 
to remove the boxes in numerical order starting with box 1 we would have a legal Young diagram at each step, 
we obtain a partially labeled Young diagram with shape $R$, which represents a subspace carrying an 
irreducible representation of the $S_n\times (S_1)^m$ subgroup. See Appendix \ref{partdiagram} for further discussion.

To build an operator which projects from the carrier space of the $S_{n+m}$ irreducible representation $R$
to the carrier space of an $S_n\times S_m$ irreducible representation $(r,s)j$, we now need to assemble the
partially labeled Young diagrams (which already carry a representation $r$ of $S_n$) in such a way that the
resulting linear combinations carry an irreducible representation of $S_n\times S_m$. We turn to this task
in the next subsection.

\subsection{Basic Idea for Young diagrams with $p$ rows}
\label{prows}

We will consider Young diagrams built using $n+m\sim O(N)$ boxes and with $p$ rows.
Thus, for the generic diagram, each row has $O(N)$ boxes. We set $m=\alpha N$ with $\alpha\ll 1$.
After labeling the $m$ boxes, two labeled boxes with labels $i$ and $j$, that are in different 
rows, will have associated factors $c_i$ and $c_j$ respectively, with $c_i-c_j\sim O(N)$. 

Consider the $S_m$ subgroup of $S_{n+m}$ which acts on the labeled boxes. We can obtain a matrix representation of
this action by thinking about the partially labeled Young diagrams as Young-Yamonouchi
states. As discussed in Appendix \ref{simpleyoung}, the fact that $c_i-c_j\sim O(N)$ for boxes in different rows implies a 
significant simplification in the representations of $S_m$. When adjacent permutations $(i,i+1)$ act on labeled
boxes that belong to the same row, the Young diagram is unchanged and when acting on labeled
boxes that belong to the different rows, the labeled boxes are swapped. 

\begin{figure}[h]
          \centering
          {\epsfig{file=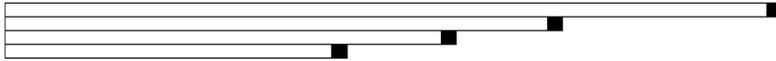,width=10.5cm,height=1.0cm}}
          \caption{An example of a Young diagram with $p=4$ rows. The rows are shown; the columns are not shown. There are
                   $O(N)$ boxes in each row. The $m$ numbered boxes have been colored black.
                   The difference in factors associated to any two numbered boxes that are in different rows is $O(N)$. This is easily
                   seen by recalling that the difference in the factors counts the number of boxes one needs to step
                   through to move between the two boxes. The difference in the number of boxes in any two rows is generically $O(N)$ so that
                   to move from one of the black tips to another one, generically, one needs to step through $O(N)$ boxes.}
 \end{figure}

If we have a Young diagram with $p$ rows and we label $m$ boxes in all possible ways consistent with the rule of the previous subsection, 
we find a total of $p^m$ possible partially labeled Young diagrams. We associate a particular $p$-dimensional vector 
to each box that is labeled. This gives a total of $m$ vectors $\vec{v}(i)$ with $i=1,2,\cdots,m$. 
We will denote the components of these vectors as $\vec{v}(i)_n$ where $n=1,...,p$. If box $i$ is pulled from the $j^{\rm th}$ 
row we have
$$
\vec{v}(i)_n=\delta_{nj}\, .
$$
For each index $i$ (equivalently, for each labeled box) we have a vector space $V_p$. Taking the tensor product of these spaces 
we obtain a set of $p^m$ dimensional vectors, of the form
$$
\vec{v}(1)\otimes\vec{v}(2)\otimes\vec{v}(3)\otimes\cdots \otimes\vec{v}(m-1)\otimes\vec{v}(m)  \, .
$$
Call the vector space spanned by these vectors $V_p^{\otimes m}$.
When we talk about vectors of the above form we will say that ``vector $\vec{v}(i)$ occupies the $i^{\rm th}$ slot.'' 
The matrix action of $S_m$ on the partially labeled Young diagrams described above implies the following action on $V_p^{\otimes m}$
$$
\sigma\cdot\left( \vec{v}(1)\otimes\vec{v}(2)\otimes\cdots \otimes\vec{v}(m)\right)  =
\vec{v}\left(\sigma(1)\right)\otimes\vec{v}\left(\sigma(2)\right)\otimes \cdots 
\otimes\vec{v}\left(\sigma(m)\right) \, .
$$
Thus, $\sigma\in S_m$ will move the vector in the $i^{\rm th}$ slot to the $\sigma(i)^{\rm th}$ slot, but does not
change its value. We can also define an action of $U(p)$ on $V_p^{\otimes m}$ 
$$
U\cdot\left( \vec{v}(1)\otimes\vec{v}(2)\otimes\cdots \otimes\vec{v}(m)\right)  =
D(U)\vec{v}\left(1\right)\otimes D(U)\vec{v}\left(2\right)\otimes \cdots 
\otimes D(U)\vec{v}\left(m\right) \, ,
$$
where $D(U)$ is the $p\times p$ unitary matrix representing group element $U\in U(p)$ 
in the fundamental representation. Thus, $U\in U(p)$ will change the
value of the vector in the $i^{\rm th}$ slot but it will not move it to a different slot. It acts in exactly the same way
on each slot. It is quite clear that these are commuting actions of $U(p)$ and $S_m$ on $V_p^{\otimes m}$
\begin{eqnarray} 
\nonumber
U\cdot\left(\sigma\cdot\left(\vec{v}(1)\otimes\dots\otimes\vec{v}(m)\right)\right)&&=
U\cdot\left(\vec{v}\left(\sigma(1)\right)\otimes\dots\otimes\vec{v}\left(\sigma(m)\right)\right)\\
\nonumber
&&=D(U)\vec{v}\left(\sigma(1)\right)\otimes\dots\otimes D(U)\vec{v}\left(\sigma(m)\right)\\
\nonumber
&&=\sigma\cdot\left( D(U)\vec{v}\left(1\right)\otimes\dots\otimes D(U)\vec{v}\left(m\right)\right)\\
\nonumber
&&=\sigma \cdot\left(U\cdot\left(\vec{v}(1)\otimes\dots\otimes\vec{v}(m)\right)\right)
\end{eqnarray}
and consequently by Schur-Weyl duality the space can be organized as\footnote{Part of what is behind Shur-Weyl duality is
simple and familiar: any two operators that commute can be simultaneously diagonalized.} \cite{Fulton}
\begin{equation}
  V_p^{\otimes m} = \oplus_{s} V_s^{U(p)}\otimes V_s^{S_m}\, ,
\label{SWDuality}
\end{equation}
where the sum runs over all Young diagrams built from $m$ boxes and each has at most $p$ rows. One consequence of this formula is 
that
$$
  p^m=\sum_{s} {\rm Dim}(s)\, d_s
$$
where ${\rm Dim}(s)$ is the dimension of $s$ as an irreducible representation of $U(p)$ and $d_s$ is the dimension
of $s$ as an irreducible representation of $S_m$. The reader is invited to check a few examples herself. Thus,
by identifying states with good $U(p)$ labels we have identified states with good $S_m$ labels. Therefore
an important consequence of (\ref{SWDuality}) is that it provides an efficient method to construct the projectors 
which are used to define the restricted Schur polynomials\footnote{The reader will be familiar with the usual use
of Schur-Weyl duality, to construct projectors onto good $U(p)$ irreducible representations using the Young symmetrizers
i.e. by symmetrizing and antisymmetrizing indices on a tensor. We are turning this argument on its head by using the
irreducible representations of the unitary group to build symmetric group projectors. Bear in mind that the details
of our Schur-Weyl duality are different to the usual construction.}. 

{\vskip 0.25cm}

\noindent
{\bf Key Idea:} {\sl Using Schur-Weyl duality it follows that the symmetric group operators $P_{R\to (r,s)jk}$ carry good
                     $U(p)$ labels (where $p$ is the number of rows in $R$) and, consequently, can be constructed using
                     nothing more than $U(p)$ group theory.}

{\vskip 0.25cm}

A necessary step towards building the projectors entails constructing a dictionary between the original labels
$R,(r,s)jk$ of the restricted Schur polynomial $\chi_{R,(r,s)jk}$ and the new $U(p)$ labels. Exactly the same Young 
diagram $s$ that originally specifies an $S_m$ irreducible representation, specifies a $U(p)$ irreducible representation. 
The Young diagram $r$ is included among the new labels and it still specifies an irreducible representation of
$S_n$. The final label is the choice of a state from the carrier space of $U(p)$ representation $s$. The $\Delta$ 
weight of this state (see Appendix \ref{weights})
tells us how boxes were removed from $R$ to obtain $r$. This point deserves some explanation. Label the state
chosen from the carrier space $s$ by its Gelfand-Tsetlin pattern. This state can be put into one-to-one
correspondence with a semi-standard Young tableau and this correspondence plays a central role. Consider for example 
the $U(3)$ state with Gelfand-Tsetlin pattern
$$
\left[ 
\begin {array}{ccccc} 
    4  &            &    3    &           &      3    \\\noalign{\medskip}
       &     3      &         &     2     &           \\\noalign{\medskip}
       &            &    2    &           &      
\end {array} \right]\, .
$$
The uppermost row of the pattern gives the shape of the Young diagram. Each row (starting from the bottom row)
tells us how to distribute 1s, then 2s and so on till the semi standard Young tableau is obtained. This connection
is reviewed in detail in Appendix \ref{semistandard}. For the Gelfand-Tsetlin pattern shown above the semi-standard Young tableau is
$$
\left[ 
\begin {array}{ccccc} 
    *  &            &    *    &           &      *    \\\noalign{\medskip}
       &     *      &         &     *     &           \\\noalign{\medskip}
       &            &    2    &           &      
\end {array} \right]\leftrightarrow  \young(11**,***,***)\quad
\left[ 
\begin {array}{ccccc} 
    *  &            &    *    &           &      *    \\\noalign{\medskip}
       &     3      &         &     2     &           \\\noalign{\medskip}
       &            &    2    &           &      
\end {array} \right]\leftrightarrow  \young(112*,22*,***)\quad
\left[ 
\begin {array}{ccccc} 
    4  &            &    3    &           &      3    \\\noalign{\medskip}
       &     3      &         &     2     &           \\\noalign{\medskip}
       &            &    2    &           &      
\end {array} \right]\leftrightarrow  \young(1123,223,333)\, .
$$
Each row in the pattern corresponds to a particular number in the semi standard tableau. From the definition of the 
Gelfand-Tsetlin pattern, we also know that each row in the pattern corresponds to a particular subgroup in the chain
of subgroups $U(1)\subset U(2)\subset \cdots \subset U(p-1)\subset U(p)$. So, from the point of view of the semi-standard
Young tableau or of the Gelfand-Tsetlin pattern, going to the $U(p-1)$ subgroup implies that we consider a subgroup that 
does not act on one of the numbers appearing in the semi-standard tableau. What does it mean to consider a $U(p-1)$ subgroup
of our action of $U(p)$ on the boxes that have been removed from $R$? Recall that the particular state that is assigned to 
each removed box depends on the row it was removed from. Thus going to a $U(p-1)$ subgroup corresponds to considering
a subgroup that does not act on the boxes belonging to a particular row. Clearly then, the numbers in the semi-standard
tableau can be identified with the row from which the corresponding box has been removed from $R$. Recall that the $\Delta$
weight is a sequence of integers $\Delta (M)=(\delta_n(M),\delta_{n-1}(M),\cdots\delta_1(M))$. The number of boxes labeled
$i$ which is the number of boxes removed from row $i$ of $R$ to produce $r$, is given by $\delta_i(M)$. Thus, given $r$ and 
the delta weight we can reconstruct $R$. 

There is a subtlety that needs to be discussed. Two states that belong to the same $U(p)$ representation and have the same 
$\Delta$ weight correspond to the same set of labels $R,(r,s)$. Consequently, we find that $(r,s)$ can be subduced more
than once in the carrier space of $R$. These multiplicities only arise for $p\ge 3$ and hence were not treated in \cite{bhw}.
Our analysis here shows that this multiplicity index is easily organized using the $U(p)$ representations: The number
of states having the same $\Delta$ weight is called the inner multiplicity of the state $I(\Delta(M))$. In this case, we label
each state with a multiplicity index which runs from 1 to $I(\Delta(M))$\footnote{An alternative approach to resolving 
these multiplicities has been outlined in \cite{Kimura:2008wy}. The idea is to consider elements in the group algebra $CS_{n+m}$ 
which are invariant under conjugation by $CS_n\times CS_m$. The Cartan subalgebra of these elements are the natural generalization 
of the Jucys-Murphy elements which define a Cartan subalgebra for $S_n$\cite{Okunkov}. 
The multiplicities will be labeled by the eigenvalues of this Cartan subalgebra\cite{Kimura:2008wy}.}.
These multiplicities have been resolved by the $U(p)$ state labels. 
Finally note that each $U(p)$ representation $s$ will also appear with a particular multiplicity. However, thanks to Schur-Weyl
duality, we know that this multiplicity is organized by the $S_m$ representation $s$.

{\vskip 0.25cm}

\noindent
{\bf Key Idea:} {\sl The Gelfand-Tsetlin patterns of $U(p)$ provide a non-degenerate set of multiplicity labels $jk$ for the
                     symmetric group operators $P_{R\to (r,s)jk}$.}

{\vskip 0.25cm}

In summary then we trade the labels
$$
\begin{array}{cc}
R &{\rm an\,\, irreducible\,\, representation\,\, of}\,\, S_{n+m}\\ 
r &{\rm an\,\, irreducible\,\, representation\,\, of}\,\, S_n\\
s &{\rm an\,\, irreducible\,\, representation\,\, of}\,\, S_m\\
j & {\rm multiplicity\, label\, resolving\, copies\,of}\, (r,s)
\end{array}
$$
for the new labels 
$$
\begin{array}{cc}
r &{\rm an\,\, irreducible\,\, representation\,\, of}\,\, S_n\\
s &{\rm an\,\, irreducible\,\, representation\,\, of}\,\, U(p)\\
M^i &{\rm a\,\, state\,\, in\,\, the\,\, carrier\,\, space\,\, of}\,\, s\,\, {\rm where}\\
    &i {\rm\,\, runs\,\, over\,\, inner\,\, multiplicity}\, . 
\end{array}
$$

At this point we have identified an orthonormal set of states spanning any particular carrier space $(r,s)j$ of the $S_n\times S_m$
subgroup. It is now a trivial task to write down the corresponding projector. 

\subsection{From $S_n\times (S_1)^m$ to $S_n\times S_m$}

We can now write the symmetric group operator used to define the restricted Schur polynomial as
$$
  P_{R\to (r,s)jk}=\sum_{\alpha =1}^{d_s}|s,M^j ,\alpha\rangle\langle s,M^k,\alpha |\otimes {\bf I}_{r}\, ,
$$
where, by Schur-Weyl duality, the multiplicity label $\alpha$ for the $U(p)$ states is organized by the irreducible 
representation $s$ of the symmetric group $S_m$. The indices $j$ and $k$ pick out states $M$ that have a particular
$\Delta$ weight and hence range over $1,2,...,I(\Delta(M))$. 
The components $\delta_i$ of the particular $\Delta$ that must be used are equal to the number of 
boxes removed from row $i$ of $R$ to produce $r$. ${\bf I}_{r}$ is simply the identity matrix in the carrier space
of the $S_n$ irreducible representation labeled by $r$.

We will end this subsection with a few examples. The labels
$$
R={\tiny \yng(10,5)},\quad r={\tiny\yng(8,3)},\quad s={\tiny \yng(2,2)}
$$
become
$$
r={\tiny\yng(8,3)},\quad s={\tiny\yng(2,2)},\quad M=\left[
\begin{array}{ccc}
2 &  &2\\
  &2 &
\end{array}
\right]
$$
For this example $\Delta = (2,2)$ because 2 boxes are removed from the first row and two from the second row of $R$ to produce $r$.
The first row of $M$ is read off $s$ and the second row is chosen to obtain the correct $\Delta$. 
The inner multiplicity for this case is 1, so that there is a single possible projection operator. For our second example
consider the labels
$$
R={\tiny \yng(15,10,5)},\quad r={\tiny\yng(14,9,4)},\quad s={\tiny \yng(2,1)}\, .
$$
The new labels are
$$
r={\tiny\yng(14,9,4)},\quad s={\tiny\yng(2,1)}
$$
and
$$
M_1=
\left[ 
\begin {array}{ccccc} 
    2  &            &    1    &           &      0    \\\noalign{\medskip}
       &     1      &         &     1     &           \\\noalign{\medskip}
       &            &    1    &           &      
\end {array} \right]\quad
M_2=
\left[ 
\begin {array}{ccccc} 
    2  &            &    1    &           &      0    \\\noalign{\medskip}
       &     2     &         &     0    &           \\\noalign{\medskip}
       &            &    1    &           &      
\end {array} \right]
$$
For this example $\Delta=(1,1,1)$ because one box is removed from each row. The inner multiplicity is 2. The two possible
Gelfand-Tsetlin patterns are shown. Thus, for the $R,(r,s)$ labels given, one can construct a total of four possible restricted 
Schur polynomials. This second example is discussed in detail in Appendix \ref{exampleprojector} where the 
allowed operators $P_{R\to (r,s)jk}$ are explicitly constructed.

\subsection{Young Diagrams with $p$ Columns}
\label{columns}

We will consider Young diagrams with a total of $p$ columns. In this case, boxes that are in different 
columns, will again have associated factors with $c_i-c_j\sim O(N)$. As discussed in Appendix \ref{simpleyoung}, the 
fact that $c_i-c_j\sim O(N)$ for boxes in different rows again implies a 
significant simplification in the representations of $S_m$. When adjacent permutations $(i,i+1)$ act on labeled
boxes that belong to the same column, the Young diagram changes sign and when acting on labeled
boxes that belong to the different columns, the labeled boxes are swapped. This change in sign for the case that boxes
belong to the same column is the only difference to what was considered in section \ref{prows}.

\begin{figure}[h]
          \centering
          {\epsfig{file=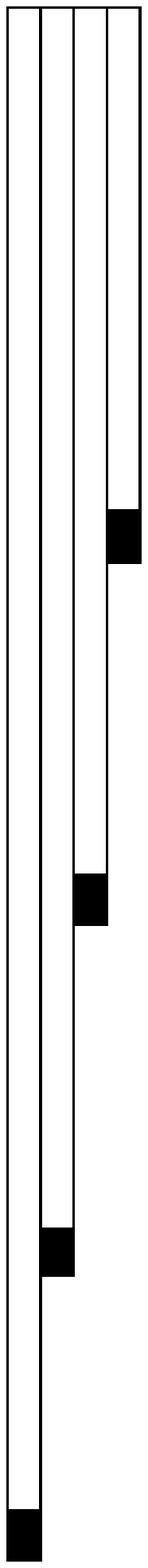,width=1.0cm,height=7.5cm}}
          \caption{An example of a Young diagram with $p=4$ columns. The columns are shown; the rows are not shown. There are
                   $O(N)$ boxes in each column. The $m$ numbered boxes have been colored black.
                   The difference in factors associated to any two boxes that are in different columns is $O(N)$.}
\end{figure}

The number of states that can be obtained when $m$ boxes are labeled is again $p^m$ and we again associate a $p$-dimensional
vector to each box that is labeled. This again allows us to put partially labeled Young diagrams into one-to-one correspondence
with vectors in $V_p^{\otimes m}$. In this case however, we will include some additional phases when we identify vectors
in $V_p^{\otimes m}$ with partially labeled Young diagrams. These extra phases occur precisely because adjacent permutations 
$(i,i+1)$ acting on labeled boxes that belong to the same column flip the sign of the Young diagram. Choose any specific state 
with a particular set of labels. This state plays the role of a reference state. Any other state with the same boxes labeled 
but with a different assignment of the labels can be obtained by acting on the reference state with adjacent permutations
$(i,i+1)$. Further, the only adjacent permutation $(i,i+1)$ that we are allowed to apply to the reference state to reach
any other given state have boxes labeled $i$ and $i+1$ in different columns when $(i,i+1)$ acts. 
If we act with $q$ adjacent permutations of this type to get from the reference state to another distinct state, it is assigned a
phase $(-1)^q$. See Appendix \ref{examplesphereprojector} for an explicit example.
With this choice for the phases, it is easy to see that the action of $S_m$ on the partially labeled Young diagrams
induces the following action on $V_p^{\otimes m}$
$$
\sigma\cdot\left( \vec{v}(1)\otimes\vec{v}(2)\otimes\cdots \otimes\vec{v}(m)\right)  =
{\rm sgn}(\sigma)\vec{v}\left(\sigma(1)\right)\otimes\vec{v}\left(\sigma(2)\right)\otimes \cdots 
\otimes\vec{v}\left(\sigma(m)\right) \, ,
$$
where ${\rm sgn}(\sigma)$ denotes the signature of permutation $\sigma$: it is +1 for even permutations and -1
for odd permutations\footnote{Recall that a permutation is even (odd) if it can be written as a product of an
even (odd) number of two cycles.}.
Thus, $\sigma\in S_m$ will move the vector in the $i^{\rm th}$ slot to the $\sigma(i)^{\rm th}$ slot and may change the
overall phase. We can also define an action of $U(p)$ on $V_p^{\otimes m}$ 
$$
U\cdot\left( \vec{v}(1)\otimes\vec{v}(2)\otimes\cdots \otimes\vec{v}(m)\right)  =
D(U)\vec{v}\left(1\right)\otimes D(U)\vec{v}\left(2\right)\otimes \cdots 
\otimes D(U)\vec{v}\left(m\right) \, ,
$$
where $D(U)$ is the $p\times p$ unitary matrix representing group element $U\in U(p)$. Thus, $U\in U(p)$ will change the
value of the vector in the $i^{\rm th}$ slot but it will not move it to a different slot. It acts in exactly the same way
on each slot. It is quite clear that again these are commuting actions of $U(p)$ and $S_m$ on $V_p^{\otimes m}$
\begin{eqnarray} 
\nonumber
U\cdot\left(\sigma\cdot\left(\vec{v}(1)\otimes\dots\otimes\vec{v}(m)\right)\right)&&=
U\cdot\,{\rm sgn}(\sigma)\left(\vec{v}\left(\sigma(1)\right)\otimes\dots\otimes\vec{v}\left(\sigma(m)\right)\right)\\
\nonumber
&&={\rm sgn}(\sigma) D(U)\vec{v}\left(\sigma(1)\right)\otimes\dots\otimes D(U)\vec{v}\left(\sigma(m)\right)\\
\nonumber
&&=\sigma\cdot\left( D(U)\vec{v}\left(1\right)\otimes\dots\otimes D(U)\vec{v}\left(m\right)\right)\\
\nonumber
&&=\sigma \cdot\left(U\cdot\left(\vec{v}(1)\otimes\dots\otimes\vec{v}(m)\right)\right)
\end{eqnarray}
and consequently by Schur-Weyl duality we can again use $U(p)$ to organize the multiplicity label of the $S_m$ irreducible
representations. In this case, the space can be organized as
\begin{equation}
  V_p^{\otimes m} = \oplus_{s} V_{s^T}^{U(p)}\otimes V_s^{S_m}\, ,
\label{SWDuality2}
\end{equation}
where $s^T$ is obtained by exchanging row and columns in $s$. The discussion from here on is identical to the case of $p$ rows.
The reader is invited to consult Appendix \ref{examplesphereprojector} for a concrete 
example of a projector constructed using this Schur-Weyl duality. 

\section{Action of the Dilatation Operator}

The action of the one loop dilatation operator in the $SU(2)$ sector\cite{Beisert:2003tq}
$$
D = - g_{\rm YM}^2 {\rm Tr}\,\big[ Y,Z\big]\big[ \partial_Y ,\partial_Z\big]
$$
on the restricted Schur polynomial has been studied in \cite{Koch:2010gp,VinceKate,bhw}.
We will find it convenient to work with operators normalized to give a unit two point function.
The normalized operators $O_{R,(r,s)}(Z,Y)$ can be obtained from
$$
\chi_{R,(r,s)jk}(Z,Y)=\sqrt{f_R \, {\rm hooks}_R\over {\rm hooks}_r\, {\rm hooks}_s}O_{R,(r,s)jk}(Z,Y)\, .
$$
In terms of these normalized operators (see Appendix \ref{derivedilat}), \cite{VinceKate} found
$$
DO_{R,(r,s)jk}(Z,Y)=\sum_{T,(t,u)lq} N_{R,(r,s)jk;T,(t,u)lq}O_{T,(t,u)lq}(Z,Y)
$$
\begin{eqnarray}
\label{dilat}
N_{R,(r,s)jk;T,(t,u)lq}&&= - g_{YM}^2\sum_{R'}{c_{RR'} d_T n m\over d_{R'} d_t d_u (n+m)}
\sqrt{f_T \, {\rm hooks}_T\, {\rm hooks}_r \, {\rm hooks}_s \over f_R \, {\rm hooks}_R\, {\rm hooks}_t\, {\rm hooks}_u}\times
\\
\nonumber
&&\times{\rm Tr}\Big(\Big[ \Gamma_R((1,m+1)),P_{R\to (r,s)jk}\Big]I_{R'\, T'}\Big[\Gamma_T((1,m+1)lm),P_{T\to (t,u)ql}\Big]I_{T'\, R'}\Big) \, .
\end{eqnarray}
$c_{RR'}$ is the factor of the corner box removed from Young diagram $R$ to obtain diagram $R'$, and similarly $T'$ is a Young
diagram obtained from $T$ by removing a box. The intertwiner $I_{AB}$ is a map from the carrier space of irreducible representation 
$A$ to the carrier space of irreducible representation $B$. Consequently, Schur's Lemma implies that
$A$ and $B$ must be Young diagrams of the same shape for a non-zero intertwiner. The 
intertwiner operators relevant for our study are described in Appendix \ref{intertwiners}. It turns out that the product of the intertwiners
with $\Gamma_R (1,m+1)$ can be expressed as a matrix acting on the first slot of $V_p^{\otimes m}$. Thus, evaluating the action of
the dilatation operator reduces to evaluating the trace of a product of matrices, which are either the operators $P_{R\to (r,s)jk}$,
$P_{T\to (t,u)lq}$ or matrices acting on the first slot of $V_p^{\otimes m}$. The simplest way to evaluate this trace is to decompose
(with the help of the known Clebsch-Gordon coefficients given in Appendix \ref{Clebschs}) the states in $V_p^{\otimes m}$ into direct product of states,
where the first state in the direct product lives in $V_p$ (which is a copy of the carrier space of the defining representation of $U(p)$
and corresponds to the first slot) and the second state in the direct product lives in $V_p^{\otimes\, m-1}$ (corresponding to the remaining 
slots). The complete details of this computation are given in Appendix \ref{dilatationoperator}.

\subsection{System of Two Giant Gravitons}

Operators dual to a system of two giant gravitons are labeled by Young diagrams with two rows (for AdS giants) or two
columns (for sphere giants). The third label $s$ in the restricted Schur polynomial $\chi_{R,(r,s)}$ is thus replaced
by Gelfand-Tsetlin patterns for $U(2)$. Since the sum of the two numbers in the first row is equal to the number of 
impurities $m$, which is fixed, the Young diagram $s$ can be traded for two independent numbers. These two numbers specify both
the weight $\Delta$ and $s$. The Young diagram $r$ is given by specifying the number of columns with two boxes per column
($b_0$) and the number of columns with one box per column ($b_1$). Thus, our operators are specified by four labels
$O(b_0,b_1,j,j^3)$. See figure \ref{fig:twocolumn} in Appendix \ref{tworows}. When acting on $O(b_0,b_1,j,j^3)$, the 
dilatation operator produces a total of 9 terms that can be grouped into three collections of three terms each. Indeed, 
in terms of
{\small
\begin{eqnarray}
\Delta O(b_{0},b_{1},j,j^3) &&=\sqrt{(N+b_{0})(N+b_{0}+b_{1})}(O(b_{0}+1,b_{1}-2,j,j^3)+O(b_{0}-1,b_{1}+2,j,j^3))\nonumber \\
 &&-(2N+2b_{0}+b_{1})O(b_{0},b_{1},j,j^3)\label{nicecombo}
\end{eqnarray}
}
the dilatation operator is
\begin{eqnarray}\label{recursionj3}
 && DO(b_{0},b_{1},j,j^3)=g_{YM}^2 \left[-{1\over 2}\left( m-{(m+2)(j^3)^2\over j(j+1)}\right)
\Delta O(b_{0},b_{1},j,j^3)\right.\nonumber \\
&& + \sqrt{(m + 2j + 4)(m - 2j)\over (2j + 1)(2j+3)} {(j+j^3 +
1)(j-j^3 + 1) \over 2(j + 1)}
 \Delta O(b_{0},b_{1},j+1,j^3)
\nonumber \\
&& \left. +\sqrt{(m + 2j + 2)(m - 2j +2)\over (2j + 1)(2j-1)}
{(j+j^3 )(j-j^3 ) \over 2 j} \Delta O(b_{0},b_{1},j-1,j^3)
\right]
\end{eqnarray}
This reproduces the result of \cite{bhw} and is a nice check of our method. Notice that the dilatation operator does not change the 
$j^3$ label of the operator it acts on. The general statement, true for a system of $p$ giant gravitons is that dilatation
operator does not change the weight $\Delta$ of the operator it acts on. For the case of giant gravitons labeled by Young
diagrams with two long columns denote the relevant operators $Q(b_0,b_1,j,j^3)$. The dilatation operator has a very similar action
\begin{eqnarray}\label{recursionjj3}
 && DQ(b_0,b_1,j,j^3)=g_{YM}^2 \left[-{1\over 2}\left( m-{(m+2)(j^3)^2\over j(j+1)}\right)
\Delta Q(b_{0},b_{1},j,j^3)\right.\nonumber \\
&& + \sqrt{(m + 2j + 4)(m - 2j)\over (2j + 1)(2j+3)} {(j+j^3 +
1)(j-j^3 + 1) \over 2(j + 1)}
 \Delta Q(b_{0},b_{1},j+1,j^3)
\nonumber \\
&& \left. +\sqrt{(m + 2j + 2)(m - 2j +2)\over (2j + 1)(2j-1)}
{(j+j^3 )(j-j^3 ) \over 2 j} \Delta Q(b_{0},b_{1},j-1,j^3)
\right]
\end{eqnarray}
where
\begin{eqnarray}
\Delta Q(b_{0},b_{1},j,j^3) &&=\sqrt{(N-b_{0})(N-b_{0}-b_{1})}(Q(b_{0}+1,b_{1}-2,j,j^3)+Q(b_{0}-1,b_{1}+2,j,j^3))\nonumber \\
 &&-(2N-2b_{0}-b_{1})Q(b_{0},b_{1},j,j^3).\label{secondcombo}
\end{eqnarray}
Notice that the sphere giant and AdS giant cases are related by replacing expressions like $N+b_0$ with $N-b_0$.

\subsection{System of Three Giant Gravitons}

In this case our operators are labeled by Young diagrams with three rows (for AdS giants) or three columns (for sphere giants). 
The third label $s$ and multiplicity labels $j,k$
in $\chi_{R,(r,s),jk}$ are thus traded for Gelfand-Tsetlin patterns for $U(3)$. Similar to the two giant case,
since the sum of the three numbers in the first row is equal to the number of impurities $m$, which is fixed, $s$ can be traded 
for five independent numbers and these specify the weight $\Delta$, multiplicity labels $j,k$ and $s$. 
The Young diagram $r$ is given by specifying the 
number of columns with three boxes per column ($=b_0$), the number of columns with two boxes per column ($=b_1$) and the number 
of columns with one box per column ($b_2$). Since the number of boxes in $r$ is given by $n=3b_0+2b_1+b_2$, and since $n$ is
fixed we need not specify $b_0$ - it is determined once $b_1$ and $b_2$ are given. Thus, accounting for inner multiplicity,
our operators are specified by a total of 10 labels. Although the general expression can be computed using our methods, we have decided
to focus on two special cases. For the first case we study $m=3$ impurities and $\Delta =(1,1,1)$. There are a total of 6 possible
labels $s$ giving 6 possible operators $O_i(b_1,b_2)$. These operators are defined in detail in Appendix  \ref{explicitdilatationoperator}.
The action of the dilatation operator is given by
\begin{equation}
D O_i(b_1,b_2)=-g_{YM}^2\left(M^{(12)}_{ij}\Delta_{12}O_j (b_1,b_2)+M^{(13)}_{ij}\Delta_{13}O_j (b_1,b_2)+M^{(23)}_{ij}\Delta_{12}O_j (b_1,b_2)\right)
\label{frstexmpl}
\end{equation}
where
{\footnotesize
$$
M^{(12)}=\left[ 
\begin {array}{cccccc} 
    {2\over 3}      &  0                  & -{2\over 3\sqrt{2}} &   {1\over\sqrt{6}}   &   {1\over\sqrt{6}} & 0                     \\\noalign{\medskip}
      0             &  {2\over 3}         &    0                &  -{1\over\sqrt{6}}   &  -{1\over\sqrt{6}} & -{2\over 3\sqrt{2}}   \\\noalign{\medskip}
-{2\over 3\sqrt{2}} &  0                  & {1\over 3}          &  -{1\over 2\sqrt{3}} & -{1\over 2\sqrt{3}}&   0                   \\\noalign{\medskip}
 {1\over\sqrt{6}}   & -{1\over\sqrt{6}}   &-{1\over 2\sqrt{3}}  &       1              &       0            &  {1\over 2\sqrt{3}}   \\\noalign{\medskip}
 {1\over\sqrt{6}}   & -{1\over\sqrt{6}}   &-{1\over 2\sqrt{3}}  &       0              &       1            &  {1\over 2\sqrt{3}}   \\\noalign{\medskip}
      0             & -{2\over 3\sqrt{2}} &      0              &  {1\over 2\sqrt{3}}  & {1\over 2\sqrt{3}} &  {1\over 3}           \\\noalign{\medskip}
\end {array} \right]
\quad
M^{(13)}=\left[ 
\begin {array}{cccccc} 
    {2\over 3}      &  0                  & -{2\over 3\sqrt{2}} &  -{1\over\sqrt{6}}   &  -{1\over\sqrt{6}}  & 0                     \\\noalign{\medskip}
      0             &  {2\over 3}         &    0                &   {1\over\sqrt{6}}   &   {1\over\sqrt{6}}  & -{2\over 3\sqrt{2}}   \\\noalign{\medskip}
-{2\over 3\sqrt{2}} &  0                  & {1\over 3}          &   {1\over 2\sqrt{3}} &  {1\over 2\sqrt{3}} &   0                   \\\noalign{\medskip}
-{1\over\sqrt{6}}   &  {1\over\sqrt{6}}   & {1\over 2\sqrt{3}}  &       1              &       0             & -{1\over 2\sqrt{3}}   \\\noalign{\medskip}
-{1\over\sqrt{6}}   &  {1\over\sqrt{6}}   & {1\over 2\sqrt{3}}  &       0              &       1             & -{1\over 2\sqrt{3}}   \\\noalign{\medskip}
      0             & -{2\over 3\sqrt{2}} &      0              & -{1\over 2\sqrt{3}}  & -{1\over 2\sqrt{3}} &  {1\over 3}           \\\noalign{\medskip}
\end {array} \right]
$$
}
{\footnotesize
$$
M^{(23)}=\left[ 
\begin {array}{cccccc} 
    {2\over 3}      &  0                  &  {1\over 3\sqrt{2}} &       0              &        0           & 0                     \\\noalign{\medskip}
      0             &  {2\over 3}         & -{1\over\sqrt{2}}   &       0              &        0           & -{2\over 3\sqrt{2}}   \\\noalign{\medskip}
 {1\over 3\sqrt{2}} & -{1\over\sqrt{2}}   & {5\over 6}          &       0              &        0           &   0                   \\\noalign{\medskip}
      0             &   0                 &      0              &     {1\over 2}       &       -{1\over 2}  &   0                   \\\noalign{\medskip}
      0             &   0                 &      0              &     -{1\over 2}      &       {1\over 2}   &   0                   \\\noalign{\medskip}
 -{1\over\sqrt{2}}  &  {1\over 3\sqrt{2}} &     -{1\over 2}     &       0              &        0           &  {5\over 6}           \\\noalign{\medskip}
\end {array} \right]\, .
$$
}
and
{\footnotesize
\begin{eqnarray}
\label{D12}
&&\Delta_{12} O(b_1,b_2,j,k,j^3,k^3,l^3)=-(2N+2b_0+2b_1+b_2)O(b_1,b_2,j,k,j^3,k^3,l^3)+\\
&&\sqrt{(N+b_0+b_1)(N+b_0+b_1+b_2)}\left(O(b_1-1,b_2+2,j,k,j^3,k^3,l^3)+O(b_1+1,b_2-2,j,k,j^3,k^3,l^3)\right),
\nonumber
\end{eqnarray}
{\vskip 0.15cm}
\begin{eqnarray}
\label{D13}
&&\Delta_{13} O(b_1,b_2,j,k,j^3,k^3,l^3)=-(2N+2b_0+b_1+b_2)O(b_1,b_2,j,k,j^3,k^3,l^3)\\
\nonumber
&&\sqrt{(N+b_0)(N+b_0+b_1+b_2)}\left(O(b_1-1,b_2-1,j,k,j^3,k^3,l^3)+O(b_1+1,b_2+1,j,k,j^3,k^3,l^3)\right),
\end{eqnarray}
{\vskip 0.15cm}
\begin{eqnarray}
\label{D23}
&&\Delta_{23} O(b_1,b_2,j,k,j^3,k^3,l^3)=-(2N+2b_0+b_1)O(b_1,b_2,j,k,j^3,k^3,l^3)+\\
\nonumber
&&\sqrt{(N+b_0)(N+b_0+b_1)}\left(O(b_1-2,b_2+1,j,k,j^3,k^3,l^3)+O(b_1+2,b_2-1,j,k,j^3,k^3,l^3)\right).
\end{eqnarray}
}

The second special case we consider is the sector with $j^3=O(1)$ and the remaining quantum numbers ($j,k,k^3,l^3$ and $m$) are all order $N$. 
The action of the dilatation operator simplifies considerably in this limit because it leaves the $j^3$ quantum number fixed. Given $j,k,m,j^3$ 
and the weight $\Delta =(n_1,n_2,n_3)$, we easily obtain
$$
  k^3={m-3n_1-3j^3+2j+k\over 3},\qquad l^3={m-3n_2+3j^3+k-j\over 3}\, .
$$
Thus, after specifying $\Delta$ and $j^3$ the $k^3,l^3$ labels are fixed and our operators can be labeled by four quantum numbers $O(b_1,b_2,j,k)$.
The dilatation operators produces 45 terms when acting on $O(b_1,b_2,j,k)$, which can be grouped into 5 collections of 9 terms each
{\footnotesize
\begin{eqnarray}
\label{intdil}
  DO(b_1,b_2,j,k)&&=-g_{YM}^2\left[
  {k^3(j+k-k^3)(k-k^3-l^3)\over 3(j+k)^2 (k-k^3)}\Delta^{(a)}\Delta_{12} O(b_1,b_2,j,k)\right.\\
&&+{l^3 k^3 (j+k-k^3)\over 3(j+k)^2 (k-k^3)}\Delta^{(a)}\Delta_{13} O(b_1,b_2,j,k)
  -{l^3k^3(k-k^3-l^3)(j+k-k^3)\over 3(j+k)^2 (k-k^3)^2}\Delta^{(a)}\Delta_{23} O(b_1,b_2,j,k)
\nonumber\\
&&\left.
  +{l^3(k-k^3-l^3)(j+k-k^3)\over 3(j+k) (k-k^3)^2}\Delta^{(b)}\Delta_{23} O(b_1,b_2,j,k)
  +{k^3l^3(k-k^3-l^3)\over 3(j+k)(k-k^3)^2}\Delta^{(c)}\Delta_{23} O(b_1,b_2,j,k)\right]
\nonumber
\end{eqnarray}
}
where
{\footnotesize
$$
 \Delta^{(a)}O(b_1,b_2,j,k) = (2m+j-k) O(b_1,b_2,j,k)
$$
$$
  -\sqrt{(m+2j+k)(m-j-2k)}\left(O(b_1,b_2,j-1,k-1)+O(b_1,b_2,j+1,k+1)\right)
$$
{\vskip 0.15cm}
$$
 \Delta^{(b)}O(b_1,b_2,j,k) = (2m-2j-k) O(b_1,b_2,j,k)
$$
$$
  -\sqrt{(m-j-2k)(m-j+k)}\left(O(b_1,b_2,j+1,k-2)+O(b_1,b_2,j-1,k+2)\right)
$$
{\vskip 0.15cm}
$$
 \Delta^{(c)}O(b_1,b_2,j,k) = (2m+j+2k) O(b_1,b_2,j,k)
$$
$$
  -\sqrt{(m+2j+k)(m-j+k)}\left(O(b_1,b_2,j-2,k+1)+O(b_1,b_2,j+2,k-1)\right)
$$
}
For these two examples, the sphere giant and AdS gaint cases are again related by replacing expressions like $N+b_0$ with $N-b_0$.

\section{Diagonalization of the Dilatation Operator}

The dilatation operator when acting on two giant systems has already been diagonalized in \cite{bhw}. 
We start with a quick review of this material because it is relevant for the multiple giant systems we consider next.
Make the following ansatz for the operators of good scaling dimension\footnote{$f(b_0,b_1)$ is not a function of
$b_0$ and $b_1$ separately because $2b_0+b_1$ is fixed equal to the number of $Z$s.}
$$
  O_{p,n}=\sum_{b_1}\, f(b_0,b_1)\, O_{p,j^3}(b_0,b_1)\,=\sum_{j,b_1}\, C_{p,j^3}(j)\, f(b_0,b_1)\,
  O_{j,j^3}(b_0,b_1)\, .
$$
Solving the eigenproblem 
$$
   DO(p,n)=\kappa O(p,n)
$$
where $\kappa$ is the one loop anomalous dimension, amounts to solving the recursion relations
\begin{eqnarray}
 && -\alpha_{p,j^3}C_{p,j^3}(j)=
  \sqrt{(m + 2j + 4)(m - 2j)\over (2j + 1)(2j+3)} {(j+j^3 + 1)(j-j^3 + 1) \over 2(j + 1)} C_{p,j^3}(j+1)
\nonumber \\
 && \sqrt{(m + 2j + 2)(m - 2j +2)\over (2j + 1)(2j-1)} {(j+j^3 )(j-j^3 ) \over 2 j}C_{p,j^3}(j-1)
  -{1\over 2}\left( m-{(m+2)(j^3)^2\over j(j+1)}\right)C_{p,j^3}(j)\, .\nonumber \\
\end{eqnarray}
and
\begin{eqnarray}
&&-\alpha_{p,j^3} g_{YM}^{2}[\sqrt{(N+b_{0})(N+b_{0}+b_{1})}(f(b_{0}-1,b_{1}+2)+f(b_{0}+1,b_{1}-2))\nonumber \\
&&-(2N+2b_{0}+b_{1})f(b_{0},b_{1})]=\kappa f(b_{0}, b_{1})\, .
\label{Kp}
\end{eqnarray}
These recursion relations are solved by 
\begin{equation}
C_{p,j^3}(j)=(-1)^{{m\over 2}-p}\left({m\over
2}\right)!\sqrt{(2j+1)\over \left({m\over
2}-j\right)!\left({m\over 2}+j+1\right)!}
{}_3F_2\left({}^{|j^3|-j,j+|j^3|+1,-p}_{|j^3|-{m\over 2},1};1\right)
\label{universalefuncs}
\end{equation}
and
\begin{equation}
f(b_0,b_1)={(-1)^n  ({1\over2})^{N+b_0+{b_1 \over
2}}}\sqrt{\left(^{2N+2b_0+b_1}_{N+b_0+b_1}\right)\left(^{2N+2b_0+b_1}_{\quad\quad
n}\right)}{}_2F_1({}^{-(N+b_0+b_1),-n}_{-(2N+2b_0+b_1)};2)
\label{alllevels}
\end{equation}
where the range of $j$ and $p$ are $|j^3| \le j\le\frac{m}{2}$, $0\le p\le\frac{m}{2}-|j^3|$, and
the associated eigenvalues are
$$ 
  -\alpha_{p,j^3} = -2p = 0,-2,-4,...,-(m-2|j^3|)
$$
and
$$
\kappa = 4n\alpha_{p,j^3}g_{YM}^2 = 8png_{YM}^2\qquad n=0,1,2,... \, .
$$
Since our quantum numbers are very large, one might also consider examining the above recursion relations 
in a continuum limit where one would
expect them to become differential equations. This is indeed the case\cite{bhw}. Consider first (\ref{alllevels}).
Introduce the continuous variable $\rho=\frac{2b_{1}}{\sqrt{N+b_{0}}}$ and replace $f(b_0,b_1)$ with $f(\rho)$. Now, expand
$$
\sqrt{(N+b_0+b_1)(N+b_0)}=(N+b_0)\left( 1+{1\over 2}{b_1\over
N+b_0}-{1\over 8}{b_1^2\over (N+b_0)^2}+....\right)
$$
and
$$
f\left(\rho-{1\over\sqrt{N+b_0}}\right)=f(\rho)-{1\over\sqrt{N+b_0}}{\partial f\over\partial \rho}
+{1\over 2(N+b_0)}{\partial^2 f\over\partial\rho^2}+...
$$
These expansions are only valid if $b_1\ll N+b_0$, which is certainly not always the case. However, for eigenfunctions with
all of their support in the small $\rho$ region the continuum limit of the recursion relation will give accurate answers. The 
recursion relation becomes
\begin{equation}
(2\alpha_{p,j^3} g_{YM}^{2}){1 \over 2} \left[-{\partial^2 \over
\partial {\rho}^{2}}+\rho^{2}\right]f(\rho)=\kappa f(\rho)\label{reccc}
\end{equation}
which is a harmonic oscillator with frequency 
$2\alpha_{p,j^3} g_{YM}^{2}$. We should only keep half of the oscillator states because 
the lengths of the rows (or columns) of the Young diagram are non-increasing, which implies that $b_1\ge 0$ and hence that $\rho\ge 0$.
Only wave functions that vanish at $\rho=0$ are allowed solutions. Thus, the energy spacing of the half oscillator states
is $4\alpha_{p,j^3} g_{YM}^{2}$. Clearly the description of the 
coefficients $f(b_{0},b_{1})$ obtained by solving (\ref{reccc}) will be accurate 
for the low lying oscillator eigenstates. Any operators corresponding to a finite energy state is accurately described. 
%
%
%
%
%
%

A few comments are in order. The solutions of the discrete recursion relations can be compared to the solution of the continuum 
differential equations. The agreement is perfect\cite{bhw}. Although the solution of our discrete recursion relation is in complete
agreement with the solution of the corresponding differential equation obtained by taking a continuum limit, notice that the
solution of the recursion relation does not make any additional assumptions. To obtain our differential equation we assumed that 
$b_1\ll N+b_0$. Thus, although solving the differential equation is easier, the solution is not as general.

Consider now the action of the dilatation operator when acting on three giant systems. We study the $\Delta =(1,1,1)$ example first.
It is a simple matter to check that the matrices $M^{(12)}$, $M^{(13)}$ and $M^{(23)}$ appearing in (\ref{frstexmpl}) commute and hence
can be simultaneously diagonalized. The result is the following 6 decoupled equations
{\small
\begin{eqnarray}
\label{easyweights}
  DO_{I}(b_1,b_2)&&=-2g_{YM}^2\Delta_{23}O_{I}(b_1,b_2),\qquad DO_{II}=-2g_{YM}^2\Delta_{12}O_{II}(b_1,b_2),\\
\nonumber
  DO_{III}(b_1,b_2)&&=-2g_{YM}^2\Delta_{13}O_{III}(b_1,b_2),\qquad 
  DO_{VI}(b_1,b_2)=-g_{YM}^2(\Delta_{23}+\Delta_{12}+\Delta_{13}) O_{VI}(b_1,b_2),\\
\nonumber
  DO_{V}(b_1,b_2)&&=-g_{YM}^2(\Delta_{23}+\Delta_{12}+\Delta_{13}) O_{V}(b_1,b_2),\qquad 
  DO_{IV}(b_1,b_2)=0.
\end{eqnarray}
}
Taking a continuum limit, assuming that $b_1,b_2\ll N+b_0$ we find
$$
  \Delta_{12}O(b_1,b_2)\to \left({\partial\over\partial x}-2{\partial\over\partial y}\right)^2 O(x,y) - {y^2\over 4}O(x,y)
$$
$$
  \Delta_{13}O(b_1,b_2)\to \left({\partial\over\partial x}+{\partial\over\partial y}\right)^2 O(x,y) - {(x+y)^2\over 4}O(x,y)
$$
$$
  \Delta_{23}O(b_1,b_2)\to \left(2{\partial\over\partial x}-{\partial\over\partial y}\right)^2 O(x,y) - {x^2\over 4}O(x,y)
$$
where $x=b_1/\sqrt{N+b_0}$ and $y=b_2/\sqrt{N+b_0}$. These all correspond to oscillators with an energy level spacing of\footnote{For example, 
for the oscillator corresponding to $\Delta_{12}$ we have $H={1\over 2}(aa^\dagger+a^\dagger a)$, $[a,a^\dagger]=2$, 
$a={\partial\over \partial x}-2{\partial\over\partial y}+{y\over 2}$ and 
$a^\dagger=-{\partial\over \partial x}+2{\partial\over\partial y}+{y\over 2}$.} $2$. However, again
because $b_1,b_2>0$ we keep only half the states and hence obtain oscillators with a level spacing of $4$. The corresponding eigenvalues of
the dilatation operator are $8ng_{YM}^2$ with $n$ an integer. This is remarkably consistent with what we found for the anomalous dimensions
for the two giant system. Of course, a very important difference is that since these oscillators live in a two dimensional space, there will
be an infinite discrete degeneracy in each level. Finally, it is also straight forward to show that
$$
  \Delta_{23}+\Delta_{12}+\Delta_{13} = 3{\partial^2\over\partial x^{+}{}^2}-{3\over 4}(x^+)^2+
                                        9{\partial^2\over\partial x^{-}{}^2}-{1\over 4}(x^-)^2
$$
where
$$ x^+={x+y\over\sqrt{2}},\qquad x^-={x-y\over\sqrt{2}}\, . $$
After rescaling the $x^-\to\sqrt{3}x^-$ we obtain a rotation invariant 2d harmonic oscillator with an energy level spacing of 3. Again
because $b_1,b_2>0$ we keep only half the states and hence obtain oscillators with a level spacing of 6. The corresponding eigenvalues of
the dilatation operator are $6ng_{YM}^2$ with $n$ an integer.

It is interesting to ask if we can diagonalize (\ref{easyweights}) directly without taking a continuum limit, since the resulting
spectrum is not computed with the assumption $b_1,b_2\sim \sqrt{N+b_0}$. Consider first the equation for $O_{II}(b_1,b_2)$. It is clear that
$\Delta_{12}$ does not change the value of $b_0$. In addition, the dilatation operator does not change the number of $Z$s in our
operator, so that $n_Z=3b_0+2b_2+b_1$ is fixed. This motivates the ansatz
$$
   O=\sum_{b_1}f(b_1,b_2)O_{II}(b_1,b_2)\Big|_{b_2=n_Z-3b_0-2b_1}
$$
Requiring that $DO=2g_{YM}^2\alpha_n O$ we obtain the recursion relation\footnote{Notice that we have replaced
$N+b_0+b_1+b_2\to N+b_0+b_1+b_2+1$ under the square root in the second term on the left hand side and we have replaced
$N+b_0+b_1\to N+b_0+b_1+1$ under the square root in the third term on the left hand side. We can do this with 
negligible error in the large $N$ limit.}
$$
-(2N+2b_0+2b_1+b_2)f_n(b_1,b_2)+\sqrt{(N+b_0+b_1)(N+b_0+b_1+b_2+1)}f_n(b_1-1,b_2+2)
$$
$$
+\sqrt{(N+b_0+b_1+1)(N+b_0+b_1+b_2)}f_n(b_1+1,b_2-2)=2g_{YM}^2\alpha_n f_n(b_1,b_2)
$$
where in the above equation $b_2=n_Z-3b_0-2b_1$.
Using the results of Appendix \ref{labelrelations}, it is a simple matter to verify that this recursion relation is solved by
$$
f_n=(-1)^{n}\left({1\over 2}\right)^{N+b_0+b_1+{b_2\over 2}}
\sqrt{
\left(^{2N+2b_0+2b_1+b_2}_{N+b_0+b_1+b_2}\right)
\left(^{2N+2b_0+2b_1+b_2}_{n}\right)
}
{}_2F_1\left({}^{-(N+b_0+b_1+b_2),-n}_{-(2N+2b_0+2b_1+b_2)};2\right)
$$
$$
2g_{YM}^2\alpha_n = 4ng_{YM}^2,\qquad n=0,1,2,\cdots, {\rm int}\left({n_Z-3b_0\over 2}\right)
$$
where $n_Z$ is the number of $Z$s in the restricted Schur polynomial, $b_0$ is fixed,
$b_2=n_Z-3b_0-2b_1$ and int($\cdot$) is the integer
part of the number in braces. Again, only half the states are retained because $b_1,b_2>0$ so that we
finally obtain a spacing of $8ng_{YM}^2$ - in perfect agreement with what we found above.
Notice that we obtain a set of eigenfunctions for each value of $b_0$, so that at infinite $N$ we have
an infinite degeneracy at each level.

The equation for $O_{III}(b_1,b_2)$ can be solved in the same way. We find
$$
f_n(b_0,b_1)=(-1)^{n}\left({1\over 2}\right)^{N+b_0+{b_1+b_2\over 2}}
\sqrt{
\left(^{2N+2b_0+b_1+b_2}_{N+b_0+b_1+b_2}\right)
\left(^{2N+2b_0+b_1+b_2}_{n}\right)
}
{}_2F_1\left({}^{-(N+b_0+b_1+b_2),-n}_{-(2N+2b_0+b_1+b_2)};2\right)
$$
$$
  n=0,1,....,{\rm min}(J,n_Z-2J)
$$
where $J=b_0+b_1$ is fixed, $b_2=n_Z-3b_0-2b_1$ and min($a,b$) is the smallest of the two integers
$a$ and $b$. Only half the states are retained because $b_1,b_2>0$ and we
again obtain a spacing of $8ng_{YM}^2$.
Notice that we obtain a set of eigenfunctions for each value of $J$, so that at infinite 
$N$ we again have an infinite degeneracy at each level.
For $O_{I}(b_1,b_2)$ we find
$$
f_n(b_0,b_1)=(-1)^{n}\left({1\over 2}\right)^{N+b_0+{b_1\over 2}}
\sqrt{
\left(^{2N+2b_0+b_1}_{N+b_0+b_1}\right)
\left(^{2N+2b_0+b_1}_{n}\right)
}
{}_2F_1\left({}^{-(N+b_0+b_1),-n}_{-(2N+2b_0+b_1)};2\right)
$$
$$
  n=0,1,....,{\rm int}\left({n_Z-J\over 2}\right)
$$
where $J=b_0+b_1+b_2$ is fixed and $b_2=n_Z-3b_0-2b_1$. Only half the states are retained because $b_1,b_2>0$ and we
again obtain a spacing of $8ng_{YM}^2$. Notice that we obtain a set of eigenfunctions for each value of $J$, so that at infinite 
$N$ we again have an infinite degeneracy at each level. It would be interesting to solve the recursion relations
arising from $O_V(b_1,b_2)$ and $O_{VI}(b_1,b_2)$. We will not do so here.

%
%
%
%
%
%
%
%
%
%
%
%
%
%
%

We now turn to the $j^3=O(1)$ example. We have already studied the continuum limit of the operators $\Delta_{12}$, $\Delta_{13}$, and
$\Delta_{23}$. In addition to these three operators, we will also need the continuum limit of $\Delta^{(a)}$, $\Delta^{(b)}$ and
$\Delta^{(c)}$. Taking $j,k\ll m$ and defining the continuum variables $w=k/\sqrt{m}$, $z=j/\sqrt{m}$ it is straight forward to obtain
$$
\Delta^{(a)}O(j,k)\to \left({\partial \over\partial w}+{\partial\over\partial z}\right)^2-{9\over 4}(z+w)^2
$$
$$
\Delta^{(b)}O(j,k)\to \left({\partial \over\partial z}-2{\partial\over\partial w}\right)^2-{9\over 4}w^2
$$
$$
\Delta^{(c)}O(j,k)\to \left({\partial \over\partial w}-2{\partial\over\partial z}\right)^2-{9\over 4}z^2\, .
$$
These all correspond to oscillators with an energy level spacing of $3$. Once again, because $j,k>0$, only half the states are
valid solutions implying a final level spacing of $6$. Finally, we need to consider the continuum limit of the coefficients 
appearing in (\ref{intdil}). Things simplify very nicely if we focus on those operators for which $\Delta = (n,n,n_3)$ and
$n_3\gg n$. In this case, we find
$$
   k^3=l^3={m\over 3}-n
$$
so that after taking the continuum limit (\ref{intdil}) becomes
{\footnotesize
$$
  DO(w,x,y,z)=g_{YM}^2 {(k^3)^2\over 3(j+k)^2}\left[
9\left( {\partial\over\partial x}-{\partial\over\partial y}\right)^2 -{(x-y)^2\over 4}
\right]
\left[
\left( {\partial\over\partial w}+{\partial\over\partial z}\right)^2 -9{(z+w)^2\over 4}
\right]
O(w,x,y,z)
$$
}
which is a direct product of harmonic oscillators! Although many interesting questions could be pursued at this point, we
will not do so here.

Finally, we have studied the action of the dilatation operator when acting on four giant systems. We will report the result for a four
giant system with four impurities and $\Delta =(1,1,1,1)$. There are a total of 24 operators that can be defined. The action
of the dilatation operator when acting on these 24 operators can be written in terms of (only the labels of the Young diagram
for the $Z$s is shown; the $b_i$ are again the difference in the lengths of the rows)
{\footnotesize
\begin{eqnarray}
\label{BD12}
&&\Delta_{12} O(b_1,b_2,b_3)=-(2N+2b_0+2b_1+2b_2+b_3)O(b_1,b_2,b_3)+\\
&&\sqrt{(N+b_0+b_1+b_2)(N+b_0+b_1+b_2+b_3)}\left(O(b_1,b_2+1,b_3-2)+O(b_1,b_2-1,b_3+2)\right),
\nonumber
\end{eqnarray}
{\vskip 0.1cm}
\begin{eqnarray}
\label{BD13}
&&\Delta_{13} O(b_1,b_2)=-(2N+2b_0+2b_1+b_2+b_3)O(b_1,b_2,b_3) + \\
\nonumber
&&\sqrt{(N+b_0+b_1)(N+b_0+b_1+b_2+b_3)}\left(O(b_1+1,b_2-1,b_3-1)+O(b_1-1,b_2+1,b_3+1)\right),
\end{eqnarray}
{\vskip 0.1cm}
\begin{eqnarray}
\label{BD14}
&&\Delta_{14} O(b_1,b_2,b_3)=-(2N+2b_0+b_1+b_2+b_3)O(b_1,b_2,b_3)+\\
\nonumber
&&\sqrt{(N+b_0)(N+b_0+b_1+b_2+b_3)}\left(O(b_1-1,b_2,b_3-1)+O(b_1+1,b_2,b_3+1)\right).
\end{eqnarray}
{\vskip 0.1cm}
\begin{eqnarray}
\label{BD23}
&&\Delta_{23} O(b_1,b_2,b_3)=-(2N+2b_0+2b_1+b_2)O(b_1,b_2,b_3)+\\
\nonumber
&&\sqrt{(N+b_0+b_1)(N+b_0+b_1+b_2)}\left(O(b_1+1,b_2-2,b_3+1)+O(b_1-1,b_2+2,b_3-1)\right).
\end{eqnarray}
{\vskip 0.1cm}
\begin{eqnarray}
\label{BD24}
&&\Delta_{24} O(b_1,b_2,b_3)=-(2N+2b_0+b_1+b_2)O(b_1,b_2,b_3)+\\
\nonumber
&&\sqrt{(N+b_0)(N+b_0+b_1+b_2)}\left(O(b_1-1,b_2-1,b_3+1)+O(b_1+1,b_2+1,b_3-1)\right).
\end{eqnarray}
{\vskip 0.1cm}
\begin{eqnarray}
\label{BD34}
&&\Delta_{34} O(b_1,b_2,b_3)=-(2N+2b_0+b_1)O(b_1,b_2,b_3)+\\
\nonumber
&&\sqrt{(N+b_0)(N+b_0+b_1)}\left(O(b_1-2,b_2+1,b_3)+O(b_1+2,b_2-1,b_3)\right).
\end{eqnarray}
}
After diagonalizing on the impurity labels we obtain the following decoupled problems: One BPS state
\begin{equation}
  DO(b_1,b_2,b_3)=0\, ,
\label{BPS}
\end{equation}
six operators with two rows participating
\begin{equation}
  DO(b_1,b_2,b_3)=-2g_{YM}^2\Delta_{ij}O(b_1,b_2,b_3),\quad (ij)=\{(12),(13),(14),(23),(24),(34)\}\, ,
\label{2rows}
\end{equation}
four doubly degenerate operators with three rows participating (so each equation appears twice) giving eight more operators
\begin{equation}
  DO(b_1,b_2,b_3)=-g_{YM}^2(\Delta_{12}+\Delta_{13}+\Delta_{23})O(b_1,b_2,b_3),\quad {\rm plus\,\, 3\,\, more}\, ,
\label{3rows}
\end{equation}
six operators of the type
\begin{equation}
  DO(b_1,b_2,b_3)=-g_{YM}^2(\Delta_{12}+\Delta_{23}+\Delta_{34}+\Delta_{14})O(b_1,b_2,b_3),\quad {\rm plus\,\, 5\,\, more}\, ,
\label{newtype}
\end{equation}
and finally three operators of the type
\begin{equation}
  DO(b_1,b_2,b_3)=-2g_{YM}^2(\Delta_{12}+\Delta_{34})O(b_1,b_2,b_3),\quad {\rm plus\,\, 2\,\, more}\, .
\label{anothernewtype}
\end{equation}
The equations (\ref{BPS}), (\ref{2rows}) and (\ref{3rows}) can be solved with a very simple extension of what was done for the three
giant system.

\section{Summary and Important Lessons}

Technology for working with restricted Schur polynomials has been 
developed\cite{Balasubramanian:2004nb,de Mello Koch:2007uu,de Mello Koch:2007uv,Bekker:2007ea,Bhattacharyya:2008rb,Bhattacharyya:2008rc,Koch:2010gp,VinceKate,bhw}
and is now at the stage where it is becoming useful.
In this article we have further added to this technology by describing a new version 
of Schur-Weyl duality that provides a powerful approach to the computation
and manipulation of the symmetric group operators appearing in the restricted Schur polynomials. Using this new technology we
have shown that it is straight forward to evaluate the action of the one loop dilatation operator on restricted Schur polynomials.
We studied the spectrum of one loop anomalous dimensions on restricted Schur polynomials that have $p$ long columns or rows.
For $p=3,4$ we have obtained the spectrum explicitly in a number of examples, and have shown that it is identical to the spectrum
of decoupled harmonic oscillators. This generalizes results obtained in \cite{Koch:2010gp,VinceKate,bhw}. The articles
\cite{Koch:2010gp,VinceKate,bhw} provided very strong evidence that the one loop dilatation operator acting on restricted
Schur polynomials with two long rows or columns is integrable. In this article we have found evidence that the dilatation
operator when acting on restricted Schur polynomials with $p$ long rows or columns is an integrable system. To obtain this
action we had to sum much more than just the planar diagrams so that {\it integrability in ${\cal N}=4$ super Yang-Mills
theory is not just a feature of the planar limit, but extends to other large $N$ but non-planar limits}.

The operators we have studied are dual to giant gravitons in the AdS$_5\times$S$^5$ background. These giant gravitons
have a world volume whose spatial component is topologically an S$^3$. The excitations of the giant graviton will correspond
to vibrational excitations of this S$^3$. At the quantum level, the energy in any particular vibrational mode will be quantized
and consequently, the free theory of giant gravitons should be a collection of decoupled oscillators, which provides a rather natural
interpretation of the oscillators we have found. 

\begin{figure}[h]
          \centering
          {\epsfig{file=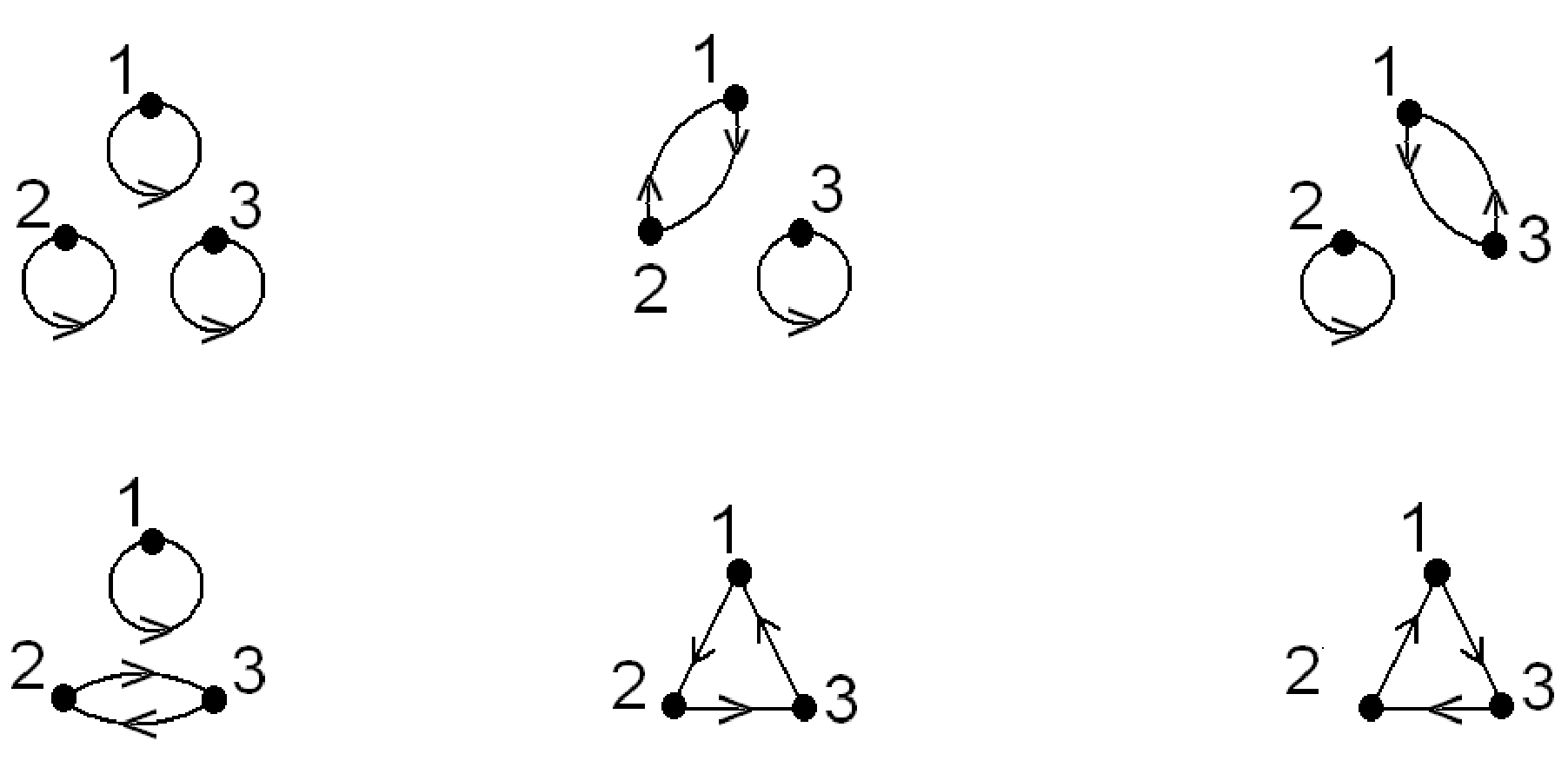,width=9.0cm,height=5.0cm}}
          \caption{A schematic representation of the possible excitations of a three giant system that are consistent with the Gauss Law.
                   Each giant graviton is represented by a labeled point. Lines represent open strings.}\label{fig:configs}
\end{figure}

Giant gravitons are D-branes. Attaching open strings to a D-brane provides a
concrete way to describe excitations. Are these open strings visible in our work?
Recall that, since the giant graviton has a compact world volume, the Gauss Law implies 
that the total charge on the giant's world volume must vanish. When enumerating
the possible stringy excitation states of a system of giant gravitons, only those states
consistent with the Gauss Law should be retained. In \cite{Balasubramanian:2004nb}, restricted
Schur polynomials corresponding to giants with ``string words'' attached were constructed and, remarkably,
the number of possible operators that could be defined in the gauge theory matches the number
of stringy excitation states of the system of giant gravitons. In this study we have replaced
open strings words with impurities $Y$, which does not modify the counting argument of \cite{Balasubramanian:2004nb}.
Our results add something new and significant to this story: not only does the counting of states match with that expected
from the Gauss Law, but, as we now explain, the structure of the action of the dilatation 
on restricted Schur polynomials itself is closely related to the Gauss Law. Consider
the three giant system with $\Delta = (1,1,1)$. For this $\Delta$ we have three impurities and hence we 
consider open string configurations with 3 open strings participating. There are three rows in the Young
diagrams, corresponding to three giant gravitons. Draw each giant graviton as a solid dot as shown in 
figure \ref{fig:configs}. The Gauss Law constraint then becomes the condition that there are an equal number of
open strings coming to each particular dot as there are leaving the particular dot. We find six possible open string 
configurations consistent with the Gauss Law as shown in figure \ref{fig:configs}. 
Our results suggest that the action of the one loop
dilatation operator is also coded into these diagrams. For each figure associated a factor of $\Delta_{ij}$ for a string
stretching between dots $i$ and $j$\footnote{$\Delta_{ij}$ in general is the natural generalization of the operators we 
defined in section 3, with boxes moving between rows $i$ and $j$.}. Since $\Delta_{ij}=\Delta_{ji}$, the last two
figures shown translate into the same equation, but because the string orientations are different they do represent
different states. A string starting and ending on the same dot does not contribute a $\Delta$.
Once the complete set of $\Delta_{ij}$ are read off the diagram, the action of the dilatation operator is given by
summing them and multiplying by $-g_{YM}^2$. Thus, the first diagram shown translates into

$$
  DO(b_1,b_2)=0\, .
$$
The last two diagrams both give
$$
  DO(b_1,b_2)=-g_{YM}^2(\Delta_{23}+\Delta_{12}+\Delta_{13}) O(b_1,b_2)\, .
$$
Finally, the remaining three diagrams give
$$
DO(b_1,b_2)=-2g_{YM}^2\Delta_{12}O(b_1,b_2),\quad  DO(b_1,b_2)=-2g_{YM}^2\Delta_{13}O(b_1,b_2),
$$
$$
  DO(b_1,b_2)=-2g_{YM}^2\Delta_{23}O(b_1,b_2)\, .
$$
This is exactly the action we finally obtained in (\ref{easyweights})!
The reader is invited to check that this matching between the possible open string configurations and 
the action of the dilatation operator continues for the four giant system with $\Delta = (1,1,1,1)$.
These two examples remove exactly one box from each row. However, the connection to the Gauss Law is
general. It is easy to check that it is consistent with the exact two row results obtained in \cite{Koch:2010gp,VinceKate,bhw}.   
In Appendix \ref{gausslawexample} we have given a summary of another detailed computation we have performed: a three giant
system with $\Delta=(3,2,1)$. The Gauss Law description is again perfect.
This connection provides a remarkably simple and general way of describing the action of
the one loop dilatation operator in the large $N$ but non-planar limit. For example, we
learn that the action of the dilatation operator is given by summing a collection
of operators $\Delta_{ij}$, each appearing some integer $n_{ij}$ number of times
$$
   DO(b_1,b_2)=-g_{YM}^2\sum_{ij}\, n_{ij}\Delta_{ij}\, O(b_1,b_2)\, .
$$
In Appendix \ref{continuumlimit} the action of this operator in a natural continuum limit
is studied and found to take the form
$$
  -g_{YM}^2\sum_{ij}\, n_{ij}\Delta_{ij}\to 
g_{YM}^2\sum_I D_I\left[-{\partial^2\over\partial x_I^2}+{x_I^2\over 4}\right]\, .
$$
Thus, at one loop and in this continuum limit, the dilatation operator reduces to an infinite
set of decoupled oscillators. The open string excitations of the $p$ giant graviton system are, at low
energy, described by a Yang-Mills theory with $U(p)$ gauge group. It seems natural to identify the $U(p)$ which played a
central role in our new Schur-Weyl duality with this gauge group.

Although we have written most of our formulas for Young diagrams with $p$ long rows, there is a straight forward
relation to the case with $p$ long columns - see section \ref{longcolmns}.
Further, although we have focused on the $SU(2)$ sector of the theory, it is not difficult to add another impurity flavor.
Indeed, a remarkable and surprising result of \cite{steph} which studied the $p=2$ case, is the fact that projectors from 
$S_{n+m+p}$ to $S_n\times S_m\times S_p$ can be constructed by taking a direct product of two $SU(2)$ projectors.
We have checked that this extends to the general case of projectors from 
$S_{n_1+n_2+...+n_k}$ to $S_{n_1}\times S_{n_2}\times\cdots\times S_{n_k}$, and for general $p$. 
This is presumably closely related to the math result \cite{generaldunkl}.

The Gauss Law constraint is an exact statement about the worldvolume physics of giant gravitons. 
For this reason we are optimistic that the connection we have found between the Gauss Law constraint and the action of
the one loop dilatation operator persists to higher loops.
Clearly despite the enormous number of diagrams that need to be summed to construct this large $N$ but non-planar limit, 
we are finding evidence that a simple integrable system emerges in the end!

{\vskip 0.5cm}

\noindent
{\it Acknowledgements:}
We would like to thank Tom Brown, Kevin Goldstein, Norman Ives, Jeff Murugan, Jurgis Pasukonis,
Sanjaye Ramgoolam, Stephanie Smith and Michael Stephanou 
for pleasant discussions and/or 
helpful correspondence. The work of DG is supported in part by a Claude Leon Fellowship.
This work is based upon research supported by the South African Research Chairs
Initiative of the Department of Science and Technology and National Research Foundation.
Any opinion, findings and conclusions or recommendations expressed in this material
are those of the authors and therefore the NRF and DST do not accept any liability
with regard thereto.

\appendix

\section{Elementary Facts from $U(p)$ Representation Theory}
\label{Unrep}

In this appendix we collect the background $U(p)$ representation theory needed to
understand our construction and diagonalization of the dilatation operator. 
There are many excellent references for this material. We have found \cite{Barut,Vilenkin}
useful. See also \cite{jan} for an extremely useful Clebsch-Gordan calculator.

\subsection{The Lie Algebra $u(p)$}

It is simpler to study the Lie algebra $u(p)$ instead of the group $U(p)$ itself.
Most results obtained for representations of $u(p)$ carry over to $U(p)$. 
In particular, the Clebsch-Gordan coefficients (which play a central role in our construction)
of their representations are identical.

The structure of the $u(p)$ algebra is easily illustrated using a specific basis.
Let $E_{ij}$ with $1\le i,j\le p$ be the matrix
$$
  (E_{ij})_{rs} = \delta_{ir}\delta_{js} \, ,
$$
so that it has only one non-zero matrix element. A convenient basis for 
the Lie algebra is generated by the matrices 
$$
  iE_{kk},\qquad 1\le k\le p,
$$
$$
 i(E_{k,k-1} + E_{k-1,k}), \qquad E_{k,k-1} - E_{k-1,k}, \qquad 1 < k \le p\, .
$$
$u(p)$ is spanned by real linear combinations of these matrices.
The restriction of any irreducible representation of $GL(p,C)$ onto
the subgroup $U(p)$ is also irreducible. Thus the carrier space of the
irreducible representations of $U(p)$ share the same basis as the 
irreducible representations of $GL(p,C)$ and consequently, a labeling for
$gl (p,C)$ irreducible representations is also a labeling for $u(p)$ irreducible representations. 
%
%
%
%
%
%
%
%
%
%

\subsection{Gelfand-Tsetlin Patterns}

Gelfand and Tsetlin have introduced a powerful labeling for $u(p)$ irreducible 
representations and the basis states of their carrier spaces\cite{gf}.
This labeling chooses basis states that are simultaneous eigenstates of all the 
matrices $J^{(l)}_z$, and further, explicit formulas are known for the matrix elements 
of the $J^{(l)}_\pm$ with respect to these basis states.

An inequivalent irreducible representation for $GL(p,C)$ is uniquely
given by specifying the sequence of $p$ integers
\begin{equation}
{\bf m} = (m_{1p},m_{2p}, \ldots, m_{pp}),
\end{equation}
satisfying $m_{kp} \ge m_{k+1,p}$ for $1 \le k \le p-1$.
Through out this article we call this sequence the weight of the irreducible representation.
The restriction of this irreducible representation onto the subgroup $GL(p-1,C)$ is reducible.
It decomposes into a direct sum of $GL(p-1,C)$ irreducible representations with highest
weights
\begin{equation}
{\bf m}' = (m_{1,p-1},m_{2,p-1}, \ldots, m_{p-1,p-1}),
\end{equation}
for which the ``betweenness'' conditions
$$
m_{kp} \ge m_{k,p-1} \ge m_{k+1,p}\qquad {\rm for}\qquad 1 \le k \le p-1
$$
hold. The carrier spaces of the $GL(p,C)$ irreducible representations now give rise to
(after restricting to the $GL(p-1,C)$ subgroup) $GL(p-1,C)$ irreducible representations.
We can keep repeating this procedure until we get to $GL(1,C)$ which has one-dimensional
carrier spaces. The Gelfand-Tsetlin labeling exploits this sequence of subgroups to 
label the basis states using what are called {\sl Gelfand-Tsetlin patterns}. These are 
triangular arrangements of integers, denoted by $M$, with the structure
$$
{\footnotesize
M=\left[ 
\begin {array}{ccccccc} 
m_{1p} &            & m_{2p}  & \dots     & m_{p-1,p} &             & m_{pp}\\\noalign{\medskip}
       & m_{1,p-1}  &         & m_{2,p-1} & \dots     & m_{p-1,p-1} &       \\\noalign{\medskip}
       &            & \dots   &\dots      &  \dots    &             &       \\\noalign{\medskip}
       &            &m_{12}   &           &m_{22}     &             &       \\\noalign{\medskip}
       &            &         &m_{11}     &           &             &
\end {array} \right] }
$$
The top row contains the weight that specifies the irreducible representation of the state and the entries 
of lower rows are subject to the betweenness condition. Thus, the lower rows give the sequence of irreducible
representations our state belongs to as we pass through successive restrictions from $GL(p,C)$ to 
$GL(p-1,C)$ to ... to $GL(1,C)$. 
The dimension of an irreducible representation with weight ${\bf m}$ is equal to the number of valid
Gelfand-Testlin patterns having ${\bf m}$ as their top row. 

\subsection{$\Sigma$ and $\Delta$ Weights}
\label{weights}

We make extensive use of two weights in our construction: $\Sigma$-weights and $\Delta$ weights. Define
the row sum
$$
   \sigma_l(M)=\sum_{k=1}^l m_{k,l}\, .
$$
The sequence of row sums defines the sigma weight
$$
  \Sigma (M) =(\sigma_p(M),\sigma_{p-1}(M),\cdots,\sigma_1(M))\, .
$$
The sigma weights do not provide a unique label for the states in the carrier space.
Indeed, it is possible that $\Sigma (M)=\Sigma (M')$ but $M\ne M'$. The number of
states $\vec{v}(M)$ in the carrier space that have the same $\Sigma$ weight
$\Sigma=\Sigma (M)$ is called the inner multiplicity $I(\Sigma)$ of the state. The inner multiplicity
plays an important role in determining how many restricted Schur polynomials can be defined.
The $\Delta$ weights are defined in terms of differences between row sums  
$$
  \Delta (M) =(\sigma_p(M)-\sigma_{p-1}(M),\sigma_{p-1}(M)-\sigma_{p-2}(M),\cdots ,\sigma_1(M)-\sigma_0(M))
$$
$$
     \equiv (\delta_p(M),\delta_{p-1}(M),\cdots\delta_1(M))
$$
where $\sigma_0\equiv 0$. We could also ask how many states in the carrier space have the same $\Delta$,
denoted $I(\Delta)$. It is clear that $I(\Delta)=I(\Sigma)$.

The $\Delta$ weights play an important role in determining how the three Young diagram labels $R,(r,s)$ of the restricted Schur polynomials 
$\chi_{R,(r,s)jk}$ translate into a set of $U(p)$ labels. It tells us how boxes were removed from $R$ to obtain $r$. Further, the multiplicity 
labels $jk$ of the restricted Schur polynomial each run over the inner multiplicity. 

\subsection{Relation between Gelfand-Tsetlin Patterns and Young Diagrams}
\label{semistandard}

There is a one-to-one correspondence between $\Sigma$ weights and
Young diagrams, and between Gelfand-Tsetlin patterns and semi-standard Young
tableaux. The language of semi-standard Young tableau is a key ingredient
in understanding how the three Young diagram labels $R,(r,s)$ of the restricted 
Schur polynomials $\chi_{R,(r,s)jk}$ translate into the $U(p)$ language, so we will 
review this connection here.

Recall that a Young diagram is an arrangement of boxes in rows and columns
in a single, contiguous cluster of boxes such that the left borders of all rows are aligned and
each row is not longer than the one above. The empty Young diagram consisting of no boxes is a valid
Young diagram. For a $u(p)$ irreducible representation there are at most $p$ rows.
Every Young diagram uniquely labels a $u(p)$ irreducible representation.

A (semi-standard) Young tableau is a Young diagram, with labeled boxes.
The rules for labeling are that each box contains a single integer between 1 
and $p$ inclusive, the numbers in each row of boxes weakly increase from left to right
(each number is equal to or larger than the one to its left) and
the numbers in each column strictly increase from top to bottom
(each number is strictly larger than the one above it).

The basis states of a $u(p)$ representation identified
by a given Young diagram $D$ can be uniquely labeled by the set of all 
semi-standard Young tableaux. The dimension of a carrier space labeled 
by a Young diagram is equal to the number of valid Young tableaux with 
the same shape as the Young diagram.

Each Gelfand-Tsetlin pattern $M$ corresponds to a unique Young tableau.
We will now explain how to construct the Young tableau given a Gelfand-Tsetlin
pattern. Each step in the procedure is illustrated with a concrete example
given by the following Gelfand-Tsetlin pattern
$$
{\footnotesize
\left[ 
\begin {array}{ccccccc} 
    4  &            &    3    &           &      1    &             &   1   \\\noalign{\medskip}
       &     3      &         &     2     &           &    1        &       \\\noalign{\medskip}
       &            &    3    &           &      2    &             &       \\\noalign{\medskip}
       &            &         &     2     &           &             &
\end {array} \right]\, .}
$$
Start with an empty Young diagram (no labels). The first line of the Gelfand-Tsetlin
pattern tells you the shape of the Young diagram - $m_{in}$ is the number of boxes in row
$i$. Thus, the information specifying the irreducible representation resides in the topmost row
of the pattern. 
%
%
The last row of the Gelfand-Tsetlin pattern tells us which boxes are labeled with a $1$. Imagine
superposing the smaller Young diagram defined by the last row of the pattern onto the full Young diagram,
so that the topmost and leftmost boxes of the two are identified. Label all boxes of this smaller
Young diagram with a $1$. For the example we consider
$$
  {\footnotesize \young(11\,\,,\,\,\,,\,,\,)\, .}
$$
The second last row of the pattern tells us which boxes are labeled with a $2$.
Again superpose the smaller Young diagram defined by the second last row of the pattern onto the full Young diagram
and again identify the topmost and leftmost boxes of the two. Label all empty boxes of this smaller
Young diagram with a $2$. For the example we consider
$$
  {\footnotesize \young(112\,,22\,,\,,\,)\, .}
$$
Keep repeating this procedure until you have used the first row to identify the boxes labeled $p$.
The result is a semi-standard Young tableau. The semi standard Young tableau for the example we consider is
$$
 {\footnotesize  \young(1124,224,3,4)\, .}
$$
The number of boxes containing the number $l$ in tableau row $k$ is given by 
$m_{kl} - m_{k,l-1}$ and we set $m_{kl} \equiv 0$ if $k > l$.
The converse process of transcribing a semi-standard Young tableau to a Gelfand-Tsetlin pattern
is now obvious.

The components $\delta_l(M)$ of the $\Delta$ weight of a Gelfand-Tsetlin pattern $M$, 
is the number of boxes containing $l$ in the tableau corresponding to $M$.
Thus, the tableau corresponding to two patterns with the same $\Delta$ weight
contain the same set of entries (i.e. the same number of $l$-boxes) but arranged 
in different ways. One interpretation for the inner multiplicity is that it 
simply counts the number of ways to arrange the relevant fixed set of entries in the
tableau.

\subsection{Clebsch-Gordon Coefficients}
\label{Clebschs}

Let $R$ and $S$ be two irreducible unitary representations of the group $U(p)$. The tensor product of these
representations decomposes into a direct sum of irreducible components
\begin{equation}
   R\otimes S =\sum_T\oplus \nu (T) T \, .
\label{dprod}
\end{equation}
In general a particular irreducible representation $T$ can appear more than once in the product $R\otimes S$.
The integer $\nu (T)$ indicates the multiplicity of $T$ in this decomposition. For the applications we have in
mind, we will need the direct product of an arbitrary representation with weight ${\bf m}_n$ with the defining
representation which has weight $(1,{\bf 0})$. In this case
all multiplicities are equal to 1 and we need not worry about tracking multiplicities. Use the notation
${\bf m}_R$ to denote the weight of irreducible representation $R$ and $M_R$ to denote the Gelfand-Tsetlin
pattern for a particular state in the carrier space of this irreducible representation. There are two natural bases 
for $R\otimes S$. The first is simply obtained by taking the direct product of the states spanning the carrier
spaces of $R$ and $S$. The states in this basis are labeled, using a bra/ket notation, as\footnote{When discussing
and using the Clebsch-Gordan coefficients, we prefer to use a bra/ket notation. In our previous notation we could
write this basis vector as $\vec{v}(M_R)\otimes\vec{v}(M_S)$.}
$$
     |{\bf m}_R,M_R; {\bf m}_S,M_S\rangle\, .
$$
The second natural basis is given as a direct sum over the bases of the carrier spaces for the irreducible representations
$T$ appearing in the sum on the right hand side of (\ref{dprod}). The states in this basis are labeled as\footnote{In general
one would also need to include a multiplicity label among the labels for these states.} 
$$
     |{\bf m}_T,M_T\rangle
$$
where $T$ runs over all irreducible representations appearing in the sum on the right hand side of (\ref{dprod}). The Clebsch-Gordan
coefficients supply the transformation matrix which takes us between the two bases. They are written as the overlap
$$
   \langle {\bf m}_R,M_R; {\bf m}_S,M_S |{\bf m}_T,M_T\rangle\, .
$$
From now on we will drop the $R,S,T$ labels which are actually redundant since the particular irreducible representations we consider
are uniquely labeled by the weight which is recorded in the first row of the corresponding Gelfand-Tsetlin patterns.
It is known that we can write the Clebsch-Gordan coefficients of $U(p)$ in terms of the Clebsch-Gordan coefficients of $U(p-1)$
as\footnote{Again, we are using the fact that for our applications multiple copies of the same representation are absent. In general one needs to 
worry about multiplicities.}
$$
\langle {\bf m}_p,M; {\bf m}'_p,M' |{\bf m}''_p ,M''\rangle
=
{\footnotesize \left(
\begin{array}{c}
 {\bf m}_{p} \\\noalign{\medskip}
 {\bf m}_{p-1}
\end{array}
\begin{array}{c}
 {\bf m}'_{p} \\\noalign{\medskip}
 {\bf m}'_{p-1}
\end{array}
\Big|
\begin{array}{c}
{\bf m}''_{p} \\\noalign{\medskip}
{\bf m}''_{p-1}
\end{array}
\right)}
%
\langle {\bf m}_{p-1},M_1; {\bf m}'_{p-1},M_1' |{\bf m}_{p-1}'',M_1''\rangle\, .
$$
On the right hand side we have the Clebsch-Gordan coefficients of the group $U(p-1)$ and on the left hand side we have the Clebsch-Gordan
coefficients of the group $U(p)$. The weights ${\bf m}_p,{\bf m}_p',{\bf m}_p''$ label irreducible representations of $U(p)$, while 
weights ${\bf m}_{p-1},{\bf m}_{p-1}',{\bf m}_{p-1}''$ label irreducible representations of $U(p-1)$. The Gelfand-Tsetlin patterns
$M_1,M_1'$ and $M_1''$ are obtained from $M,M'$ and $M''$ respectively by removing the first row. Thus, the weights
${\bf m}_{p-1},{\bf m}_{p-1}',{\bf m}_{p-1}''$ correspond with the second rows in $M,M'$ and $M''$. The coefficients
{\footnotesize $\left(
\begin{array}{c}
 {\bf m}_{p} \\\noalign{\medskip}
 {\bf m}_{p-1}
\end{array}
\begin{array}{c}
 {\bf m}'_{p} \\\noalign{\medskip}
 {\bf m}'_{p-1}
\end{array}
\Big|
\begin{array}{c}
{\bf m}''_{p} \\\noalign{\medskip}
{\bf m}''_{p-1}
\end{array}
\right)$} are called the scalar factors of the Clebsch-Gordan coefficients
$\langle {\bf m}_p,M; {\bf m}'_p,M' |{\bf m}''_p,M''\rangle$. Applying the above factorization to the chain of subgroups
referenced by the Gelfand-Tsetlin pattern, we obtain
$$
\langle {\bf m}_p,M; {\bf m}'_p,M' |{\bf m}''_p,M''\rangle
=
{\footnotesize \left(
\begin{array}{c}
 {\bf m}_{p} \\\noalign{\medskip}
 {\bf m}_{p-1}
\end{array}
\begin{array}{c}
 {\bf m}'_{p} \\\noalign{\medskip}
 {\bf m}'_{p-1}
\end{array}
\Big|
\begin{array}{c}
{\bf m}''_{p} \\\noalign{\medskip}
{\bf m}''_{p-1}
\end{array}
\right)
%
\left(
\begin{array}{c}
 {\bf m}_{p-1} \\\noalign{\medskip}
 {\bf m}_{p-2}
\end{array}
\begin{array}{c}
 {\bf m}'_{p-1} \\\noalign{\medskip}
 {\bf m}'_{p-2}
\end{array}
\Big|
\begin{array}{c}
{\bf m}''_{p-1} \\\noalign{\medskip}
{\bf m}''_{p-2}
\end{array}
\right)
\left(
\begin{array}{c}
 {\bf m}_{p-2} \\\noalign{\medskip}
 {\bf m}_{p-3}
\end{array}
\begin{array}{c}
 {\bf m}'_{p-2} \\\noalign{\medskip}
 {\bf m}'_{p-3}
\end{array}
\Big|
\begin{array}{c}
{\bf m}''_{p-2} \\\noalign{\medskip}
{\bf m}''_{p-3}
\end{array}
\right)
\cdots}
$$
Thus, the Clebsch-Gordan coefficients can be written as a product of scalar factors.

There is a selection rule for the Clebsch-Gordan coefficients. The Clebsch-Gordan coefficients
vanish unless
$$
\sum_{i=1}^j m_{ij} +
\sum_{i=1}^j m_{ij}'=
\sum_{i=1}^j m_{ij}''\qquad j=1,2,...,p\, .
$$

The only Clebsch-Gordan coefficient that we will need for our applications come from taking the product
of some general representation ${\bf m}_p$ with the fundamental representation. The weight of the fundamental
representation is $(1,0,...,0)$ with $p-1$ 0s appearing. The product we consider has been studied and the 
following result is known
\begin{equation}
\label{tprod}
   {\bf m}_p\otimes (1,{\bf 0})= \sum_{i=1}^m {\bf m}_p^{+i}\, .
\end{equation}
where ${\bf m}_p^{+i}$ is obtained from ${\bf m}_p$ by replacing $m_{ip}$ by $m_{ip}+1$.
Of course, if this replacement does not lead to a valid Gelfand-Tsetlin pattern there is no
corresponding representation. The term with the illegal pattern should be dropped from the right hand side
of (\ref{tprod}). From (\ref{tprod}) we see that multiple copies of the same irreducible representation
are absent on the right hand side. We have made use of this repeatedly in this subsection.
These Clebsch-Gordan coefficients factor into products of scalar factors of the form
{\footnotesize
$$
\left(
\begin{array}{c}
 {\bf m}_{p} \\\noalign{\medskip}
 {\bf m}_{p-1}
\end{array}
\begin{array}{c}
 (1,{\bf 0})_p \\\noalign{\medskip}
 (1,{\bf 0})_{p-1}
\end{array}
\Big|
\begin{array}{c}
{\bf m}^{+i}_{p} \\\noalign{\medskip}
{\bf m}^{+j}_{p-1}
\end{array}
\right)\qquad{\rm or}\qquad
\left(
\begin{array}{c}
 {\bf m}_{p} \\\noalign{\medskip}
 {\bf m}_{p-1}
\end{array}
\begin{array}{c}
 (1,{\bf 0})_p \\\noalign{\medskip}
 (0,{\bf 0})_{p-1}
\end{array}
\Big|
\begin{array}{c}
{\bf m}^{+i}_{p} \\\noalign{\medskip}
{\bf m}_{p-1}
\end{array}
\right)\, .$$
}
Explicit formulas for these scalar factors are known
$$
{\footnotesize
\left(
\begin{array}{c}
 {\bf m}_{p} \\\noalign{\medskip}
 {\bf m}_{p-1}
\end{array}
\begin{array}{c}
 (1,{\bf 0})_p \\\noalign{\medskip}
 (1,{\bf 0})_{p-1}
\end{array}
\Big|
\begin{array}{c}
{\bf m}^{+i}_{p} \\\noalign{\medskip}
{\bf m}^{+j}_{p-1}
\end{array}
\right)}=S(i,j)\left|
{\prod_{k\ne j}^{p-1}(l_{k,p-1}-l_{ip}-1)\prod_{k\ne i}^p(l_{kp}-l_{j,p-1})\over
 \prod_{k\ne i}^p(l_{kp}-l_{ip})\prod_{k\ne j}^{p-1}(l_{k,p-1}-l_{j,p-1}-1)}
\right|^{1\over 2}
$$
$$
{\footnotesize
\left(
\begin{array}{c}
 {\bf m}_{p} \\\noalign{\medskip}
 {\bf m}_{p-1}
\end{array}
\begin{array}{c}
 (1,{\bf 0})_p \\\noalign{\medskip}
 (0,{\bf 0})_{p-1}
\end{array}
\Big|
\begin{array}{c}
{\bf m}^{+i}_{p} \\\noalign{\medskip}
{\bf m}_{p-1}
\end{array}
\right)} = 
\left|
\prod_{j=1}^{p-1} (l_{j,p-1}-l_{ip}-1)\over \prod_{j\ne i}^p(l_{jp}-l_{ip})\right|^{1\over 2}
$$
where $l_{sk}=m_{sk}-s$, $S(i,j)=1$ if $i\le j$ and $S(i,j)=-1$ if $i>j$.

\subsection{Explicit Association of labeled Young Diagrams and Gelfand-Tsetlin Patterns}

The association we spell out in this section is at the heart of our new Schur-Weyl duality and it demonstrates
how we associate an action of $U(p)$ to a Young diagram with $p$ rows or columns.
First consider the case of a Young diagram with $O(1)$ rows and $O(N)$ columns. This situation is relevant for
the description of AdS giant gravitons. We consider Young diagrams in which a certain number of boxes are labeled. 
To keep the argument general assume that the Young diagram has $p$ rows. These labeled boxes are put into a 
one-to-one correspondence with $p$-dimensional vectors. If box $i$ appears in the $q^{\rm th}$ row it is associated
to vector with components
$$
    \vec{v}(i)_k=\delta_{kq}\, .
$$
These states live in the carrier space of the fundamental representation of $U(p)$. In this subsection we would like to
clearly spell out the Gelfand-Testlin pattern labeling of these vectors. We will spell out our conventions for $U(3)$.
The generalization to any $p$ is trivial. Our conventions are
$$
{\footnotesize
  \young(\,\,\,\,\,\,\,\,\,\,\,\,\,\,\,1,\,\,\,\,\,\,\,\,\,,\,\,\,\,\,)
\leftrightarrow\left[
\begin{array}{ccccc}
1  &   &0   &   &0\\\noalign{\medskip}
   & 1 &    & 0 & \\\noalign{\medskip}
   &   &1   &   & \\\noalign{\medskip}
\end{array}
\right]}
$$
$$
{\footnotesize
  \young(\,\,\,\,\,\,\,\,\,\,\,\,\,\,\,\,,\,\,\,\,\,\,\,\,1,\,\,\,\,\,)
\leftrightarrow
\left[
\begin{array}{ccccc}
1  &   &0   &   &0\\\noalign{\medskip}
   & 1 &    & 0 & \\\noalign{\medskip}
   &   &0   &   & \\\noalign{\medskip}
\end{array}
\right]}
$$
$$
{\footnotesize
  \young(\,\,\,\,\,\,\,\,\,\,\,\,\,\,\,\,,\,\,\,\,\,\,\,\,\,,\,\,\,\,1)
\leftrightarrow
\left[
\begin{array}{ccccc}
1  &   &0   &   &0\\\noalign{\medskip}
   & 0 &    & 0 & \\\noalign{\medskip}
   &   &0   &   & \\\noalign{\medskip}
\end{array}
\right]}
$$
The particular label (the $1$ in this case) is irrelevant - its the row the label appears in that determines the pattern.

For the case of Young diagrams with $O(N)$ rows and $O(1)$ columns we have
$$
{\footnotesize 
\young(\,\,\,,\,\,\,,\,\,1,\,\,,\,\,,\,\,,\,,\,,\,)
\leftrightarrow
\left[
\begin{array}{ccccc}
1  &   &0   &   &0\\\noalign{\medskip}
   & 1 &    & 0 & \\\noalign{\medskip}
   &   &1   &   & \\\noalign{\medskip}
\end{array}
\right],\,
%
\young(\,\,\,,\,\,\,,\,\,\,,\,\,,\,\,,\,1,\,,\,,\,)
\leftrightarrow
\left[
\begin{array}{ccccc}
1  &   &0   &   &0\\\noalign{\medskip}
   & 1 &    & 0 & \\\noalign{\medskip}
   &   &0   &   & \\\noalign{\medskip}
\end{array}
\right],\,
%
\young(\,\,\,,\,\,\,,\,\,\,,\,\,,\,\,,\,\,,\,,\,,1)
\leftrightarrow
\left[
\begin{array}{ccccc}
1  &   &0   &   &0\\\noalign{\medskip}
   & 0 &    & 0 & \\\noalign{\medskip}
   &   &0   &   & \\\noalign{\medskip}
\end{array}
\right]}
$$
This situation is relevant for the description of sphere giant gravitons.
Note that in addition to specifying the above correspondence between Gelfand-Tsetlin
patterns and labeled Young diagrams, one also needs to assign the phases of the different
states carefully. For a discussion see section \ref{columns}.

\subsection{Last Remarks}
\label{lastremarks}

A box in row $i$ and column $j$ has a factor equal to $N-i+j$.
To obtain the hook length associated to a given box, draw a line starting from the given
box towards the bottom of the page until you exit the Young diagram, and another line
starting from the same box towards the right until you again exit the diagram. These
two lines form an elbow - the hook.
The hook length for the given box is obtained by counting the number of boxes the
elbow belonging to the box passes through. Here is a Young diagram with the hook
lengths filled in
$$ \young(531,31,1) $$
For Young diagram $R$ we denote the product of the hook lengths by ${\rm hooks}_R$.

\section{Elementary Facts from $S_n$ Representation Theory}
\label{Snrep}

The complete set of irreducible representations of $S_n$ are uniquely labeled by Young diagrams with $n$ boxes.
From this Young diagram we can construct both a basis for the carrier space of the representation as well as the
matrices representing the group elements. We will review these constructions in this Appendix. A useful reference for
this material is \cite{hammermesh}.

\subsection{Young-Yamonouchi Basis}

The elements of this basis are labeled by numbered Young diagrams - a Young tableau. For a Young diagram with
$n$ boxes, each box in the tableau is labeled with a unique integer $i$ with $1\le i\le n$. In our conventions this numbering is done in
such a way that if all boxes with labels less than $k$ with $k < n$ are dropped, a valid Young diagram remains. As an example, if we 
consider the irreducible representation of $S_4$ corresponding to
$$
{\footnotesize  \yng(2,2)}
$$
then the allowed labels are
$$
{\footnotesize  \young(43,21)\qquad \young(42,31)\, .}
$$
Examples of labels that are not allowed include
$$
{\footnotesize  \young(41,32)\qquad\young(12,34)\qquad \young(13,24)\, .}
$$
For any given Young diagram the number of valid labels is equal to the dimension of the irreducible representation and each
label corresponds to a vector in the basis for the carrier space. This basis is orthonormal so that, for example
$$
  \left\langle{\footnotesize \young(43,21)}\right|\left.{\footnotesize \young(43,21)}\right\rangle =1,\qquad 
  \left\langle{\footnotesize\young(43,21)}\right|\left.{\footnotesize\young(42,31)}\right\rangle =0\, .
$$

\subsection{Young's Orthogonal Representation}

A rule for constructing the matrices representing the elements of the symmetric group is easily given by specifying 
the action of the group elements on the Young-Yamonouchi basis. The rule is only stated for ``adjacent permutations''
which correspond to cycles of the form $(i,i+1)$. This is enough because these adjacent permutations generate the complete 
group. To state the rule it is helpful to associate to each box a factor\footnote{This number is also commonly called the ``weight''
of the box. Here we will refer to it as the factor since we do not want to confuse it with the weight of the Gelfand-Tsetlin pattern.}. 
The factor of a box in the $i^{\rm th}$ row and the $j^{\rm th}$ column is given by $K-i+j$. Here $K$ is an arbitrary integer that will 
not appear in any final results. We will denote the factor of the box labeled $l$ by $c_l$. Let $\hat{T}$ denote a Young tableau
corresponding to Young diagram $T$ and let $\hat{T}_{ij}$ denote exactly the same tableau, but with boxes $i$ and $j$ swapped. The rule for the
action of the group elements on the basis vectors of the carrier space is
$$
  \Gamma_T\left( (i,i+1)\right)\left| \hat{T}\right\rangle = {1\over c_i-c_{i+1}}\left| \hat{T}\right\rangle +
  \sqrt{1-{1\over (c_{i}-c_{i+1})^2}}\left| \hat{T}_{i,i+1}\right\rangle\, .
$$

\subsection{Partially labeled Young diagrams}
\label{partdiagram}

Consider a Young diagram containing $n+m$ boxes so that it labels an irreducible representations of $S_{n+m}$. We will often consider
``partially labeled'' Young diagrams, which are obtained by labeling $m$ boxes. The remaining $n$ boxes are not labeled. We only consider
labelings which have the property that if all boxes with labels $\le i$ are dropped, the remaining boxes are still arranged in a legal
Young diagram. We refer to this as a ``sensible labeling''.
What is the interpretation of these partially labeled
Young diagrams? To make the discussion concrete, we will develop the discussion using an explicit example. For the example we consider
take $n=m=3$ and use the following partially labeled Young diagram
\begin{equation}
\label{partlabel}
 {\footnotesize \young(\,\,1,\,2,3)\, .}
\end{equation}
If the labeling is completed, this partially labeled diagram will give rise to a number of Young tableau. For our present example
two tableau are obtained
$$
{\footnotesize  \young(651,42,3)\qquad \young(641,52,3)\, .}
$$
Each of these represents a vector in the carrier space of the $S_6$ irreducible representation labeled by the Young diagram
${\tiny \yng(3,2,1)}$. Thus, a partially labeled Young diagram stands for a collection of states. Next, note that the subspace
formed by this collection of states is invariant (you don't get transformed out of the subspace) under the action of the $S_3$
subgroup which acts on the boxes labeled 4,5 and 6. Thus, this subspace is a representation of $S_3$. In fact, it is easy to see 
that it is the irreducible representation labeled by ${\tiny \yng(2,1)}$. This Young diagram can be obtained by dropping all the 
labeled boxes in (\ref{partlabel}). From this example we can now extract the general rule:

{\vskip 0.5cm}

\noindent
{\bf Key Idea:}
{\sl A partially labeled Young diagram that has $n+m$ boxes, $m$ of which are labeled, stands for a collection of states which
furnish the basis for an irreducible representation of $S_n\times (S_1)^m$. The Young diagram that labels the representation
of the $S_n$ subgroup is given by dropping all labeled boxes.}

{\vskip 0.5cm}

Finally, note that the only representations $r$ that are subduced by $R$ are those with Young diagrams that can be obtained by 
pulling boxes off $R$. This follows immediately from the well known subduction rule for the symmetric group which states that
an irreducible representation of $S_n$ labeled by Young diagram $R$ with $n$ boxes will subduce all possible representations 
$R'_i$ of $S_{n-1}$, where $R_i'$ is obtained by removing any box of $R$ that can be removed such the we are left with a
valid Young diagram after removal. Each such irreducible representation of the subgroup is subduced once.

\subsection{Simplifying Young's Orthogonal Representation}
\label{simpleyoung}

In this section we would like to consider a collection of partially labeled Young diagrams. 
A total of $m$ boxes are labeled, with a unique integer $i$ ($1\le i\le m$) appearing in each box.
The set of boxes to be removed are the same for every partially labeled Young diagram.
The set of partially labeled Young diagrams we consider is given by including
all possible ways in which the $m$ boxes in the Young diagrams can sensibly be labeled. 
We can consider the action of the $S_m$ subgroup which acts on the labeled boxes.
This action will mix these partially labeled Young diagrams.

We will consider Young diagrams with $p$ rows built out of $O(N)$ boxes. For the generic
operator we consider, the difference in the length between any two rows will be $O(N)$. If
we consider the case $m=\gamma N$ with $\gamma\sim O(N^0)\ll 1$, any two labeled boxes
($i$ and $j$ say) that are not in the same row will have factors that obey $|c_i-c_j|\sim O(N)$. 
Young's orthogonal representation is particularly useful because it simplifies dramatically 
in this situation. Indeed, if the boxes $i$ and $i+1$ are in the same
row, $i+1$ must sit in the next box to the left of $i$ so that
\begin{eqnarray}
\Gamma_R \left( (i,i+1)\right)|{\rm same\, row\, state}\rangle
= |{\rm same\, row\, state}\rangle\, .
\end{eqnarray}
The same state appears on both sides of this last equation.
If $i$ and $i+1$ are in different rows, then $c_i-c_{i+1}$ must itself be $O(N)$. In this case, 
at large $N$ replace ${1\over c_i-c_{i+1}}=O(b_1^{-1})$ by 0 and
$\sqrt{1-{1\over (c_i-c_{i+1})^2}}=1-O(b_1^{-1})$ by 1 so that
\begin{eqnarray}
\Gamma_R \left( (i,i+1)\right)|{\rm different\, row\,
state}\rangle = |{\rm swapped\, different\, row\, state}\rangle\, .
\end{eqnarray}
The notation in this last equation is indicating two things: $i$ and $i+1$ are in different rows and the states on the two
sides of the equation differ by swapping the $i$ and $i+1$ labels. An example illustrating these rules is
$$
{\footnotesize
\Gamma_R\left( (1,2)\right)\left|
\young(\,\,\,\,\,\,\,\,\,\,321,\,\,\,\,\,)\right\rangle =\left|
\young(\,\,\,\,\,\,\,\,\,\,321,\,\,\,\,\,)\right\rangle }
$$
$$
{\footnotesize
\Gamma_R\left((1,2)\right)\left|
\young(\,\,\,\,\,\,\,\,\,\,32,\,\,\,\,\,1)\right\rangle 
=\left| \young(\,\,\,\,\,\,\,\,\,\,31,\,\,\,\,\,2)\right\rangle\, .}
$$

We will also consider Young diagrams with $p$ columns built out of $O(N)$ boxes. For the generic
operator we consider, the difference in the length between any two columns will be $O(N)$. Since
we consider the case $m=\gamma N$ with $\gamma\sim O(N^0)\ll 1$, any two labeled boxes
($i$ and $j$ say) that are not in the same column will again have factors that obey $|c_i-c_j|\sim O(N)$. 
If the boxes $i$ and $i+1$ are in the same column, $i+1$ must sit above $i$ so that
\begin{eqnarray}
\Gamma_R \left( (i,i+1)\right)|{\rm same\, column\, state}\rangle
= -|{\rm same\, column\, state}\rangle\, .
\end{eqnarray}
The same state appears on both sides of this last equation.
If $i$ and $i+1$ are in different columns, then $c_i-c_{i+1}$ must itself be $O(N)$. In this case, 
at large $N$ again replace ${1\over c_i-c_{i+1}}=O(b_1^{-1})$ by 0 and
$\sqrt{1-{1\over (c_i-c_{i+1})^2}}=1-O(b_1^{-1})$ by 1 so that
\begin{eqnarray}
\Gamma_R \left( (i,i+1)\right)|{\rm different\, column\,
state}\rangle = |{\rm swapped\, different\, column\, state}\rangle\, .
\end{eqnarray}
%
An example illustrating these rules is:
{\footnotesize
$$
\Gamma_R \left(
(1,2)\right)\left.\Big|\young(\,\,,\,\,,\,1,\,,\,,\,,3,2)\right\rangle
=\left.\Big|\young(\,\,,\,\,,\,2,\,,\,,\,,3,1)\right\rangle
\qquad\qquad \Gamma_R \left(
(1,2)\right)\left.\Big|\young(\,\,,\,\,,\,\,,\,,\,,3,2,1)\right\rangle
=-\left.\Big|\young(\,\,,\,\,,\,\,,\,,\,,3,2,1)\right\rangle
$$
}
Thus, the representations of the symmetric group simplify dramatically in this limit.

\section{Examples of Projectors}
\label{examplesofprojectors}

In this section we will compute some projectors using the new method outlined in this article.
This is done to both check the nuts and bolts of the construction and to make the arguments presented concrete.
Indeed, the main technical new result is the understanding that we can use $U(p)$ group theory to construct a
basis for the carrier space of an irreducible representation of an $S_n\times S_m$ subgroup from the carrier space 
of an irreducible representation $R$ of $S_{n+m}$, when $R$ has $p$ long rows or long columns. 
In this Appendix we give concrete results illustrating these facts.

\subsection{A Three Row Example using $U(3)$}
\label{exampleprojector}

Consider the following three row Young diagram
$$
{\footnotesize
  \young(\,\,\,\,\,\,\,\,\,\,\,\,\,\,*,\,\,\,\,\,\,\,\,\,*,\,\,\,\,*)\, .}
$$
The starred boxes are to be removed. 
There are six possible ways to distribute the labels $1,2,3$ between these boxes.
One possible representation that can be suduced has $r$ as given above
but with the starred boxes removed and $s={\tiny \yng(2,1)}$. To build the projector $P_{R\to (r,s)jk}$ we need to build the
projector onto the $U(3)$ irreducible representation labeled by $s={\tiny \yng(2,1)}$. Further, since one box is pulled off
each row, the relevant $U(3)$ states have a $\Delta$ weight of $(1,1,1)$. This representation is 8 dimensional
and the corresponding Gelfand-Tsetlin patterns are
$$
{\footnotesize
\left[ 
\begin {array}{ccccc} 
    2  &            &    1    &           &      0    \\\noalign{\medskip}
       &     2      &         &     1     &           \\\noalign{\medskip}
       &            &    2    &           &      
\end {array} \right]
\quad 
\left[ 
\begin {array}{ccccc} 
    2  &            &    1    &           &      0    \\\noalign{\medskip}
       &     2      &         &     1     &           \\\noalign{\medskip}
       &            &    1    &           &      
\end {array} \right]
\quad 
\left[ 
\begin {array}{ccccc} 
    2  &            &    1    &           &      0    \\\noalign{\medskip}
       &     2      &         &     0     &           \\\noalign{\medskip}
       &            &    2    &           &      
\end {array} \right]
\quad 
\left[ 
\begin {array}{ccccc} 
    2  &            &    1    &           &      0    \\\noalign{\medskip}
       &     2      &         &     0     &           \\\noalign{\medskip}
       &            &    1    &           &      
\end {array} \right]
}
$$
$$
{\footnotesize
\left[ 
\begin {array}{ccccc} 
    2  &            &    1    &           &      0    \\\noalign{\medskip}
       &     2      &         &     0     &           \\\noalign{\medskip}
       &            &    0    &           &      
\end {array} \right]
\quad 
\left[ 
\begin {array}{ccccc} 
    2  &            &    1    &           &      0    \\\noalign{\medskip}
       &     1      &         &     1     &           \\\noalign{\medskip}
       &            &    1    &           &      
\end {array} \right]
\quad 
\left[ 
\begin {array}{ccccc} 
    2  &            &    1    &           &      0    \\\noalign{\medskip}
       &     1      &         &     0     &           \\\noalign{\medskip}
       &            &    1    &           &      
\end {array} \right]
\quad 
\left[ 
\begin {array}{ccccc} 
    2  &            &    1    &           &      0    \\\noalign{\medskip}
       &     1      &         &     0     &           \\\noalign{\medskip}
       &            &    0    &           &      
\end {array} \right]
}
$$
The fourth and sixth states in the above list have the correct $\Delta$ weight, so that for weight $\Delta =(1,1,1)$
we have inner multiplicity $I(\Delta )=2$. The fact that there are two states
with the correct $\Delta$ weight implies that this particular $(r,s)$ is subduced twice from the carrier space of $R$.
This in turn implies that there are four possible projection operators and hence four possible restricted Schur
polynomials that can be defined.

To build the projector we need to take linear combinations of the above subspaces in such a way that the resulting
combination is an invariant subspace of $S_n\times S_m$ (here $m=3$) and further that this invariant subspace carries
the correct irreducible representation of $S_n\times S_m$. To streamline our notation for the six subspaces we work
with, we will set
$$
{\footnotesize
\left| a,b,c\right\rangle =
\young(\,\,\,\,\,\,\,\,\,\,\,\,\,\,a,\,\,\,\,\,\,\,\,\,b,\,\,\,\,c) }\, .
$$
The $U(3)$ action is defined on the labeled boxes. The box labeled 1 is always in the first slot of the tensor product;
its position inside the ket tells you what row (and hence what $U(3)$ state) it is in. Notice that all reference to the
carrier space of $r_n$ is omitted. This is perfectly consistent because this subspace is common to all the subspaces we
consider and it plays no role in the problem of finding good $S_m$ invariant subspaces. Thus, for example,
$$
{\footnotesize \left|1,2,3\right\rangle =
\left[ 
\begin {array}{ccccc} 
    1  &            &    0    &           &      0    \\\noalign{\medskip}
       &     1      &         &     0     &           \\\noalign{\medskip}
       &            &    1    &           &      
\end {array} \right]
\otimes
\left[ 
\begin {array}{ccccc} 
    1  &            &    0    &           &      0    \\\noalign{\medskip}
       &     1      &         &     0     &           \\\noalign{\medskip}
       &            &    0    &           &      
\end {array} \right]
\otimes
\left[ 
\begin {array}{ccccc} 
    1  &            &    0    &           &      0    \\\noalign{\medskip}
       &     0      &         &     0     &           \\\noalign{\medskip}
       &            &    0    &           &      
\end {array} \right]}
$$ 
and
$$
{\footnotesize
\left|2,1,3\right\rangle =
\left[ 
\begin {array}{ccccc} 
    1  &            &    0    &           &      0    \\\noalign{\medskip}
       &     1      &         &     0     &           \\\noalign{\medskip}
       &            &    0    &           &      
\end {array} \right]
\otimes
\left[ 
\begin {array}{ccccc} 
    1  &            &    0    &           &      0    \\\noalign{\medskip}
       &     1      &         &     0     &           \\\noalign{\medskip}
       &            &    1    &           &      
\end {array} \right]
\otimes
\left[ 
\begin {array}{ccccc} 
    1  &            &    0    &           &      0    \\\noalign{\medskip}
       &     0      &         &     0     &           \\\noalign{\medskip}
       &            &    0    &           &      
\end {array} \right]}
$$ 
Using the Clebsch-Gordan coefficients given in section \ref{Clebschs} we easily find that the subspaces
considered above break up into subspaces labeled by states from $U(3)$ representations. For example 
{\footnotesize
$$
\left|1,2,3\right\rangle =
{1\over\sqrt{6}}
\left[ 
\begin {array}{ccccc} 
    1  &            &    1    &           &      1    \\\noalign{\medskip}
       &     1      &         &     1     &           \\\noalign{\medskip}
       &            &    1    &           &      
\end {array} \right]
-{1\over\sqrt{12}}
\left[ 
\begin {array}{ccccc} 
    2  &            &    1    &           &      0    \\\noalign{\medskip}
       &     1      &         &     1     &           \\\noalign{\medskip}
       &            &    1    &           &      
\end {array} \right]^{(1)}
+{1\over 2}
\left[ 
\begin {array}{ccccc} 
    2  &            &    1    &           &      0    \\\noalign{\medskip}
       &     2      &         &     0     &           \\\noalign{\medskip}
       &            &    1    &           &      
\end {array} \right]^{(2)}
$$
$$
+{1\over 2}
\left[ 
\begin {array}{ccccc} 
    2  &            &    1    &           &      0    \\\noalign{\medskip}
       &     1      &         &     1     &           \\\noalign{\medskip}
       &            &    1    &           &      
\end {array} \right]^{(2)}
+{1\over\sqrt{12}}
\left[ 
\begin {array}{ccccc} 
    2  &            &    1    &           &      0    \\\noalign{\medskip}
       &     2      &         &     0     &           \\\noalign{\medskip}
       &            &    1    &           &      
\end {array} \right]^{(1)}
+{1\over\sqrt{6}}
\left[ 
\begin {array}{ccccc} 
    3  &            &    0    &           &      0    \\\noalign{\medskip}
       &     2      &         &     0     &           \\\noalign{\medskip}
       &            &    1    &           &      
\end {array} \right]\, .
$$
}
It is a simple matter to compute the decomposition for the remaining 5 subspaces.
Given these results it is straight forward to write down the two possible sets of states that carry the $S_m$ irreducible
representation ${\tiny \yng(2,1)}$\footnote{To verify these formulas yourself, you should use the following action for the 
symmetric group: $\sigma|a,b,c\rangle=|\sigma(a),\sigma(b),\sigma(c)\rangle$.}
{\footnotesize
$$
\left|\yng(2,1),1\right\rangle^{(1)} = \left[ 
\begin {array}{ccccc} 
    2  &            &    1    &           &      0    \\\noalign{\medskip}
       &     1      &         &     1     &           \\\noalign{\medskip}
       &            &    1    &           &      
\end {array} \right]^{(1)}
%
\begin {array}{c} 
={1\over \sqrt{12}}\left(-|1,2,3\rangle +|2,1,3\rangle -|3,1,2\rangle\right.    \\\noalign{\medskip}
 \left. +|1,3,2\rangle +2|2,3,1\rangle -2|3,2,1\rangle \right)
\end {array}
$$
{\vskip 0.35cm}
$$
\left|\yng(2,1),2\right\rangle^{(1)} = \left[ 
\begin {array}{ccccc} 
    2  &            &    1    &           &      0    \\\noalign{\medskip}
       &     1      &         &     1     &           \\\noalign{\medskip}
       &            &    1    &           &      
\end {array} \right]^{(2)}
%
={1\over 2}\left( |1,2,3\rangle - |2,1,3\rangle - |3,1,2\rangle + |1,3,2\rangle \right)
$$
{\vskip 0.35cm}
and
$$
\left|\yng(2,1),1\right\rangle^{(2)} = \left[ 
\begin {array}{ccccc} 
    2  &            &    1    &           &      0    \\\noalign{\medskip}
       &     2      &         &     0     &           \\\noalign{\medskip}
       &            &    1    &           &      
\end {array} \right]^{(2)}
%
={1\over 2}\left( |1,2,3\rangle + |2,1,3\rangle - |3,1,2\rangle - |1,3,2\rangle \right)
$$
{\vskip 0.35cm}
$$
\left|\yng(2,1),2\right\rangle^{(2)} = \left[ 
\begin {array}{ccccc} 
    2  &            &    1    &           &      0    \\\noalign{\medskip}
       &     2      &         &     0     &           \\\noalign{\medskip}
       &            &    1    &           &      
\end {array} \right]^{(1)}
%
\begin {array}{c}
={1\over \sqrt{12}}\left( |1,2,3\rangle + |2,1,3\rangle + |3,1,2\rangle\right. \\\noalign{\medskip}
\left. + |1,3,2\rangle - 2 |2,3,1\rangle - 2 |3,2,1\rangle \right)
\end {array}
$$
}
The superscripts on the kets on the right hand sides of these equations are multiplicity labels and 
the integer inside each ket indexes states in the carrier space.
These formulas have all been obtained using the Clebsch-Gordan coefficients of $U(3)$ - we have not used any symmetric group
theory. However, as a consequence of Schur-Weyl duality, we claim that the above states fill out representations of $S_3$. This
is easily verified. The four possible projectors that can be defined are now given by
$$
{\footnotesize
  P_{R\to (r,{\tiny \yng(2,1)}),ij}=\,\sum_{k=1}^{2}\,\, \left|\yng(2,1),k\right\rangle^{(i)\qquad(j)}\left\langle\yng(2,1),k\right|\, .}
$$

\subsection{A Four Column Example using $U(4)$}
\label{examplesphereprojector}

Consider the following four column Young diagram
$$
{\footnotesize
  \young(\,\,\,\,,\,\,\,\,,\,\,\,*,\,\,\,*,\,\,\,*,\,\,\,,\,\,*,\,\,,\,\,,\,,\,)\, .}
$$
The starred boxes are to be removed. There are four possible ways to distribute the labels $1,2,3,4$ between these boxes. One possible
$S_n\times S_m$ irreducible representation that can be subduced has $r$ as given above but with the starred boxes removed and 
$s={\tiny\yng(2,1,1)}$. The build the corresponding projector we need to build the projector onto the $U(4)$ irreducible representation
labeled by $s^T={\tiny \yng(3,1)}$. Since we pull three boxes off the right most column and one box off the neighboring column, the
states we are interested in will have a $\Delta$ weight of $(0,0,1,3)$. For this example, we will need to assign nontrivial phases
between the states in $V_p^{\otimes m}$ and the Young diagrams. The four possible ways to distribute the labels are
$$
{\footnotesize
  \young(\,\,\,\,,\,\,\,4,\,\,\,3,\,\,\,2,\,\,\,,\,\,1,\,\,,\,\,,\,,\,)\quad
  \young(\,\,\,\,,\,\,\,4,\,\,\,3,\,\,\,1,\,\,\,,\,\,2,\,\,,\,\,,\,,\,)\quad
  \young(\,\,\,\,,\,\,\,4,\,\,\,2,\,\,\,1,\,\,\,,\,\,3,\,\,,\,\,,\,,\,)\quad
  \young(\,\,\,\,,\,\,\,3,\,\,\,2,\,\,\,1,\,\,\,,\,\,4,\,\,,\,\,,\,,\,)\, .}
$$
Take the first state shown as the reference state. To get the second state from the first we need to act with $(12)$, so that the second
state has a phase of $-1$. The get the third state from the first we need to act with $(12)$ and then with $(23)$, so that it has a phase
of 1. Finally, to get the fourth state from the first we need to act with $(12)$ and then $(23)$ and then $(34)$ giving a phase of $-1$.
Writing our states as
$$
{\footnotesize  \left|a,b,c,d\right\rangle =
  \young(\,\,\,\,,\,\,\,d,\,\,\,c,\,\,\,b,\,\,\,,\,\,a,\,\,,\,\,,\,,\,)\, .
}
$$
we have
{\footnotesize
$$
\left|1,2,3,4\right\rangle =
\left[ 
\begin {array}{ccccccc} 
    1  &            &    0    &           &      0   &        &   0 \\\noalign{\medskip}
       &     1      &         &     0     &          &    0   &     \\\noalign{\medskip}
       &            &    1    &           &      0   &        &     \\\noalign{\medskip}
       &            &         &     0     &          &        &
\end {array} \right]
\otimes
\left[ 
\begin {array}{ccccccc} 
    1  &            &    0    &           &      0   &        &   0 \\\noalign{\medskip}
       &     1      &         &     0     &          &    0   &     \\\noalign{\medskip}
       &            &    1    &           &      0   &        &     \\\noalign{\medskip}
       &            &         &     1     &          &        &
\end {array} \right]
\otimes
\left[ 
\begin {array}{ccccccc} 
    1  &            &    0    &           &      0   &        &   0 \\\noalign{\medskip}
       &     1      &         &     0     &          &    0   &     \\\noalign{\medskip}
       &            &    1    &           &      0   &        &     \\\noalign{\medskip}
       &            &         &     1     &          &        &
\end {array} \right]
\otimes
\left[ 
\begin {array}{ccccccc} 
    1  &            &    0    &           &      0   &        &   0 \\\noalign{\medskip}
       &     1      &         &     0     &          &    0   &     \\\noalign{\medskip}
       &            &    1    &           &      0   &        &     \\\noalign{\medskip}
       &            &         &     1     &          &        &
\end {array} \right]
$$
$$
=-{\sqrt{3}\over 2}
\left[ 
\begin {array}{ccccccc} 
    3  &            &    1    &           &      0   &        &   0 \\\noalign{\medskip}
       &     3      &         &     1     &          &    0   &     \\\noalign{\medskip}
       &            &    3    &           &      1   &        &     \\\noalign{\medskip}
       &            &         &     3     &          &        &
\end {array} \right]^{(1)}
+{1\over 2}
\left[ 
\begin {array}{ccccccc} 
    4  &            &    0    &           &      0   &        &   0 \\\noalign{\medskip}
       &     4      &         &     0     &          &    0   &     \\\noalign{\medskip}
       &            &    4    &           &      0   &        &     \\\noalign{\medskip}
       &            &         &     3     &          &        &
\end {array} \right]\, ,
$$
$$
\left|2,1,3,4\right\rangle =
-\left[ 
\begin {array}{ccccccc} 
    1  &            &    0    &           &      0   &        &   0 \\\noalign{\medskip}
       &     1      &         &     0     &          &    0   &     \\\noalign{\medskip}
       &            &    1    &           &      0   &        &     \\\noalign{\medskip}
       &            &         &     1     &          &        &
\end {array} \right]
\otimes
\left[ 
\begin {array}{ccccccc} 
    1  &            &    0    &           &      0   &        &   0 \\\noalign{\medskip}
       &     1      &         &     0     &          &    0   &     \\\noalign{\medskip}
       &            &    1    &           &      0   &        &     \\\noalign{\medskip}
       &            &         &     0     &          &        &
\end {array} \right]
\otimes
\left[ 
\begin {array}{ccccccc} 
    1  &            &    0    &           &      0   &        &   0 \\\noalign{\medskip}
       &     1      &         &     0     &          &    0   &     \\\noalign{\medskip}
       &            &    1    &           &      0   &        &     \\\noalign{\medskip}
       &            &         &     1     &          &        &
\end {array} \right]
\otimes
\left[ 
\begin {array}{ccccccc} 
    1  &            &    0    &           &      0   &        &   0 \\\noalign{\medskip}
       &     1      &         &     0     &          &    0   &     \\\noalign{\medskip}
       &            &    1    &           &      0   &        &     \\\noalign{\medskip}
       &            &         &     1     &          &        &
\end {array} \right]
$$
$$
=\sqrt{2\over 3}
\left[ 
\begin {array}{ccccccc} 
    3  &            &    1    &           &      0   &        &   0 \\\noalign{\medskip}
       &     3      &         &     1     &          &    0   &     \\\noalign{\medskip}
       &            &    3    &           &      1   &        &     \\\noalign{\medskip}
       &            &         &     3     &          &        &
\end {array} \right]^{(2)}
+{1\over\sqrt{12}}
\left[ 
\begin {array}{ccccccc} 
    3  &            &    1    &           &      0   &        &   0 \\\noalign{\medskip}
       &     3      &         &     1     &          &    0   &     \\\noalign{\medskip}
       &            &    3    &           &      1   &        &     \\\noalign{\medskip}
       &            &         &     3     &          &        &
\end {array} \right]^{(1)}
+{1\over 2}
\left[ 
\begin {array}{ccccccc} 
    4  &            &    0    &           &      0   &        &   0 \\\noalign{\medskip}
       &     4      &         &     0     &          &    0   &     \\\noalign{\medskip}
       &            &    4    &           &      0   &        &     \\\noalign{\medskip}
       &            &         &     3     &          &        &
\end {array} \right]
$$
}
plus two more.
Given these results, it is a simple matter to write down the states that carry the $S_m$ irreducible 
representation ${\tiny \yng(2,1,1)}$\footnote{To verify these formulas yourself, don't forget that the action for the 
symmetric group now includes the phase ${\rm sgn}(\sigma)$.}
{\footnotesize
$$
\left|\yng(2,1,1),1\right\rangle = 
\left[ 
\begin {array}{ccccccc} 
    3  &            &    1    &           &      0   &        &   0 \\\noalign{\medskip}
       &     3      &         &     1     &          &    0   &     \\\noalign{\medskip}
       &            &    3    &           &      1   &        &     \\\noalign{\medskip}
       &            &         &     3     &          &        &
\end {array} \right]^{(1)}
={1\over\sqrt{12}}\left(
-3|1,2,3,4\rangle + |2,1,3,4\rangle + |3,1,2,4\rangle + |4,1,2,3\rangle \right)
$$
$$
\left|\yng(2,1,1),2\right\rangle = 
\left[ 
\begin {array}{ccccccc} 
    3  &            &    1    &           &      0   &        &   0 \\\noalign{\medskip}
       &     3      &         &     1     &          &    0   &     \\\noalign{\medskip}
       &            &    3    &           &      1   &        &     \\\noalign{\medskip}
       &            &         &     3     &          &        &
\end {array} \right]^{(2)}
={1\over\sqrt{6}}\left(
 2 |2,1,3,4\rangle - |3,1,2,4\rangle - |4,1,2,3\rangle \right)
$$
$$
\left|\yng(2,1,1),3\right\rangle = 
\left[ 
\begin {array}{ccccccc} 
    3  &            &    1    &           &      0   &        &   0 \\\noalign{\medskip}
       &     3      &         &     1     &          &    0   &     \\\noalign{\medskip}
       &            &    3    &           &      1   &        &     \\\noalign{\medskip}
       &            &         &     3     &          &        &
\end {array} \right]^{(3)}
=-{1\over\sqrt{2}}\left(
 |3,1,2,4\rangle - |4,1,2,3\rangle \right)
$$
}
These formulas use only the Clebsch-Gordan coefficients of $U(4)$. It is again easy to verify that the above states fill out 
the representation ${\tiny \yng(2,1,1)}$ of $S_4$.

%

\section{Evaluation of the Dilatation Operator}
\label{dilatationoperator}

In this Appendix we collect the details of the evaluation of the dilatation operator. In the next subsection we review the derivation
of the action of the dilatation operator given in \cite{VinceKate} emphasizing those features important for our discussion. We then describe
how to explicitely evaluate this action. Our discussion is developed using restricted Schur polynomials labeled with Young diagrams that have
$O(1)$ long rows. The discussion for restricted Schur polynomials labeled with Young diagrams that have $O(1)$ long columns is very similair
so we will simply sketch how the result is obtained.

\subsection{Dilatation Operator in the $SU(2)$ sector}
\label{derivedilat}

The one loop dilatation operator in the $SU(2)$ sector\cite{Beisert:2003tq} of ${\cal N}=4$ super Yang Mills theory is
$$ 
D = - g_{\rm YM}^2 {\rm Tr}\,\big[ Y,Z\big]\big[ \partial_Y ,\partial_Z\big]\, .
$$
Acting on a restricted Schur polynomial we obtain\footnote{Our index conventions are $(YZ)^i_k = Y^i_j Z^j_k$.}  
{\small
\begin{eqnarray}
\label{basicreslt}
D \chi_{R,(r,s)jk}
={g_{\rm YM}^2\over (n-1)!(m-1)!}\sum_{\psi\in S_{n+m}}
{\rm Tr}_{(r,s)jk}\left(\Gamma_R ((1,m+1) \psi -\psi (1,m+1))\right)\times\cr
 \times\delta^{i_{1}}_{i_{\psi (1)}} Y^{i_2}_{i_{\psi (2)}}
\cdots
Y^{i_{m}}_{i_{\psi (m)}}
(YZ-ZY)_{i_{\psi (m+1)}}^{i_{m+1}}
Z^{i_{m+2}}_{i_{\psi (m+2)}}\cdots Z^{i_{n+m}}_{i_{\psi (n+m)}}
\, .
\end{eqnarray}
}
As a consequence of the $\delta^{i_{1}}_{i_{\psi (1)}}$ appearing in the summand,
the sum over $\psi$ runs only over permutations for which $\psi (1)=1$.
To perform the sum over $\psi$, write the sum over $S_{n+m}$ as a sum over cosets of the $S_{n+m-1}$ subgroup
obtained by keeping those permutations that satisfy $\psi (1)=1$. The result follows immediately
from the reduction rule for Schur polynomials (see \cite{de Mello Koch:2004ws} and appendix C of \cite{de Mello Koch:2007uu})
{\small
\begin{eqnarray}
\nonumber
D \,\chi_{R,(r,s)jk}=
{g_{\rm YM}^2\over (n-1)!(m-1)!}\sum_{\psi\in S_{n+m-1}}\sum_{R'}c_{RR'}\,
{\rm Tr}_{(r,s)jk}\Big(\Gamma_R\left((1,m+1)\right)\Gamma_{R'}(\psi) \cr
-\Gamma_{R'}(\psi) \Gamma_R\left((1,m+1)\right)\Big) 
\, Y^{i_2}_{i_{\psi (2)}}
\cdots
Y^{i_{m}}_{i_{\psi (m)}}
(YZ-ZY)_{i_{\psi (m+1)}}^{i_{m+1}}
Z^{i_{m+2}}_{i_{\psi (m+2)}}\cdots Z^{i_{n+m}}_{i_{\psi (n+m)}}
\, .
\end{eqnarray}
}
The sum over $R'$ runs over all Young diagrams that can be obtained from $R$ by dropping a single box; $c_{RR'}$ is the 
factor of the box that must be removed from $R$ to obtain $R'$. The appearance of $\Gamma_R\left((1,m+1)\right)$ is very
natural. $\Gamma_R\left((1,m+1)\right)$ is not an element of the $S_n\times S_m$ subgroup - it mixes indices belonging
to $Z$s and indices belonging to $Y$s. The dilatation operator has derivatives with respect to $Z$ and $Y$ in the same 
trace and so does indeed naturally mix $Z$s and $Y$s. We will make use of the following notation
$$
{\rm Tr} (\sigma Z^{\otimes n}Y^{\otimes m})=
Z^{i_1}_{i_{\sigma(1)}}\cdots Z^{i_n}_{i_{\sigma(n)}}Y^{i_{n+1}}_{i_{\sigma(n+1)}}\cdots Y^{i_{n+m}}_{i_{\sigma(n+m)}}\, .
$$
Now, use the identities (bear in mind that $\psi (1)=1$)
$$
Y^{i_2}_{i_{\psi (2)}} \cdots Y^{i_{m}}_{i_{\psi (m)}} (YZ-ZY)_{i_{\psi (m+1)}}^{i_{m+1}}
Z^{i_{m+2}}_{i_{\psi (m+2)}}\cdots Z^{i_{n+m}}_{i_{\psi (n+m)}}
={\rm Tr}\left(\Big((1,m+1)\, \psi -\psi\,(1,m+1)\Big)Z^{\otimes n}Y^{\otimes m} \right)
$$
and (this identity is proved in \cite{Bhattacharyya:2008rc})
$$
{\rm Tr} (\sigma Z^{\otimes n}Y^{\otimes m})=\sum_{T,(t,u)lq}{d_T n! m!\over d_t d_u (n+m)!}{\rm Tr}_{(t,u)lq}(\Gamma_T(\sigma^{-1}))\chi_{T,(t,u)ql}(Z,Y)
$$
to obtain
$$
D\chi_{R,(r,s)jk}(Z,Y)=\sum_{T,(t,u)lq} M_{R,(r,s)jk;T,(t,u)lq}\chi_{T,(t,u)ql}(Z,Y)\, ,
$$
{\small
$$
M_{R,(r,s)jk;T,(t,u)lq}=g_{YM}^2\sum_{\psi\in S_{n+m-1}}\sum_{R'}
{c_{RR'} d_T n m\over d_t d_u (n+m)!}
{\rm Tr}_{(r,s)jk}\Big(\Gamma_R ((1,m+1))\Gamma_{R'}(\psi)-\Gamma_{R'}(\psi) \Gamma_R((1,m+1))\Big)\times
$$
$$
\times {\rm Tr}_{(t,u)lq}\left(\Gamma_{T'}(\psi^{-1})\Gamma_T((1,m+1)) - \Gamma_T((1,m+1))\Gamma_{T'}(\psi^{-1})\right)\, .
$$
}
The sum over $\psi$ can be evaluated using the fundamental orthogonality relation
$$
M_{R,(r,s)jk;T,(t,u)lq} = - g_{YM}^2\sum_{R'}{c_{RR'} d_T n m\over d_{R'} d_t d_u (n+m)}
{\rm Tr}\Big( \Big[ \Gamma_R((1,m+1)),P_{R\to (r,s)jk}\Big]I_{R'\, T'}\times
$$
\begin{equation}
\label{final}
\times \Big[\Gamma_T((1,m+1)),P_{T\to (t,u)ql}\Big] I_{T'\, R'}\Big)  \, .
\end{equation}
Sums of this type are discussed in detail in the next section and the intertwiners $I_{R'\, T'}$ which arise are
discussed in detail. This expression for the one loop dilatation operator is exact in $N$.

To obtain the spectrum of anomalous dimensions, we need to consider the action of the dilatation operator on normalized 
operators. The two point function for the restricted Schur polynomials (\ref{restschurtwopoint}) is not unity.
Normalized operators which do have unit two point function can be obtained from
$$
\chi_{R,(r,s)jk}(Z,Y)=\sqrt{f_R \, {\rm hooks}_R\over {\rm hooks}_r\, {\rm hooks}_s}O_{R,(r,s)jk}(Z,Y)\, .
$$
In terms of these normalized operators
$$
DO_{R,(r,s)jk}(Z,Y)=\sum_{T,(t,u)lq} N_{R,(r,s)jk;T,(t,u)ql}O_{T,(t,u)ql}(Z,Y)
$$
{\small
$$
N_{R,(r,s)jk;T,(t,u)ql}= - g_{YM}^2\sum_{R'}{c_{RR'} d_T n m\over d_{R'} d_t d_u (n+m)}
\sqrt{f_T \, {\rm hooks}_T\, {\rm hooks}_r \, {\rm hooks}_s \over f_R \, {\rm hooks}_R\, {\rm hooks}_t\, {\rm hooks}_u}\times
$$
$$
\times{\rm Tr}\Big(\Big[ \Gamma_R((1,m+1)),P_{R\to (r,s)jk}\Big]I_{R'\, T'}\Big[\Gamma_T((1,m+1)),P_{T\to (t,u)lq}\Big]I_{T'\, R'}\Big) \, .
$$
}
It is this last expression that we evaluate explicitely. The bulk of the work entails evaluating the trace. There are three objects
which appear: the symmetric group operators $P_{R\to (r,s)jk}$, the intertwiners $I_{T'\, R'}$ and the symmetric group element $\Gamma_R((1,m+1))$. 
We have already discussed the operators $P_{R\to (r,s)jk}$. The next two subsections are used to discuss $I_{T'\, R'}$ and $\Gamma_R((1,m+1))$.

\subsection{Intertwiners}
\label{intertwiners}

In this section we will consider the sum over $S_{n+m-1}$ which was performed to obtain (\ref{final}).
This will give a very explicit understanding of the intertwiners appearing in the expression for the dilatation operator.
When $S^n$ acts on $V^{\otimes n}$ $n>1$ it furnishes a reducible representation. Imagine that this includes the 
irreducible representations $R$ and $T$. Representing the action of $\sigma$ as a matrix $\Gamma(\sigma )$, 
in a suitable basis we can write
$$
\Gamma(\sigma)=\left[
\matrix{\Gamma_R(\sigma) &0 &\cdots\cr 0 &\Gamma_S(\sigma) &\cdots\cr \cdots &\cdots &\cdots}\right]\, .
$$
If we restrict ourselves to an $S_{n-1}$ subgroup of $S_n$, then in general, both $R$ and $S$ will subduce a number of
representations. Assume for the sake of this discussion that $R$ subduces $R_1'$ and $R_2'$ and that
$S$ subduces $S_1'$ and $S_2'$. This is precisely the situation that arises in the sum performed to obtain (\ref{final}).
Then, for $\sigma\in S_{n-1}$ we have
$$
\Gamma(\sigma)=
\left[
\matrix{
\Gamma_{R_1'}(\sigma) &0 &0 &0 &\cdots\cr 
0 &\Gamma_{R_2'}(\sigma) &0 &0 &\cdots\cr 
0 &0 &\Gamma_{S_1'}(\sigma) &0 &\cdots\cr
0 &0 &0 &\Gamma_{S_2'}(\sigma) &\cdots\cr
\cdots &\cdots &\cdots &\cdots &\cdots}\right]\, .
$$
Imagine that as Young diagrams $S_1'=R_1'$, that is, one of the irreducible representations subduced by $R$ is isomorphic to one of 
the representations subduced by $S$. Then, a simple application of the fundamental orthogonality relation gives
$$
\sum_{\sigma\in S_{n-1}}
\left[
\matrix{
\Gamma_{R_1'}(\sigma) &0 &0 &0 &\cdots\cr 
0 &0 &0 &0 &\cdots\cr 
0 &0 &0 &0 &\cdots\cr
0 &0 &0 &0 &\cdots\cr
\cdots &\cdots &\cdots &\cdots &\cdots}\right]_{ij}
\left[
\matrix{
0 &0 &0 &0 &\cdots\cr 
0 &0 &0 &0 &\cdots\cr 
0 &0 &\Gamma_{S_1'}(\sigma) &0 &\cdots\cr
0 &0 &0 &0 &\cdots\cr
\cdots &\cdots &\cdots &\cdots &\cdots}\right]_{ab}
$$
$$
={(n-1)!\over d_{R_1'}}\delta_{R_1'S_1'}
\left[
\matrix{
0 &0 &{\bf 1} &0 &\cdots\cr 
0 &0 &0 &0 &\cdots\cr 
0 &0 &0 &0 &\cdots\cr
0 &0 &0 &0 &\cdots\cr
\cdots &\cdots &\cdots &\cdots &\cdots}\right]_{ib}
\left[
\matrix{
0 &0 &0 &0 &\cdots\cr 
0 &0 &0 &0 &\cdots\cr 
{\bf 1} &0 &0 &0 &\cdots\cr
0 &0 &0 &0 &\cdots\cr
\cdots &\cdots &\cdots &\cdots &\cdots}\right]_{aj}
$$
$$
\equiv {(n-1)!\over d_{R_1'}}\delta_{R_1'S_1'} (I_{R_1'S_1'})_{ib}(I_{S_1'R_1'})_{aj}
$$
where the form of the intertwiners has been spelled out.
Intertwiners are maps between two isomorphic spaces. For $\sigma\in S_{n-1}$
$$
  I_{R'T'}\Gamma_{T'}(\sigma)=\Gamma_{R'}(\sigma)I_{R'T'}
$$
The box removed to obtain $R'$ and $T'$ can be removed from any corner of the Young diagram. 

It is useful to make a few comments on how the intertwiners are realized in our calculation.
Since the first box is removed from $R$ or $T$ the intertwiner acts on the first slot
of $V_p^{\otimes m}$. Now, look back at formula (\ref{basicreslt}). The delta function which 
appears freezes the $1$ index and hence the $S_{n+m-1}$ subgroup of $S_{n+m}$ is obtained by
keeping all elements of $S_{n+m}$ that leave index $1$ inert. Consequently, with our choice that
the intertwiner acts on the first slot of $V_p^{\otimes m}$, we see that the first slot corresponds
to index $i_1$. Recall that the particular vector a box corresponds to is determined by
the row/column the box belongs to. Thus, the explicit form of the intertwiner is determined
once the location of the box removed from $T$ and the box removed from $R$ are specified.
As an example, for the Young diagrams shown below we have
$$
  I_{R'T'}=E_{1,5}\otimes{\bf 1}\otimes\cdots\otimes{\bf 1}\, ,\qquad
  I_{T'R'}=E_{5,1}\otimes{\bf 1}\otimes\cdots\otimes{\bf 1}\, .
$$
\begin{figure}[h]
          \centering
          {\epsfig{file=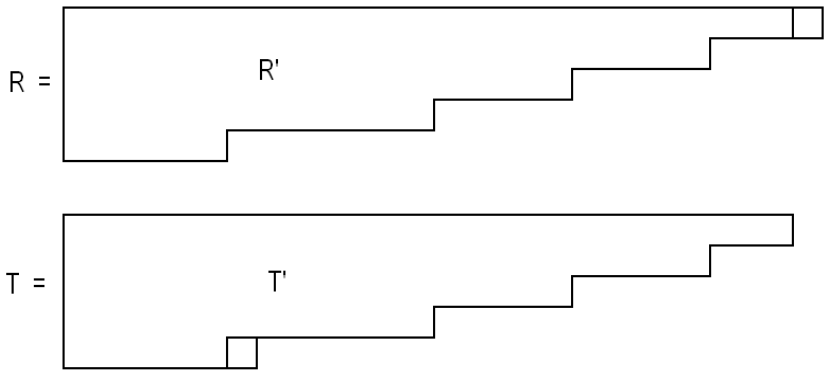,width=11.0cm,height=5.0cm}}
          \caption{A figure showing $R$ and the box that must be removed to obtain $R'$ and $T$ and the 
                   box that must be removed to obtain $T'$. As Young diagrams, $T'=R'$. $T$ and $R$ both have 5 rows.}
 \end{figure}

It is straight forward to extract the general rule from this example. Consider first the case that $R\ne T$.
To obtain $R'$ from $R$ we remove a box from row $i$ and to obtain $T'$ from $T$ we remove a box from row $j$.
In this situation we have
$$
  I_{R'T'}=E_{ij}\otimes {\bf 1}\otimes\cdots{\bf 1}\, , \qquad
  I_{T'R'}=E_{ji}\otimes {\bf 1}\otimes\cdots{\bf 1}\, .
$$
In the case that $R=T$, the box that must be removed can be removed from any row and we get a contribution to the 
dilatation operator from each possible removal. Each possible removal must be represented by a different intertwiner
and one needs to sum over all possible intertwiners. In this situation, the possible intertwiners are
$$
  I_{R'T'}=E_{kk}\otimes{\bf 1}\otimes\cdots\otimes{\bf 1}=I_{T'R'},\qquad k=1,2,\cdots,p\, .
$$

\subsection{$\Gamma_R (1,m+1)$}

This group element acts on one slot from the $Y$s and one slot from the $Z$s. The box removed from $R$ to get $R'$ is the box 
acted on by the intertwiner and it is a $Y$ box. This is one of the boxes that $\Gamma_R (1,m+1)$ acts on. The second box that
$\Gamma_R (1,m+1)$ acts on can be any box associated to the $Z$s. Up to now we have discussed the projectors and intertwiners.
These only have an action on the boxes corresponding to $Y$s and as a result, our discussion has always taken place in the
vector space $V_p^{\otimes\, m}$.  However, because $\Gamma_R (1,m+1)$ acts on a $Z$ box we must include one more slot and work
in $V_p^{\otimes\, m+1}$. The intertwiners and projectors have a trivial action on the $(m+1)^{\rm th}$ slot and hence 
the $(m+1)^{\rm th}$ slot is simply occupied with the identity. For the
rest of this Appendix we work in $V_p^{\otimes\, m+1}$ and not in $V_p^{\otimes\, m}$. Acting in $V_p^{\otimes\, m+1}$,
$\Gamma_R (1,m+1)$ has a very simple action: it simply swaps the $1^{\rm st}$ and the $(m+1)^{\rm th}$ slots. The projectors
when acting on $V_p^{\otimes\, m+1}$ are given by
$$
  {\cal P}_{R\to (r,s)ij}=p_{R,(r,s)ij}\otimes {\bf 1}
$$
where the $p\times p$ unit matrix ${\bf 1}$ acts on the $(m+1)^{\rm th}$ slot. $p_{R,(r,s)ij}$ acts only in $V_p^{\otimes\, m}$.
For comparison, the projectors appearing in the defintion of the restricted Schur polynomial are
$$
  P_{R\to (r,s)ij}=p_{R,(r,s)ij}\otimes {\bf I}_{r}
$$
where ${\bf I}_{r}$ is the identity matrix acting on the carrier space of the $S_n$ irreducible representation $r$. Below we will
make use of the obvious formula
$$
  {\bf 1} = \sum_{k=1}^p E_{kk}\, .
$$

In evaluating the dilatation operator, we will need to take products of the intertwiners and $\Gamma (1,m+1)$. These products are
easily evaluated
\begin{eqnarray}
\nonumber
\Gamma_R(1,m+1) \, E_{ij}\otimes {\bf 1}\otimes\cdots\otimes {\bf 1}&&=
\Gamma_R(1,m+1) \, \sum_{k=1}^p\, E_{ij}\otimes {\bf 1}\otimes\cdots\otimes E_{kk}\\
\nonumber
&&=\sum_{k=1}^p\, E_{kj}\otimes {\bf 1}\otimes\cdots\otimes E_{ik}
\end{eqnarray}
\begin{eqnarray}
\nonumber
E_{ij}\otimes {\bf 1}\otimes\cdots\otimes {\bf 1}\, \Gamma_R(1,m+1) &&=
\sum_{k=1}^p\, E_{ij}\otimes {\bf 1}\otimes\cdots\otimes E_{kk}\, \Gamma_R(1,m+1)\\
\nonumber
&&=\sum_{k=1}^p\, E_{ik}\otimes {\bf 1}\otimes\cdots\otimes E_{kj}
\end{eqnarray}
\begin{equation}
\nonumber
\Gamma_R(1,m+1) \, E_{ij}\otimes {\bf 1}\otimes\cdots\otimes {\bf 1}\, \Gamma_R(1,m+1) =
{\bf 1} \otimes {\bf 1}\otimes\cdots\otimes E_{ij} \, .
\end{equation}
From now on we will write the $E_{ij}$ with a superscript, indicating which slot $E_{ij}$ acts on. In this notation we have
$$
E_{ik}\otimes {\bf 1}\otimes\cdots\otimes E_{kj} = E_{ik}^{(1)} E_{kj}^{(m+1)}\, .
$$

\subsection{Dilatation Operator Coefficient}
\label{dilopco}

In this secton we explain how to evaluate the value of the coefficient
$$
g_{YM}^2{c_{RR'} d_T n m\over d_{R'} d_t d_u (n+m)}
\sqrt{f_T \, {\rm hooks}_T\, {\rm hooks}_r \, {\rm hooks}_s \over f_R \, {\rm hooks}_R\, {\rm hooks}_t\, {\rm hooks}_u}
$$
in the large $N$ limit.
The Young diagrams $R$, $T$, $r$, $t$, $s$ and $u$ each have $p$-rows. We use the symbols $R_i$, $T_i$, $r_i$, $t_i$, $s_i$ and $u_i$
$i=1,2,...,p$ to denote the number of boxes in each row respectively. We assume $p$ is fixed to be $O(1)$.
The top row (which is also the longest row) is the value $i=1$ and the bottom row (shortest row) has $i=p$. 
It is straight forward to argue that the product of hook lengths, in $r$ for example, is
$$
   {\rm hooks}_R={\prod_{i=1}^p (r_i+p-i)!\over \prod_{j<k} (r_j-r_k+k-j)}\, .
$$
For the diagrams $R$ and $T$, the row lengths $R_i$ are of order $N$. Further, $R$ and $T$ differ by at most the placement
of a single box. This implies that $R_i=T_i$ for all except two values of $i$, say $i=a,b$. For these values of $i$ we have
$$
   R_b = T_b + 1,\qquad R_a = T_a - 1\, .
$$
This implies that
$$
  {{\rm hooks}_R\over {\rm hooks}_T}= 
{(T_a-1+p-a)! (T_b+1+p-b)!\over ((T_a+p-a)!(T_b+p-b)!}
\prod_{{}^{k\ne a}_{k\ne b}}{|T_a-T_k|+|k-a|\over |T_a-1-T_k|+|k-a|}\times
$$
$$
\times
\prod_{{}^{k\ne a}_{k\ne b}}{|T_b-T_k|+|k-b|\over |T_b+1-T_k|+|k-b|}
{|T_b-T_a|+|a-b|\over |T_a-T_b-2|+|a-b|}
={R_b\over R_a}\left( 1+O(N^{-1})\right) \, .
$$
Use $R_+$ to denote the row length of the row in $R$ that is longer than the corresponding row in $T$ and let
$R_-$ denote the row length of the row in $R$ this is shorter than the corresponding row in $T$. With this notation
$$
  {{\rm hooks}_R\over{\rm hooks}_T}= {R_+\over R_-}\left( 1+O(N^{-1})\right)\, .
$$
This argument has an obvious generalization to the other hook factors ${{\rm hooks}_r\over{\rm hooks}_t}$ and
${{\rm hooks}_s\over{\rm hooks}_u}$. Now consider a Young diagram $R'$ that is obtained by removing a single box from 
Young diagram $R$. Assuming this box is removed from row $a$, we have
the following relation between the lengths of the rows in $R$ and the lengths of the rows in $R'$
$$
  R_i=R'_i\quad i\ne a,\qquad R_a=R'_a+1\, .
$$
Thus, we find
$$
  {{\rm hooks}_R\over{\rm hooks}_{R'}}={(R_a+p-a)!\over (R_a+p-1-a)!}\prod_{j\ne a}
{|R_j-R_a-1|+|a-j|\over |R_j-R_a|+|a-j|}=R_a \left( 1+O(N^{-1})\right)\, .
$$

The coefficient quoted at the start of this section is multiplied by the trace over an $(r,s)$ subspace. This trace
produces a number of order 1 multiplied by $d_{r'} d_s$. The product of the coefficient and the trace now reduces
to quantities that we have studied. Thus, we now have all the ingredients needed to estimate
the large $N$ values of the combinations of symmetric group dimensions and hook factors that appear in the dilatation
operator. Notice that both the product of the hook lengths and the dimensions of symmetric group irreducible 
representations are invariant under the flip of the Young diagram which exchanges columns and rows. Thus, these conclusions
can immediately be recycled when studying the case of $p$ long columns.

Next, recalling that $f_R$ is the product of factors in Young diagram $R$ and $R'=T'$ we learn that
$$
  c_{RR'} \sqrt{f_T \over f_R }=\sqrt{c_{RR'}c_{TT'}}
$$
where $c_{RR'}$ is the factor associated to the box that must be removed from $R$ to obtain $R'$ and
$c_{TT'}$ is the factor associated to the box that must be removed from $T$ to obtain $T'$.

\subsection{Evaluating Traces}

In this section we evaluate the trace
$$
{\cal T}={\rm Tr}\Big(\Big[ \Gamma_R((1,m+1)),P_{R\to (r,s)jk}\Big]I_{R'\, T'}\Big[\Gamma_T((1,m+1)),P_{T\to (t,u)lm}\Big]I_{T'\, R'}\Big) \, .
$$
We start by writing this trace as a sum of traces over $m+1$ slots (all the $Y$ slots plus one $Z$ slot)
times a trace over $n-1$ slots (the remaining $Z$ slots). The trace over the $n-1$ slots is over the carrier space
$R^{m+1}$ which is described by a Young diagram that can be obtained by removing $m+1$ boxes from $R$, or equivalently
by removing one box from $r$ or equivalently by removing one box from $t$ - these all give the same Young diagram
describing $R^{m+1}$.
$R^{m+1}$ has different shapes depending on where the $(m+1)^{\rm th}$ box is removed.
The results from the last subsection clearly imply that the dimension of symmetric group representation 
$R^{m+1}$, denoted $d_{R^{m+1}}$, depends on the details of this shape. 
If the $(m+1)^{\rm th}$ box is removed from row $i$ denote this dimension by $d_{R^{m+1}}^i$ 
Our general strategy is then to trace over the last $Z$ slot (the $(m+1)^{\rm th}$ slot) 
which then leaves a trace over $V_p^{\otimes\, m}$. 
This trace is then evaluated using elementary $U(p)$ representation theory.

The box removed from $R$ to obtain $R'$ is removed from the $b^{\rm th}$ row of $R$ and the box removed from $T$ to
obtain $T'$ is removed from the $a^{\rm th}$ row of $T$. After tracing over the $n-1$ $Z$ slots associated to $R^{m+1}$ 
(this produces a factor of $d_{R^{m+1}}^b$), multiplying the symmetric group elements $(1,m+1)$ with the intertwiners 
and then tracing over the $(m+1)^{\rm th}$ slot we obtain
$$
{\cal T}=-\delta_{ab}\delta_{RT}\delta_{(r,s)\,(t,u)}\delta_{jm}\delta_{kl}d_{R^{m+1}}^b\left[
{\rm Tr}_{V_p^{\otimes\, m}}\Big(P_{R\to (r,s)lk}E^{(1)}_{bb}\Big)+{\rm Tr}_{V_p^{\otimes\, m}}\Big(P_{R\to (r,s)jm}E^{(1)}_{bb}\Big)
\right]
$$
$$
+d_{R^{m+1}}^b{\rm Tr}_{V_p^{\otimes\, m}}\Big(P_{R\to (r,s)lk}E^{(1)}_{bb}P_{T\to (t,u)lm}E^{(1)}_{aa}\Big)
+d_{R^{m+1}}^b{\rm Tr}_{V_p^{\otimes\, m}}\Big(P_{R\to (r,s)lk}E^{(1)}_{aa}P_{T\to (t,u)lm}E^{(1)}_{bb}\Big)\, .
$$
We now need to evaluate the traces over ${V_p^{\otimes\, m}}$.
Towards this end, write the projector as
$$
  p_{R\to (r,s)ij}= \sum_{a=1}^{d_s} \left| M_s^i,a\right\rangle\left\langle M_s^j,a\right| \, .
$$
$M_s^i$ and $M_s^j$ label states from $U(p)$ irreducible representation $s$ which have the same $\Delta$ weight. The indices $i,j$ range
from $1,...,I(\Delta (M))$. Index $a$ is a multiplicity index that, as a consequence of Schur-Weyl duality, is organized by
representation $s$ of the symmetric group $S_m$. To evaluate the traces over ${V_p^{\otimes\, m}}$ we need to allow $E^{(1)}_{kk}$ 
to act on the state $\left| M_s^i,a\right\rangle$. The state $\left| M_s^i,a\right\rangle$ was obtained by taking a tensor product of $m$ 
copies (one for each slot) of the fundamental representation of $U(p)$. It is possible and useful to rewrite this state as a linear
combination of states which are each the tensor product of the fundamental representation for the first slot with a state
obtained by taking the tensor product of states of the remaining $m-1$ slots. This is a useful thing to do because then
$E^{(1)}_{kk}$ has a particularly simple action on each state in the linear combination. Towards this end we can write
(in the following ${\bf 0}$ stands for a string of $p-1$ $0$s)
$$
  \left| M_s^i,a\right\rangle =\sum_{M_{s'},M_{1{\bf 0}}} C^{M_s^i}_{M_{s'},M_{1{\bf 0}}} 
              \left|M_{1{\bf 0}}\right\rangle\otimes\left|M_{s'},b\right\rangle
$$
where $M_{1{\bf 0}}$ indexes states in the carrier space of the fundamental representation and $C^{M_s, i}_{M_1,M_{1{\bf 0}}}$ are 
the Clebsch Gordan coefficients (discussed in detail in Appendix \ref{Clebschs})
$$
  C^{M_s^i}_{M_{s'},M_{1{\bf 0}}}=
    \left(\left\langle M_{1{\bf 0}}\right|\otimes\left\langle M_{s'},b\right|\right) \left| M_s^i ,a\right\rangle\, .
$$
$s'$ is obtained by removing a single box from $s$. By appealing to the Schur-Weyl duality which organizes the space ${V_p^{\otimes\, m-1}}$,
we know that the multiplicity index $b$ of the state $\left|M_{s'},b\right\rangle$ is organized by the irreducible representation $s'$ of
$S_{m-1}$. This allows us to easily evaluate the action of $E^{(1)}_{kk}$: it simply projects onto the state corresponding to box 1 sitting in 
the $k^{\rm th}$ row. Evaluating the traces over ${V_p^{\otimes\, m}}$ is now straight forward.

\subsection{Long Columns}
\label{longcolmns}

Our computation of the action of the dilatation operator for restricted Schur polynomials labeled by Young diagrams that
have a total of $p$ long rows has made extensive use of the fact that we can organize the space of partially labeled
Young diagrams into $S_n\times S_m$ irreducible representations $(r,s)$ by appealing to Schur-Weyl duality. We have already 
argued that it is also possible to perform this organization when considering restricted Schur polynomials labeled by Young 
diagrams that have a total of $p$ long columns - all that is required is that we fine tune a few phases in our map between 
partially labeled Young diagrams and vectors in $V^{\otimes m}_p$. The same irreducible representations of $U(p)$ are
used for both of these organizations, and further since $d_s =d_{s^T}$, each $U(p)$ representation $s$ appears with the same
multiplicity in these two cases\footnote{Recall that $s^T$ is obtained by exchanging rows and columns in $s$.}. Consequently, 
the traces computed in the last subsection for labels with $p$ long rows
are equal to the values for labels with $p$ long columns. To obtain the action of the dilatation operator all that remains
is the computation of the coefficient discussed in \ref{dilopco}. The only quantity appearing in \ref{dilopco} which is not 
invariant under exchanging rows and columns is
$$
  c_{RR'} \sqrt{f_T \over f_R }=\sqrt{c_{RR'}c_{TT'}}
$$
This factor is the only difference between the case of $p$ long rows and $p$ long columns.
Consequently, the action of the dilatation operator on restricted Schur polynomials with $p$
long columns is obtained from its action on restricted Schur polynomials with $p$ long rows
by making substitutions of the form $N+b\to N-b$. For concrete examples of this substitution 
see the end of sections \ref{tworows} and \ref{threerows}. This generalizes the two row/column relation 
observed in \cite{bhw} to an arbitrary number of rows and columns.

This completes the evaluation of the action of the dilatation operator.

\section{Explicit Formulas for the Dilatation Operator}
\label{explicitdilatationoperator}

In this Appendix we evaluate the matrix elements $N_{R,(r,s)jk;T,(t,u)lm}$
of the dilatation operator, for the case that the Young diagram labels have either two or three rows or columns.

\subsection{Young Diagrams with Two Rows or Columns}
\label{tworows}

In this case, we will be using $U(2)$ representation theory. The Gelfand-Tsetlin patterns are extremely useful for
understanding the structure of the carrier space of a particular $U(2)$ representation. However, the betweenness conditions
make it awkward to work directly with the labels $m_{ij}$ which appear in the pattern. For this reason we will employ a new
notation: trade the $m_{ij}$ for $j,j^3$ specified by
$$
{\footnotesize
\left[ 
\begin {array}{ccc} 
    m_{12}  &            &    m_{22}  \\\noalign{\medskip}
            &   m_{11}   &           
\end {array} \right] =
\left[ 
\begin {array}{ccc} 
    m_{22}+2j  &            &    m_{22}  \\\noalign{\medskip}
            &   m_{22}+j^3+j   &           
\end {array} \right]\, .}
$$
The new labels are just the familiar angular momenta we usually use for $SU(2)$. It looks as if this trade in labels is not
well defined because we have traded three labels $m_{12},m_{22},m_{11}$ for two labels $j,j^{3}$. There is no need for concern:
recall that $m$ is fixed. Further,
$$
  m=2(m_{22}+j)
$$
so that knowing $j,j^3$ and $m$ we can indeed reconstruct $m_{12},m_{22},m_{11}$. The benefit of the new labels is that the
betweenness conditions are replaced by
$$ j=0,{1\over 2},1,{3\over 2},2,...\qquad -j\le j^3\le j
$$
which are significantly easier to handle. Write our states as kets $|j,j^3\rangle$. The Clebsch-Gordan coefficients we
need are (its simple to compute these using Appendix \ref{Clebschs})
{\footnotesize
$$
  \left\langle j-{1\over 2},j^{3}-{1\over 2}; {1\over 2},{1\over 2}|j,j^{3}\right\rangle =\sqrt{j+j^3\over 2j}\, ,\quad
%
  \left\langle j+{1\over 2},j^{3}-{1\over 2}; {1\over 2},{1\over 2}|j,j^{3}\right\rangle =-\sqrt{j-j^3+1\over 2(j+1)}\, ,
$$
$$
  \left\langle j-{1\over 2},j^{3}+{1\over 2}; {1\over 2},-{1\over 2}|j,j^{3}\right\rangle =\sqrt{j-j^3\over 2j}\, ,\quad
%
  \left\langle j+{1\over 2},j^{3}+{1\over 2}; {1\over 2},-{1\over 2}|j,j^{3}\right\rangle =\sqrt{j+j^3+1\over 2(j+1)}\, .
$$
}

Consider first the case of two rows. To specify $r$ we will specify the number of columns with $2$ boxes ($=b_0$) and the number
of columns with a single box ($=b_1$). Thus, our operators are labeled as $O(b_0,b_1,j,j^3)$.
We will evaluate the diagonal terms (that is, the terms that don't change the value of $j$) in detail and simply quote the complete result. 
To compute the diagonal term in the dilatation operator we need to evaluate
\begin{equation}
-{2g_{YM}^2 c_{RR'}r_k m\over R_k d_s}\sum_{s'} d_{s'}\left[
 (C^{M_s}_{M_{s'},M^k_{1 0}})^2 - (C^{M_s}_{M_{s'},M^k_{1 0}})^4 \right]\delta_{jl}\delta_{iq}\, .\
\label{diagelem}
\end{equation}
For the case of two rows, there are no multiplicity labels and further for each $s'$ only a single state contributes, so that
there is no sum over $M_{s'}$. Consider the contribution obtained when $R'$ is related to $R$ by removing a box from the first
row of $R$. In this case
$$
  c_{RR'}=(N+b_0+b_1)\left( 1+O\left({n_1\over N+b_0+b_1}\right)\right),\qquad
 {r_1\over R_1}=1+O\left({n_1\over b_0+b_1}\right)
$$
and
$$
  M^1_{1 0}\leftrightarrow \left| {1\over 2},{1\over 2}\right\rangle,\qquad 
  M_s   \leftrightarrow \left| j,j^3\right\rangle\, .
$$
When we pull a box from the first row of $s$ to obtain $s'$ we have
$$
  m{d_{s'}\over d_s}={{\rm hooks}_s\over {\rm hooks}_{s'}}={2j\over 2j+1}{m+2j+2\over 2},\qquad
  M_{s'}=\left| j-{1\over 2},j^3-{1\over 2}\right\rangle\, .
$$
When we pull a box from the second row of $s$ to obtain $s'$ we have
$$
  m{d_{s'}\over d_s}={{\rm hooks}_s\over {\rm hooks}_{s'}}={2j+2\over 2j+1}{m-2j\over 2},\qquad
  M_{s'}=\left| j+{1\over 2},j^3-{1\over 2}\right\rangle\, .
$$
It is now a simple matter to show that (\ref{diagelem}) evaluates to
\begin{equation}
-{g_{YM}^2\over 2}\left( m-{(m+2)(j^3)^2\over j(j+1)}\right)
\label{firsttworow}
\end{equation}

\begin{figure}[h]
          \centering
          {\epsfig{file=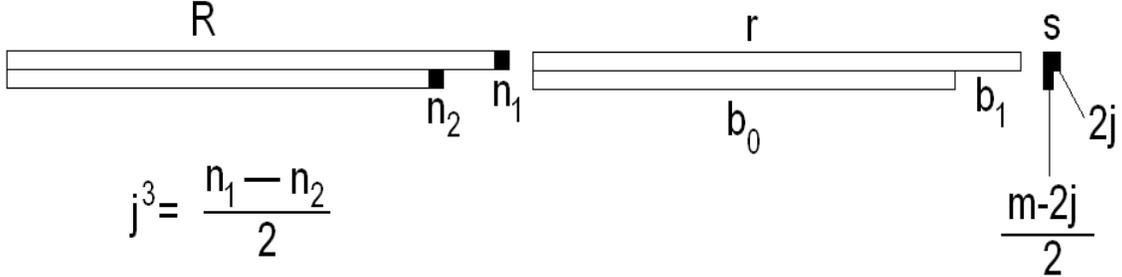,width=15.0cm,height=4.0cm}}
          \caption{This figure summarizes how to translate between the original Young Diagram labeling $O_{R,(r,s)}$ and the new
                   $O(b_0,b_1,j,j^3)$ labeling. The boxes that must be removed from $R$ to obtain $r$ have been colored black. 
                   The number of boxes to be removed from the $i^{\rm th}$ row of $R$ to obtain $r$
                   is denoted $n_i$. The label $j^3={n_1-n_2\over 2}$. In addition, $m=n_1+n_2$. The number of columns in $r$ with
                   2 boxes is $b_0$ and the number of columns with 1 box is $b_1$. The number of columns in $s$ with 2 boxes
                   is given by ${m-2j\over 2}$ and the number of columns with one box is $2j$.} \label{fig:twocolumn}
 \end{figure}

The second contribution to the diagonal terms is obtained when $R\ne T$, in which case we need to evaluate
\begin{equation}
{2g_{YM}^2\sqrt{c_{RR'}c_{TT'}}\sqrt{r_w t_x}m\over \sqrt{R_w T_x}d_u}
\sum_{s'} d_{s'}
(C^{M_s}_{\tilde{M}_{s'},M^2_{1{\bf 0}}})^2 (C^{M_s}_{M_{s'},M^1_{1 0}})^2 \, .
\label{trickyterm}
\end{equation}
When $s'$ is obtained by removing a box from the first row of $s$ we computed $m{d_{s'}\over d_s}$ above and we have
$$
 (C^{M_s}_{\tilde{M}_{s'},M^2_{1{\bf 0}}})^2 (C^{M_s}_{M_{s'},M^1_{1 0}})^2
  = \left\langle j-{1\over 2},j^{3}-{1\over 2}; {1\over 2},{1\over 2}|j,j^{3}\right\rangle^2
    \left\langle j-{1\over 2},j^{3}+{1\over 2}; {1\over 2},-{1\over 2}|j,j^{3}\right\rangle ^2\, .
$$
When $s'$ is obtained by removing a box from the second row of $s$ we computed $m{d_{s'}\over d_s}$ above and we have
$$
 (C^{M_s}_{\tilde{M}_{s'},M^2_{1{\bf 0}}})^2 (C^{M_s}_{M_{s'},M^1_{1 0}})^2
  = \left\langle j+{1\over 2},j^{3}-{1\over 2}; {1\over 2},{1\over 2}|j,j^{3}\right\rangle^2
    \left\langle j+{1\over 2},j^{3}+{1\over 2}; {1\over 2},-{1\over 2}|j,j^{3}\right\rangle ^2\, .
$$
It is now easy to see that (\ref{trickyterm}) evaluates to
\begin{equation}
{g_{YM}^2\over 2}\left( m-{(m+2)(j^3)^2\over j(j+1)}\right)\, .
\label{secondtworow}
\end{equation}
Notice that although they were computed in completely different ways (\ref{firsttworow}) and (\ref{secondtworow})
are identical up to a sign. It is due to ``accidents'' like this that the final dilatation operator depends only on
the combination given in (\ref{nicecombo}). Evaluating the remaining terms, it is now a simple matter to obtain
(\ref{recursionj3}).
This reproduces the result of \cite{bhw}. The fact that the dilatation operator does not change the 
$j^3$ label of the operator it acts on is a consequence of the fact that the $\Gamma(1,m+1)$ factor in $D$ ensures that the block
removed comes from the same row of $R$ and $r$ to produce $T$ and $t$ (in the term $\chi_{T,(t,s)}$ produced by the action of $D$ 
on $\chi_{R,(r,s)}$). This conclusion only follows in the approximation outlined in section \ref{simpleyoung} of Appendix \ref{Snrep}.
If we study the limit in which $j^3\ll j$ we obtain the significantly simpler result
\begin{eqnarray}\label{sscndrecursion_j3}
 && DO(b_{0},b_{1},j,j^3)=g_{YM}^2 \left[-{m\over 2}\Delta O(b_{0},b_{1},j,j^3)\right.\nonumber \\
&&\left. +{\sqrt{(m + 2j)(m - 2j)}\over 4}\left( \Delta O(b_{0},b_{1},j+1,j^3) + \Delta O(b_{0},b_{1},j-1,j^3)\right)
\right]
\end{eqnarray}
The system (\ref{sscndrecursion_j3}) retains the essential feature that it is again equivalent to a set of decoupled oscillators. 
When generalizing to $p>2$ rows, it is straight forward to compute the analog of (\ref{recursionj3}). The resulting expressions are quite 
lengthy and difficult to interpret. For that reason, we will focus on simplified expressions which are the analog of (\ref{sscndrecursion_j3}).
This completes our evaluation of the dilatation operator for two rows. 

Using the results of section \ref{longcolmns} we can immediately obtain the action of the dilatation operator
on restricted Schur polynomials with $p$ long columns. Transpose the Young diagram labels. In this case, for example,
the number of rows in $r$ with 2 boxes is $b_0$ and the number of rows with 1 box is $b_1$, while the number of rows 
in $s$ with 2 boxes is given by ${m-2j\over 2}$ and the number of rows with one box is $2j$. Denote the corresponding
normalized operators by $Q(b_0,b_1,j,j^3)$. The action of the dilatation
operator in this case is given in (\ref{recursionjj3}) where $\Delta Q(b_{0},b_{1},j,j^3)$ is defined in (\ref{secondcombo}).

\subsection{Young Diagrams with Three Rows or Columns}
\label{threerows}

In this case, we will be using $U(3)$ representation theory. It is again useful to trade the $m_{ij}$ appearing in the
Gelfand-Tsetlin patterns for a new set of labels $j,k,j^3,k^3,l^3$ specified by
{\footnotesize
$$
\left[ 
\begin {array}{ccccc} 
    m_{13}  &            &    m_{23}  &          &   m_{33}\\\noalign{\medskip}
            &   m_{12}   &            & m_{22}   &         \\\noalign{\medskip}
            &            &   m_{11}   &          &
\end {array} \right] =
\left[ 
\begin {array}{ccccc} 
 j+k+m_{33} &              &   k+m_{33}     &            &   m_{33}\\\noalign{\medskip}
            & j^3+k+m_{33} &                & k^3+m_{33} &         \\\noalign{\medskip}
            &              & l^3+k^3+m_{33} &            &
\end {array} \right]\, .
$$
}
It again looks like we are trading $5$ variables for $6$. However, we can again recover the value of $m_{33}$
from the value of $m$ using
$$
  m=3m_{33}+2k+j\, .
$$
The variables satisfy
$$
  j\ge 0,\quad k\ge 0,\quad j\ge j^3\ge 0,\quad k\ge k^3\ge 0,\quad k+j^3-k^3\ge l^3\ge 0,
$$
which are again much easier to handle than the betweenness conditions. We will write our states as kets
$|j,k,j^3,k^3,l^3\rangle$. The Clebsch-Gordan coefficients we will need are (its simple to compute these using
Appendix \ref{Clebschs})
{\footnotesize
$$
\left\langle j-1,k,j^3,k^3,l^3;m_1|j,k,j^3,k^3,l^3\right\rangle =
\sqrt{(j-j^3)(j+k-k^3+1)\over j(j+k+1)},
$$
$$
\left\langle j+1,k-1,j^3+1,k^3,l^3;m_1|j,k,j^3,k^3,l^3\right\rangle =
\sqrt{(j^3+1)(k-k^3)\over k(j+2)},
$$
$$
\left\langle j,k+1,j^3,k^3+1,l^3;m_1|j,k,j^3,k^3,l^3\right\rangle =
\sqrt{(k^3+1)(k+j^3+2)\over (j+k+3)(k+2)},
$$
$$
\left\langle j-1,k,j^3-1,k^3,l^3;m_2|j,k,j^3,k^3,l^3\right\rangle =
\sqrt{(j+k-k^3+1)j^3(k+j^3+1)(j^3-k^3-l^3+k)\over j(j+k+1)(k+j^3-k^3+1)(j^3+k-k^3)},
$$
$$
\left\langle j-1,k,j^3,k^3-1,l^3+1;m_2|j,k,j^3,k^3,l^3\right\rangle =
\sqrt{(j-j^3)(k-k^3+1)k^3(k+j^3-k^3-l^3+1)\over j(j+k+1)(k+j^3-k^3+1)(k+j^3-k^3+2)},
$$
$$
\left\langle j+1,k-1,j^3,k^3,l^3;m_2|j,k,j^3,k^3,l^3\right\rangle =
-\sqrt{(k-k^3)(j-j^3+1)(k+j^3+1)(k+j^3-k^3-l^3)\over (j+2)k(j^3+k-k^3+1)(k+j^3-k^3)},
$$
$$
\left\langle j+1,k-1,j^3+1,k^3-1,l^3+1;m_2|j,k,j^3,k^3,l^3\right\rangle =
\sqrt{(j^3+1)(j+k-k^3+2)k^3(k+j^3-k^3-l^3+1)\over (j+2)k(k+j^3-k^3+1)(k+j^3-k^3+2)},
$$
$$
\left\langle j,k+1,j^3-1,k^3+1,l^3;m_2|j,k,j^3,k^3,l^3\right\rangle =
-\sqrt{(k^3+1)(j-j^3+1)j^3(k+j^3-k^3-l^3)\over (j+k+3)(k+2)(k+j^3-k^3+1)(k+j^3-k^3)},
$$
$$
\left\langle j,k+1,j^3,k^3,l^3+1;m_2|j,k,j^3,k^3,l^3\right\rangle =
$$
$$
-\sqrt{(k+j^3+2)(j+k-k^3+2)(k-k^3+1)(k+j^3-k^3-l^3+1)\over (j+k+3)(k+2)(k+j^3-k^3+1)(k+j^3-k^3+2)},
$$
$$
\left\langle j-1,k,j^3-1,k^3,l^3-1;m_3|j,k,j^3,k^3,l^3\right\rangle =
\sqrt{(j+k-k^3+1)j^3(k+j^3+1)l^3\over j(j+k+1)(k+j^3-k^3+1)(j^3+k-k^3)},
$$
$$
\left\langle j-1,k,j^3,k^3-1,l^3;m_3|j,k,j^3,k^3,l^3\right\rangle =
-\sqrt{(j-j^3)(k-k^3+1)k^3(l^3+1)\over j(j+k+1)(k+j^3-k^3+1)(k+j^3-k^3+2)},
$$
$$
\left\langle j+1,k-1,j^3,k^3,l^3-1;m_3|j,k,j^3,k^3,l^3\right\rangle =
-\sqrt{(k-k^3)(j-j^3+1)(k+j^3+1)l^3\over (j+2)k(j^3+k-k^3+1)(k+j^3-k^3)},
$$
$$
\left\langle j+1,k-1,j^3+1,k^3-1,l^3;m_3|j,k,j^3,k^3,l^3\right\rangle =
-\sqrt{(j^3+1)(j+k-k^3+2)k^3(l^3+1)\over (j+2)k(k+j^3-k^3+1)(k+j^3-k^3+2)},
$$
$$
\left\langle j,k+1,j^3-1,k^3+1,l^3-1;m_3|j,k,j^3,k^3,l^3\right\rangle =
-\sqrt{(k^3+1)(j-j^3+1)j^3 l^3\over (j+k+3)(k+2)(k+j^3-k^3+1)(k+j^3-k^3)},
$$
$$
\left\langle j,k+1,j^3,k^3,l^3;m_3|j,k,j^3,k^3,l^3\right\rangle =
\sqrt{(k+j^3+2)(j+k-k^3+2)(k-k^3+1)(l^3+1)\over (j+k+3)(k+2)(k+j^3-k^3+1)(k+j^3-k^3+2)},
$$
}
where
$$
 m_1=1,0,0,0,0,\quad m_2=1,0,1,0,0,\quad m_3=1,0,1,0,1\,.
$$

Consider first the case of three rows. To specify $r$ we specify the number of columns with three boxes ($b_0$), the number of columns with
two boxes ($b_1$) and the number of columns with a single box ($b_2$). Thus, our operators $O(b_1,b_2,j,k,j^3,k^3,l^3)$ carry seven labels.
To simplify the notation a little
we do not explicitly display $b_0$ since it is fixed once $b_1$ and $b_2$ are chosen by $b_0=(n-b_2-2b_1)/3$.
To obtain $r$ from $R$ we remove $n_i$ boxes from each row where
$$
  n_1={m+2j+k-3k^3-3j^3\over 3},\qquad n_2={m+k-j+3j^3-3l^3\over 3},
$$
$$
  n_3={m-j-2k+3l^3+3k^3\over 3}\, .
$$
We can read $j$, $k$ and $m$ directly from the Young diagram label $s$. One might have thought that by employing the above expressions for the
$n_i$ one could obtain a formula for $j^3,k^3,l^3$ in terms of the $n_i$. This is not possible. Indeed, this conclusion follows immediately
upon noting that
$$
  n_1+n_2+n_3 =m\, .
$$
\begin{figure}[h]
          \centering
          {\epsfig{file=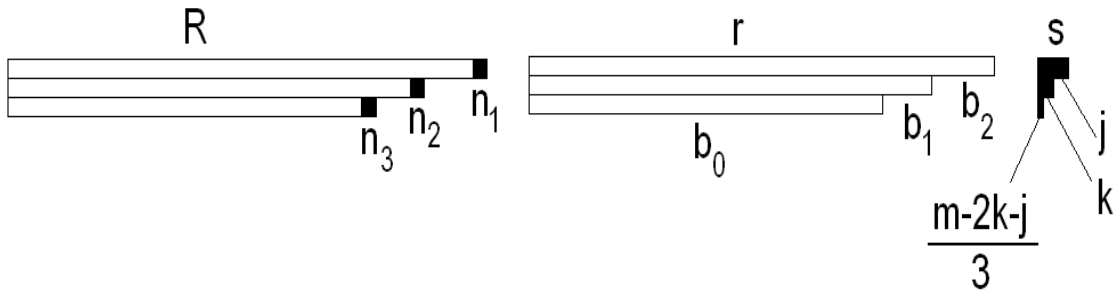,width=15.0cm,height=4.0cm}}
          \caption{This figure summarizes how to translate between the original Young Diagram labeling $O_{R,(r,s)}$ and the new
                   $O(b_1,b_2,j,k,j^3,k^3,l^3)$ labeling. The boxes that must be removed from $R$ to obtain $r$ have been colored black. 
                   The number of boxes to be removed from the $i^{\rm th}$ row of $R$ to obtain $r$
                   is denoted $n_i$. We have $m=n_1+n_2+n_3$. The number of columns in $r$ with
                   3 boxes is $b_0$, the number of columns with 2 boxes is $b_1$ and the number of columns with 1 box is $b_2$. The number 
                   of columns in $s$ with 3 boxes is given by ${m-j-2k\over 3}$, the number of columns with two boxes is $k$ and the number
                   of columns with one box is $j$.}
 \end{figure}
The reason why it is not possible to express $j^3,k^3,l^3$ in terms of the $n_i$, is simply that in all situations
where the inner multiplicity is greater than 1, there is no unique $j^3,k^3,l^3$ given the $n_i$. 
Recall that the dilatation operator, when acting on restricted Schur polynomials labeled by Young diagrams with two rows,
preserved the $j^3$ label of the operator. What is the corresponding statement that would be valid for any number of rows? In general, 
the dilatation operator preserves the $\Delta$ weight of the operator it acts on. In the two row case, preserving $j^3$ is equivalent 
to preserving the $\Delta$ weight. Further, the reason why the $\Delta$ weight is preserved can again be traced back to the factors of 
$\Gamma(1,m+1)$ appearing in the dilatation operator and again this conclusion only follows in the approximation outlined in section 
\ref{simpleyoung} of Appendix \ref{Snrep}. 
For the case of three rows it is simple to give this inner multiplicity a nice characterization: States that belong
to the same inner multiplicity multiplet
\begin{itemize}
\item{}Have the same first row in their Gelfand-Tsetlin pattern because they belong to the same $U(3)$ irreducible representation.
\item{}Have the same last row because the $\Delta$ weight is conserved.
\item{}Have the same sum of numbers in the second row of the Gelfand-Tsetlin pattern again because the $\Delta$ weight is conserved.
\end{itemize} 
This implies that states in the same inner multiplet can be written as
$$
{\footnotesize
\left[ 
\begin {array}{ccccc} 
    m_{13}  &            &    m_{23}  &          &   m_{33}\\\noalign{\medskip}
            &   m_{12}-i &            & m_{22}+i &         \\\noalign{\medskip}
            &            &   m_{11}   &          &
\end {array} \right]
}
$$
with different values of $i$ giving the different states, and that the number of states in the inner multiplet is
$$ N= {\rm max}(m_{12}-m_{11},m_{12}-m_{23},{m_{12}-m_{22}\over 2})+{\rm min}(m_{13}-m_{12},m_{22}-m_{33})+1\, ,$$
where ${\rm max}(a,b,c)$ means take the largest of $a,b,c$ and ${\rm min}(a,b)$ means take the smallest of $a,b$.

In the case of two rows, we saw that the action of the dilatation operator could be expressed entirely in terms of the combination
$\Delta O(b_0,b_1,j,j^3)$. There is a generalization of this result for $p=3$ rows: after applying the dilatation operator we only
obtain the linear combinations\footnote{The combination $\Delta_{ij}$ is relevant for terms in the dilatation operator which allow
a box to move between rows $i$ and $j$.} $\Delta_{12} O(b_1,b_2,j,k,j^3,k^3,l^3)$ (defined in (\ref{D12}))
$\Delta_{13} O(b_1,b_2,j,k,j^3,k^3,l^3)$ (defined in (\ref{D13})) and $\Delta_{23} O(b_1,b_2,j,k,j^3,k^3,l^3)$ 
(defined in (\ref{D23})).

To see how this comes about, consider for example the off diagonal terms in the dilatation operator.
%
%
The terms multiplying (as an example) {\small $(N+b_0+b_1+b_2)$} come multiplied by
$$
{\small  \left\langle M_1| E_{11} |M_2\right\rangle \left\langle M_2| E_{11} |M_1\right\rangle\, ,}
$$
the terms multiplying {\small $\sqrt{(N+b_0+b_1+b_2)(N+b_0+b_1)}$} come multiplied by
$$
{\small \left\langle M_1| E_{11} |M_2\right\rangle \left\langle M_2| E_{22} |M_1\right\rangle\, ,}
$$
and finally the terms multiplying {\small $\sqrt{(N+b_0+b_1+b_2)(N+b_0)}$} come multiplied by
$$
{\small \left\langle M_1| E_{11} |M_2\right\rangle \left\langle M_2| E_{33} |M_1\right\rangle\, .}
$$
If we are to have a dependence only on the $\Delta_{ij} O(b_1,b_2,j,k,j^3,k^3,l^3)$s we need the first number above 
to be minus the sum of the second two (plus some additional conditions which follow in the same way). Using 
the identity ${\bf 1}=E_{11}+E_{22}+E_{33}$ and
$\left\langle M_1|M_2\right\rangle=0$ (for the off diagonal terms in the dilatation operator $M_1$ and $M_2$ are by definition different
states) we easily find that this is indeed the case. Note also that this argument generalizes trivially to $p>3$ rows.

{\vskip 0.4cm}

\noindent
{\bf Some Explicit Examples}

{\vskip 0.2cm}

\noindent
{\bf $\Delta=(1,1,1)$ States of the $m=3$ Sector:} By applying the above results, it is straight forward to evaluate the
action of the dilatation operator for the case that we have 3 $Y$ fields and we set $\Delta =(1,1,1)$. There are four possible
$U(3)$ states
{\footnotesize
$$
  \left|3,2,0,0,1\right\rangle\leftrightarrow
\left[ 
\begin {array}{ccccc} 
    3  &     &   0  &      &   0   \\\noalign{\medskip}
       &   2 &      &  0   &       \\\noalign{\medskip}
       &     &   1  &      &
\end {array} \right]\qquad
%
  \left|0,0,0,0,0\right\rangle\leftrightarrow
\left[ 
\begin {array}{ccccc} 
    1  &     &   1  &      &   1   \\\noalign{\medskip}
       &   1 &      &  1   &       \\\noalign{\medskip}
       &     &   1  &      &
\end {array} \right]
$$
$$
  \left|1,1,1,0,1\right\rangle\leftrightarrow
\left[ 
\begin {array}{ccccc} 
    2  &     &   1  &      &   0   \\\noalign{\medskip}
       &   2 &      &  0   &       \\\noalign{\medskip}
       &     &   1  &      &
\end {array} \right]
%
\qquad
  \left|1,1,0,1,0\right\rangle\leftrightarrow
\left[ 
\begin {array}{ccccc} 
    2  &     &   1  &      &   0   \\\noalign{\medskip}
       &   1 &      &  1   &       \\\noalign{\medskip}
       &     &   1  &      &
\end {array} \right]
$$
}
This example was chosen because it is the simplest case in which we have a nontrivial inner multiplicity: indeed,
the last two states belong to an inner multiplicity multiplet. This implies that there are a total of 6 symmetric
group operators 
%
$$
P_1 = \left|3,2,0,0,1\right\rangle\left\langle 3,2,0,0,1\right|\qquad
P_2 = \left|0,0,0,0,0\right\rangle\left\langle 0,0,0,0,0\right|
$$
$$
P_3^{(1,1)} = \left|1,1,1,0,1\right\rangle\left\langle 1,1,1,0,1\right|\qquad
P_3^{(1,2)} = \left|1,1,1,0,1\right\rangle\left\langle 1,1,0,1,0\right|
$$
$$
P_3^{(2,1)} = \left|1,1,0,1,0\right\rangle\left\langle 1,1,1,0,1\right|\qquad
P_3^{(2,2)} = \left|1,1,0,1,0\right\rangle\left\langle 1,1,0,1,0\right|
$$
which define 6 restricted Schur polynomials. The corresponding normalized operators will be
denoted $O_1(b_1,b_2)$, $O_2(b_1,b_2)$, $O_3(b_1,b_2)$, $O_4(b_1,b_2)$, 
$O_5(b_1,b_2)$ and $O_6(b_1,b_2)$. 
The action of the dilatation operator is given in equation (\ref{frstexmpl}).
To obtain this result we have used the exact expressions for the Clebsch-Gordan coefficients given earlier in this subsection.

{\vskip 0.2cm}

\noindent
{\bf $j^3=O(1)$ Sector:} We assume that the remaining quantum numbers ($j,k,k^3,l^3$ and $m$) are all order $N$. The Clebsch-Gordan
coefficients simplify considerably in this limit. The non-zero Clebsch-Gordan coefficients are
{\footnotesize
$$
\left\langle j-1,k,j^3,k^3,l^3;m_1|j,k,j^3,k^3,l^3\right\rangle =\sqrt{j+k-k^3\over j+k},
$$
$$
\left\langle j,k+1,j^3,k^3+1,l^3;m_1|j,k,j^3,k^3,l^3\right\rangle = \sqrt{k^3\over j+k},
$$
$$
\left\langle j-1,k,j^3,k^3-1,l^3+1;m_2|j,k,j^3,k^3,l^3\right\rangle = \sqrt{k^3(k-k^3-l^3)\over (j+k)(k-k^3)},
$$
$$
\left\langle j+1,k-1,j^3,k^3,l^3;m_2|j,k,j^3,k^3,l^3\right\rangle = -\sqrt{k-k^3-l^3\over k-k^3},
$$
$$
\left\langle j,k+1,j^3,k^3,l^3+1;m_2|j,k,j^3,k^3,l^3\right\rangle = -\sqrt{(j+k-k^3)(k-k^3-l^3)\over (j+k)(k-k^3)},
$$
$$
\left\langle j-1,k,j^3,k^3-1,l^3;m_3|j,k,j^3,k^3,l^3\right\rangle = -\sqrt{k^3l^3\over (j+k)(k-k^3)},
$$
$$
\left\langle j+1,k-1,j^3,k^3,l^3-1;m_3|j,k,j^3,k^3,l^3\right\rangle = -\sqrt{l^3\over k-k^3},
$$
$$
\left\langle j,k+1,j^3,k^3,l^3;m_3|j,k,j^3,k^3,l^3\right\rangle = \sqrt{(j+k-k^3)l^3\over (j+k)(k-k^3)}\, .
$$
}
Looking at the non-zero Clebsch-Gordan coefficients, the reason for the simplification of this limit is clear. Indeed, notice that
in the limit that we are considering the $j^3$ quantum number is fixed. This in turn implies that a single state from each inner
multiplicity multiplet participates - a considerable simplification. Indeed, if $j,k,m$ and the $\Delta$ weight $\Delta =(n_1,n_2,n_3)$
are given, then we know
$$
  k^3={m-3n_1-3j^3+2j+k\over 3},\qquad l^3={m-3n_2+3j^3+k-j\over 3}\, .
$$
Thus, after specifying $\Delta$ and $j^3$ the $k^3,l^3$ labels are not needed. For this reason we can now simplify the notation for our
operators to $O(b_1,b_2,j,k)$ for a given problem which is specified by $j^3$ and $\Delta$\footnote{The symmetric group operators used to
define the restricted Schur polynomials are $P=\sum |j,k,j^3,k^3,l^3\rangle\langle j,k,j^{3\prime},k^{3\prime},l^{3\prime}|$ where we could
have $j^3\ne j^{3\prime}$, $k^3\ne k^{3\prime}$, $l^3\ne l^{3\prime}$. For simplicity we consider only the $j^3=j^{3\prime}$ case. It is a
simple extension of our analysis to consider the general case.}. The action of the dilatation operator is
{\footnotesize
$$
  DO(b_1,b_2,j,k)=-g_{YM}^2\left[
  {k^3(j+k-k^3)(k-k^3-l^3)(2m+j-k)\over 3(j+k)^2 (k-k^3)}\Delta_{12} O(b_1,b_2,j,k)\right.
$$
$$
  +{l^3 k^3 (j+k-k^3)(2m+j-k)\over 3(j+k)^2 (k-k^3)}\Delta_{13} O(b_1,b_2,j,k)
  -{l^3k^3(k-k^3-l^3)(j+k-k^3)(2m+j-k)\over 3(j+k)^2 (k-k^3)^2}\Delta_{23} O(b_1,b_2,j,k)
$$
$$
  +{l^3(k-k^3-l^3)(j+k-k^3)(2m-2j-k)\over 3(j+k) (k-k^3)^2}\Delta_{23} O(b_1,b_2,j,k)
  +{k^3l^3(k-k^3-l^3)(2m+j+2k)\over 3(j+k)(k-k^3)^2}\Delta_{23} O(b_1,b_2,j,k)
$$
$$
  -{(j+k-k^3)k^3(k-k^3-l^3)\sqrt{(m+2j+k)(m-j-2k)}\over 3(j+k)^2 (k-k^3)}\left(
  \Delta_{12} O(b_1,b_2,j-1,k-1)+\Delta_{12} O(b_1,b_2,j+1,k+1)\right)
$$
$$
  -{l^3 k^3(j+k-k^3)\sqrt{(m+2j+k)(m-j-2k)}\over 3(j+k)^2 (k-k^3)}\left(
  \Delta_{13} O(b_1,b_2,j-1,k-1)+\Delta_{13} O(b_1,b_2,j+1,k+1)\right)
$$
$$
  +{l^3 k^3(k-k^3-l^3)(j+k-k^3)\sqrt{(m+2j+k)(m-j-2k)}\over 3(j+k)^2 (k-k^3)^2}\left(
  \Delta_{23} O(b_1,b_2,j-1,k-1)+\Delta_{23} O(b_1,b_2,j+1,k+1)\right)
$$
$$
  -{l^3(k-k^3-l^3)(j+k-k^3)\sqrt{(m-j-2k)(m-j+k)}\over 3(j+k)(k-k^3)^2}\left(
  \Delta_{23} O(b_1,b_2,j+1,k-2)+\Delta_{23} O(b_1,b_2,j-1,k+2)\right)
$$
$$
  \left. -{l^3 k^3(k-k^3-l^3)\sqrt{(m+2j+k)(m-j+k)}\over 3(j+k)(k-k^3)^2}\left(
  \Delta_{23} O(b_1,b_2,j-2,k+1)+\Delta_{23} O(b_1,b_2,j+2,k-1)\right)\right]
$$
}

\section{Recursion Relations}
\label{labelrelations}

The recursion relations needed in the diagonalization of the dilatation operator acting on restricted Schur polynomials labeled with two rows/columns
are
$$
- x {}_2F_1\left({}^{-n,-x}_{-N}\Big| {1\over p}\right) =
p(N-n) {}_2F_1\left({}^{-n-1,-x}_{-N}\Big| {1\over p}\right)
-[p(N-n)+n(1-p)]{}_2F_1\left({}^{-n,-x}_{-N}\Big| {1\over p}\right)
$$
$$
+n(1-p){}_2F_1\left({}^{-n+1,-x}_{-N}\Big| {1\over p}\right)
$$
and
{\small
$$
p{}_3F_2\left({}^{j^3-j,j+1+j^3,-p}_{1,j^3-{m\over 2}}\Big|1\right)=
{(j+j^3 + 1)(j-j^3 + 1)(m-2j) \over 2(j + 1)(2j+1)}
{}_3F_2\left({}^{-1+j^3-j,j+2+j^3,-p}_{1,j^3-{m\over 2}}\Big|1\right)
$$
$$
-\left( {m\over 2}-{(m+2)(j^3)^2\over 2j(j+1)}\right)
{}_3F_2\left({}^{j^3-j,j+1+j^3,-p}_{1,j^3-{m\over 2}}\Big|1\right)
+{(j+j^3 )(j-j^3 )(m+2j+2) \over 2 j(2j+1)}
{}_3F_2\left({}^{1+j^3-j,j+j^3,-p}_{1,j^3-{m\over 2}}\Big|1\right)
$$
}
The first relation is equation (1.10.3) in \cite{HahnRecur} and is used to obtain the $f(b_0,b_1)$.
The second relation is equivalent to equation (1.5.3) in \cite{HahnRecur} and is used to obtain the $C_{p,j^3}(j)$.

\section{Gauss Law Example}
\label{gausslawexample}

\begin{figure}[h]
          \centering
          {\epsfig{file=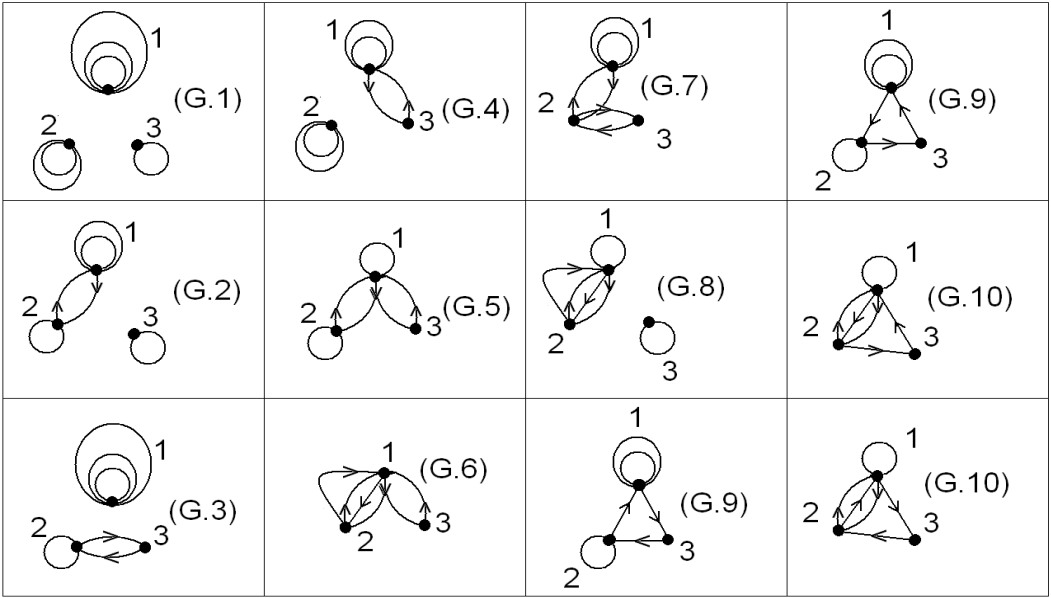,width=14.0cm,height=8.0cm}}
          \caption{The open string configurations consistent with the Gauss Law for a three giant system with $\Delta$ weight
                   $\Delta=(3,2,1)$. The figure labels match the corresponding equation.}\label{fig:glcs}
 \end{figure}

In this Appendix we will report the result of the computation of the action of the dilatation operator for restricted
Schur polynomials with three rows and $\Delta=(3,2,1)$. There are a total of 60 states that can be obtained by removing
6 boxes as specified by the $\Delta$ weight. The 6 $S_6$ irreducible representations that can be suduced are
{\tiny
$$
  \yng(3,2,1)\quad \yng(3,3)\quad \yng(4,1,1)\quad \yng(6)\quad\yng(4,2)\quad \yng(5,1)
$$
}
with the last two irreducible representations being suduced twice. Thus, there are a total of 12 operators that can be 
defined. After diagonalizing the action of the dilatation operator we find
\begin{equation}
   DO=0
\end{equation}
\begin{equation}
   DO=-2g_{YM}^2\Delta_{12}O
\end{equation}
\begin{equation}
   DO=-2g_{YM}^2\Delta_{23}O
\end{equation}
\begin{equation}
   DO=-2g_{YM}^2\Delta_{13}O
\end{equation}
\begin{equation}
   DO=-2g_{YM}^2(\Delta_{12}+\Delta_{13})O
\end{equation}
\begin{equation}
   DO=-2g_{YM}^2(2\Delta_{12}+\Delta_{13})O
\end{equation}
\begin{equation}
   DO=-2g_{YM}^2(\Delta_{12}+\Delta_{23})O
\end{equation}
\begin{equation}
   DO=-4g_{YM}^2\Delta_{12}O
\end{equation}
\begin{equation}
   DO=-g_{YM}^2(\Delta_{12}+\Delta_{13}+\Delta_{23})O
\end{equation}
\begin{equation}
   DO=-g_{YM}^2(\Delta_{13}+3\Delta_{12}+\Delta_{23})O
\end{equation}
The last two equations each appear twice. The corresponding diagrams are shown in figure \ref{fig:glcs}.

\section{Continuum Limit}
\label{continuumlimit}

In this section we will study the action of $\Delta_{ij}$ on a Young diagram with $p$ rows. The row closest to the top of the page
is row 1 and the row closest to the bottom of the page is row $p$. The number of boxes in row $i$ minus the number of boxes in
row $i+1$ is given by $b_{p-i}$. $\Delta_{ij}$ exchanges boxes between rows $i$ and $j$; we always have $i\ne j$. If
$|i-j|>1$ we have
$$
  \Delta_{ij}O(b_0,...,b_{p-1})=-(2N+\sum_{k=0}^{p-j}b_k+\sum_{q=0}^{p-i}b_q)O(b_0,...,b_{p-1})
$$
$$
+\sqrt{(N+\sum_{k=0}^{p-j}b_k)(N+\sum_{q=0}^{p-i}b_q)}\left[
O(b_0,...,b_{p-j}-1,b_{p-j+1}+1,...,b_{p-i}+1,b_{p-i+1}-1,...,b_{p-1})\right.
$$
$$
\left.
+ O(b_0,...,b_{p-j}+1,b_{p-j+1}-1,...,b_{p-i}-1,b_{p-i+1}+1,...,b_{p-1})
\right]
$$
It proves convenient to introduce the variables
$$
  l_i=\sum_{k=1}^{p-i}b_k \qquad i=1,2,...,p-1 .
$$
Making the ansatz
$$
  O=\sum_{b_0,l_i,...,l_{p-1}} f(b_0,l_1,...,l_{p-1})O(b_0,l_1,...,l_{p-1})
$$
for operators of a good scaling dimension, we find
$$
  \Delta_{ij} O = \sum_{b_0,l_i,...,l_{p-1}} f(b_0,l_1,...,l_{p-1})\Delta_{ij} O(b_0,l_1,...,l_{p-1})
                = \sum_{b_0,l_i,...,l_{p-1}}\tilde{\Delta}_{ij} f(b_0,l_1,...,l_{p-1}) O(b_0,l_1,...,l_{p-1})
$$
where\footnote{As the reader can easily check, this formula is also true when $|i-j|=1$ i.e. its completely general.}
$$
  \tilde{\Delta}_{ij} f(b_0,l_1,...,l_{p-1}) = -(2N+2b_0+l_i+l_j)f(b_0,l_1,...,l_{p-1})
$$
$$
-\sqrt{(N+b_0+l_i)(N+b_0+l_j)}\left[
f(b_0,...,l_i-1,...,l_j+1,...,l_{p-1}) + f(b_0,...,l_i+1,...,l_j-1,...,l_{p-1})
\right]\, .
$$
The continuum limit we consider takes $N+b_0\to\infty$ holding the variables
$$
  x_i={l_i\over\sqrt{N+b_0}}
$$
fixed. Using the expansions
$$
  \sqrt{(N+b_0+l_i)(N+b_0+l_j)}
%
    =N+b_0 +{x_i+x_j\over 2}\sqrt{N+b_0}-{(x_i-x_j)^2\over 8}+...
$$
and
$$
  f(b_0,...,l_i-1,...,l_j+1,...)\to f(b_0,...,x_i-{1\over\sqrt{N+b_0}},...,x_j-{1\over\sqrt{N+b_0}},...)
$$
$$
  =f(b_0,...,l_i,...,l_j,...)-{1\over\sqrt{N+b_0}}{\partial f\over\partial x_i}+{1\over\sqrt{N+b_0}}{\partial f\over\partial x_j}
+{1\over 2(N+b_0)}{\partial^2 f\over\partial x_i^2}
$$
$$
+{1\over 2(N+b_0)}{\partial^2 f\over\partial x_j^2}
-{1\over N+b_0}{\partial^2 f\over\partial x_i\partial x_j}+...
$$
we find that in the continuum limit we have
$$
  \tilde{\Delta}_{ij}f =\left({\partial\over\partial x_i}-{\partial\over\partial x_j}\right)^2 f -{(x_i-x_j)^2\over 4}f
                       =m_{ab}\left({\partial\over\partial x_a}{\partial\over\partial x_b} -{x_a x_b \over 4}\right) f
\, ,
$$
where
$$
  m_{ab}=\delta_{ai}\delta_{bi}+\delta_{aj}\delta_{bj}-\delta_{ai}\delta_{bj}-\delta_{aj}\delta_{bi}\, .
$$

In general, the action of the dilatation operator is given by summing a collection
of operators $\Delta_{ij}$, each appearing some integer $n_{ij}$ number of times
$$
   DO(b_1,b_2)=-g_{YM}^2\sum_{ij}\, n_{ij}\Delta_{ij}\, O(b_1,b_2)\, .
$$
The result that we obtained above implies that in the continuum limit we have
$$
  \sum_{ij}\, n_{ij}\Delta_{ij}\to M_{ab}\left({\partial\over\partial x_a}{\partial\over\partial x_b} -{x_a x_b \over 4}\right)\, ,
$$
where the explicite formula for $M_{ab}$ depends on the $n_{ij}$. 
In terms of the orthogonal matrix $V$ that diagonalizes $M$
$$
  V_{ik}M_{ij}V_{jl}=D_k\delta_{kl}
$$
we define the new variable $y_k=V_{ik}x_i$. Written in terms of the new $y$ variables we have
$$
  \sum_{ij}\, n_{ij}\Delta_{ij}\to \sum_a D_a \left({\partial^2\over\partial y_a^2} -{y_a^2\over 4}\right)\, ,
$$
which is (minus) the Hamiltonian of a set of decoupled oscillators. The $D_a$'s, which are the eigenvalues of $M$,
set the frequencies of the oscillators. For 
$$
  \sum_{ij}\, n_{ij}\Delta_{ij}=2\Delta_{12}\, ,
$$
we have
$$
  M=\left[\matrix{2 &-2\cr -2 &2}\right],\qquad D_1=0,\quad D_2=4\, .
$$
For
$$
  \sum_{ij}\, n_{ij}\Delta_{ij}=\Delta_{12}+\Delta_{23}+\Delta_{13}\, ,
$$
we have
$$
  M=\left[\matrix{2 &-1 &-1\cr -1 &2 &-1\cr -1 &-1 &2}\right],\qquad D_1=0,\quad D_2=3=D_3\, .
$$
These are perfectly consistent with the results given in section 4. One might wonder if the $D_i$ are always integers.
This is not the case. Indeed, for
$$
  \sum_{ij}\, n_{ij}\Delta_{ij}=\Delta_{12}+\Delta_{23}+\Delta_{34}+...+\Delta_{1d}\, ,
$$
we have
$$
  M=\left[\matrix{2 &-1 &0  &0  &\cdots &0  &-1\cr 
                 -1 &2  &-1 &0  &\cdots &0  &0 \cr 
                  0 &-1 &2  &-1 &\cdots &0  &0 \cr
                  . &.  &.  &.  &\cdots &.  &. \cr
                  . &.  &.  &.  &\cdots &.  &. \cr
                 -1 &0  &0  &0  &\cdots &-1 &2}\right]\, .
$$
In this case it is rather simple to see that the eigenvalues are
$$
  D_n=2-2\cos \left({n\pi\over d}\right),\qquad n=0,1,...,d\, .
$$
These are not, in general, integer.

\end{document}